\documentclass[12pt]{report}
\usepackage{a4,graphicx,psfrag,axodraw,amsbsy,doublespace,latexsym}
\def\ie{{\it i.e.\ }}
\def\eg{{\it e.g.\ }}
\def\etc{{\it etc.\ }}

\def\cf{{\it c.f.\ }}
\def\be{\begin{equation}}
\def\ee{\end{equation}}
\def\bea{\begin{eqnarray}}
\def\eea{\end{eqnarray}}
\def\eq#1{(\ref{#1})}

\def\q{{\bf q}}
\def\p{{\bf p}}
\def\k{{\bf k}}

\def\Pone{{\bf P}_1}
\def\Ptwo{{\bf P}_2}
\def\f{{\rm f}\,\!}
\def\g{{\rm g}\,\!}

\def\del{\delta}
\def\sig3{\sigma_3}
\def\Lam{\Lambda}
\def\lam{\lambda}
\def\Gam{\Gamma}
\def\gam{\gamma}
\def\dir{\Delta_{IR}}
\def\duv{\Delta_{UV}}
\def\clas{\phi^{cl}}

\def\half{{\textstyle{1\over2}}} 

\def\SUNN{$SU(N|N)$ }
\def\bSUNN{$\boldsymbol{SU(N|N)}$ }
\def\SUNM{$SU(N|M)$ }
\def\bSUNM{$\boldsymbol{SU(N|M)}$ }

\def\A{{\cal A}}
\def\Amu{{\cal A}_{\mu}}
\def\Bbar{\bar{B}}
\def\C{{\cal C}}
\def\Dbar{\bar{D}}
\def\H{{\cal H}}
\def\X{{\cal X}}
\def\Xdot{\dot{\cal X}}
\def\J{{\cal J}}
\def\K{{\cal K}}
\def\L{{\cal L}}
\def\P{{\cal P}}
\def\Q{{\cal Q}}
\def\S{{\cal S}}

\def\dgam{{\cal D}_{\Gamma}}

\def\c{c^{-1}}
\def\ctil{\tilde{c}^{-1}}
\def\chat{\hat{c}^{-1}}
\def\rtil{\tilde{r}}
\def\rhat{\hat{r}}

\def\vac{|0\rangle}

\def\dint{\int\!d^D\!x}
\def\Tr{{\mathbf{Tr}}}
\def\tr{{\mathrm{tr}}}
\def\str{{\mathrm{str}}}
\def\sdet{\mathrm{sdet}}
\def\wick#1#2{\hspace{-#1mm}\raisebox{-2ex}{\rule{0.02mm}{2mm}\rule{#2mm}{0.02mm}\rule{0.02mm}{2mm}}\hspace{#1mm}\hspace{-#2mm}}
\def\bigwick#1#2{\hspace{-#1mm}\raisebox{-3ex}{\rule{0.02mm}{4mm}\rule{#2mm}{0.02mm}\rule{0.02mm}{4mm}}\hspace{#1mm}\hspace{-#2mm}}
\def\one{\hbox{1\kern-.8mm l}}

\def\gap{\hspace{0.05in}}
\def\ph#1{\phantom{#1}}
\def\ds{\displaystyle}

\begin{document}

%
%
\pagestyle{empty} 
\begin{center} 
    {\LARGE UNIVERSITY OF SOUTHAMPTON} \\
\vspace{3cm} 
    {\Huge{\bf Derivative Expansions   }} \vspace{12pt} \\
    {\Huge{\bf of the }} \vspace{12pt} \\
    {\Huge{\bf Exact Renormalisation Group }} \vspace{12pt} \\
    {\Huge{\bf and }} \vspace{12pt} \\
    {\Huge{\bf \bSUNN Gauge Theory}} \vspace{12pt} \\ 
\vspace{1cm} 
    by \\
\vspace{1cm} 
    {\LARGE John Francis Tighe} \\
\vspace{1.0cm} 
    A thesis submitted for the degree of \\
\bigskip 
    Doctor of Philosophy \\
\vspace{0.7cm} 
\bigskip 
    Department of Physics and Astronomy \\
\bigskip 
July 2001
\end{center} 
\vspace{0.5cm}
%

%
\newpage 
\pagestyle{empty} 
\begin{center} 
      {\Large UNIVERSITY OF SOUTHAMPTON}  \\
\bigskip 
      \underline{\large ABSTRACT} \\
\bigskip 
      {\Large FACULTY OF SCIENCE} \\
\bigskip 
      {\Large PHYSICS} \\
\bigskip 
      \underline{\large Doctor of Philosophy} \\
\bigskip 
      {\Large Derivative Expansions  } \\
      {\Large of the Exact Renormalization Group} \\
      {\Large and \SUNN Gauge Theory} \\
\bigskip 
      {\large John Francis Tighe} \\
\end{center} 

We investigate the convergence of the derivative expansion of the
exact renormalisation group,
by using it to compute  the $\beta$
function of scalar $\lam\varphi^4$ theory.
We show that the derivative expansion of the Polchinski flow equation
converges at one loop for certain fast falling smooth cutoffs.
The derivative expansion of the Legendre flow equation trivially converges 
at one loop, but also at two loops: slowly with 
sharp cutoff (as a momentum-scale expansion), and rapidly in the case of
a smooth exponential cutoff. Finally, we show that the two loop contributions
to certain higher derivative 
operators (not involved in $\beta$) have divergent 
momentum-scale expansions for sharp cutoff, but 
the smooth exponential cutoff
gives convergent derivative expansions  
for all such operators with any number of derivatives.

In the latter part of the thesis, we address the problems of applying the
exact renormalisation group to gauge theories.  A regularisation scheme
utilising higher covariant derivatives and the spontaneous symmetry
breaking of the gauge supergroup \SUNN is introduced and it is demonstrated
to be finite to all orders of perturbation theory.


\newpage 
\pagestyle{empty} 
\begin{center} 
\vspace*{8cm} 
\hspace{-1cm}
\emph{Dedicated to my family}\\
\hspace{1cm}
\end{center}

%
%
\newpage 
\pagenumbering{roman} 
\pagestyle{plain} 
\tableofcontents 
\newpage 
\listoffigures 
\newpage 
\listoftables 
%
%
\newpage 
\chapter*{Preface} 
\addcontentsline{toc}{chapter} 
{\protect\numberline{Preface\hspace{-96pt}}} 
Original work appears in chapters three (in collaboration with
T.R.\ Morris) and five (in collaboration with S.\ Arnone, Yu.A.\ Kubyshin
and T.R.\ Morris) and has appeared in:

\newcounter{token}
\setcounter{token}{1}
\begin{tabbing}
(\roman{token}) \ \ \ \=T.R.\ Morris and J.F.\ Tighe, JHEP {\bf 08} (1999)
007. 
\stepcounter{token} \\
(\roman{token}) \> T.R.\ Morris and J.F.\ Tighe, Int.\ J.\ Mod.\ Phys.\
{\bf A16} (2001) 2905. 
\stepcounter{token} \\
(\roman{token}) \> S.\ Arnone, Yu.A.\ Kubyshin, T.R.\ Morris and J.F.\
Tighe, {\tt hep-th/0102011}.
\stepcounter{token} \\
(\roman{token}) \> S.\ Arnone, Yu.A.\ Kubyshin, T.R.\ Morris and J.F.\
Tighe, Int.\ J.\ Mod.\ Phys.\ \\ \>{\bf A16} (2001) 1989. 
\stepcounter{token} \\
(\roman{token}) \> S.\ Arnone, Yu.A.\ Kubyshin, T.R.\ Morris and J.F.\
Tighe, {\tt hep-th/0106258}.
\end{tabbing}

No claim to originality is made for the content of chapters one, two and four
which were compiled using a variety of other sources.

%
%
\newpage 
\chapter*{Acknowledgements} 
\addcontentsline{toc}{chapter}  
                {\protect\numberline{Acknowledgements\hspace{-96pt}}} 

Firstly I wish  to thank my supervisor  Tim Morris, without
whose cheerful enthusiasm, encouragement and guidance,
this thesis would not exist.  It is also a pleasure to thank my other
collaborators Stefano Arnone and Yuri Kubyshin for all their assistance.

Next, I would like to express my gratitude to  all the students, postdocs
and staff who have made the SHEP group such a friendly and pleasant
environment in which to work.  A special mention should go to Alex Dougall
and Luke Weston who always provided help, encouragement and good company.

Above all,  I would like to thank my parents and my sister for their patience
and ever present support.

\newpage

\pagenumbering{arabic}
\chapter{Introduction}\label{introduction}

Central to the acceptance of quantum field theories (QFTs) as the best 
description of physics on sub-nuclear scales, has been the deepening in
understanding of the 
process of renormalisation. Through this development,
the attitude towards the infinities that pervade QFT calculations has
shifted from the opinion that they are a disease that has to be removed by a
seemingly {\emph{ad hoc}} mathematical trick, to the view that they
are, in some sense, natural consequences of the underlying physics.  This
latter view has arisen from the insights gleaned from the formulation 
of the Wilsonian approach to the Renormalisation Group (RG)
\cite{wilson}.

At the heart of this approach lies the concept of effective field
theories.  The viewpoint of the existence of a fundamental QFT (in the
sense that it is valid for all particles at all energy scales) is
abandoned, to be replaced by an effective theory that attempts only to
describe physics up to a specified high energy cutoff.  The Lagrangian that
is suitable for this energy range is kept general in the sense that all
possible interactions consistent with the symmetries of the system are
included.  There is no longer a problem with divergences 
(we have regularised).  The issue now is that of predictive power;  is the 
theory now capable of making predictions given that it contains an infinite
number of coupling constants?  Remarkably, the answer can often be given in
the affirmative.  If it can be demonstrated that this is the case, this is 
equivalent to proving the renormalisability of the theory.

From such ideas, flow equations (that are non-perturbative in the coupling
constants) for the QFT can be derived.  However, to
make progress with calculations it is often necessary to make
approximations.  One very powerful method is that of using the derivative
expansion; a Taylor expansion in the momenta of the vertices of the QFT. An
obvious issue  which must be considered and which is addressed in this
thesis,  is under which conditions such an expansion converges. 

The extension of the Wilsonian RG from scalar field theory where it has
proved very powerful, to the more physically relevant topic, as far as
particle physics is concerned, of gauge theories (specifically Yang-Mills
theories) has been fraught with difficulty.  The main problem lies with the
incompatibility of restricting the momentum domain over which the theory is
applied with the fact that observables are invariant under internal gauge 
symmetries. 

Until recently, this obstacle has been tackled by initially breaking the
gauge invariance with the aim of re-imposing it at the end of the
calculation \cite{gbreak}.  Obviously such a method is far from ideal.
However, a 
fresh approach \cite{mor:manerg,mor:erg1,mor:erg2} enables gauge invariance
to be maintained at all stages. This is achieved by first regularising via
higher covariant derivatives, the gauge theory equivalent of a momentum
space cutoff.  It is well established though, that this cannot remove all 
divergences.  However, by adding extra gauge invariant particles known as
Pauli-Villars  fields, complete regularisation can be achieved.  It has
been postulated that a similar but more elegant mechanism can be obtained
by embedding the gauge group within a larger supersymmetric gauge group
which is then spontaneously broken to regain the low energy physics of the
original theory.  We prove that this is the case in this thesis.

This thesis falls into two main parts and is structured as follows.
Chapter \ref{wilson} is an introductory chapter and presents some of the
formalism and background necessary for dealing with the Wilsonian RG.  Two
versions of the flow equation are constructed followed by a discussion of how
renormalisability is expressed within this framework.  We conclude with a
review of some approximation methods in current use.  Chapter \ref{beta}
then considers some of the conditions necessary for one such approximation
method, the derivative expansion, to converge.  The Wilson/Polchinski
and Legendre flow equations are considered at one and two loops for the
$\beta$ function of scalar $\lam \varphi^4$ field theory for a variety of
cutoffs. Chapter \ref{super} is another introductory chapter, this time
concentrating on the problems concerned  with constructing a gauge
invariant regularisation scheme compatible with Wilsonian ideas, and some of
the group theoretical background for the scheme introduced in chapter
\ref{sunn}.  The final chapter sets up the regularisation scheme utilising
higher covariant derivatives and supersymmetric gauge groups
and demonstrates that it does indeed render the desired theory finite.

\chapter{Exact renormalisation group}\label{wilson}

\section{Wilson's renormalisation group}

The concepts that provide the basis of the exact renormalisation group were
first formulated within the context of statistical field theory by Wilson and
co-workers \cite{wilson}.  The problem of performing calculations
concerning \eg a lattice of spins is exacerbated when the system is
undergoing a continuous phase transition since the (already large) number
of degrees of freedom which are effectively interacting with one another,
diverges in this regime.  A procedure for systematically reducing the
degrees of freedom yet retaining the basic physics of the model is the
concept of blocking, first introduced by Kadanoff \cite{kadanoff}.  

This is the idea that in \eg an array of spins such
as a ferromagnet,  spins could be grouped together into blocks and treated
as if they were single spins with local interactions.\footnote{But not
necessarily just nearest neighbour interactions.}  Of course these
`new' interactions would not be exactly the same as in the original and on
short scales the new system would differ markedly from the old one.
However they exhibit the same \emph{long} distance behaviour
and it is only this behaviour that we are interested in describing.  Since
the number of degrees of freedom falls with this 
procedure,  iteration reduces them to a manageable level.  The payment  for this is that the new system will in general  
be much more complicated than the original, containing as it does many new
interactions.  However, as we shall see in the context of quantum field
theory,  we can obtain flow equations for the changes in coupling constants
of the new interactions  with the iteration of the procedure and thus
extract much useful information from this.

This thesis is concerned with the exact RG\footnote{Also referred to as the
continuum RG.}  which takes these ideas and applies them to 
quantum field theories directly in the continuum and as such, we will no
longer refer to statistical mechanics examples. As we shall see there are
a number of different (but equivalent) flow equations that can be derived
using the exact RG approach.  The work in this chapter is based upon that
of \cite{mor:approx}--\cite{mor:momexp} unless otherwise specified.

\section{Wilson/Polchinski flow equation}\label{sec:cut} 

\begin{figure}[tbh]
\begin{picture}(100,150)(-80,10)
 \psfrag{1}{$1$}
 \psfrag{0}{$0$}
 \psfrag{lam}{$\Lam$}
 \psfrag{p}{$p$}
 \psfrag{Cuv}{$C_{UV}(p^2/\Lam^2)$}
 \psfrag{Cir}{$C_{IR}(p^2/\Lam^2)$}
 \includegraphics[scale=0.4]{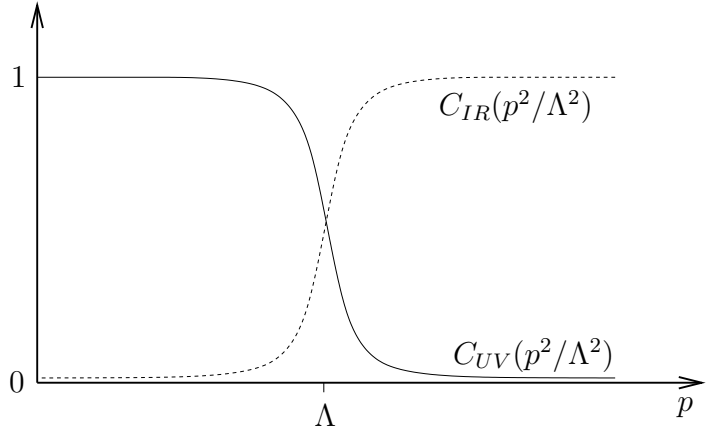}
\end{picture}
\caption{Sketch of the properties of the cutoff functions}
\label{cutoffs}
\end{figure}

The partition function for a single scalar field $\varphi$ in $D$ Euclidean
spacetime dimensions is given
by\footnote{The following shorthand is employed $a \cdot b \cdot c = \int
\! d^Dx \int \! d^Dy \, a(x) \, b(x,y) \, c(y)$, \\ \ph{the following
shorthand is employed \hspace{1cm}} $d \cdot e = \int \! d^Dx \, d(x) \,
e(x)$\\ \ph{hell}. } 
\be\label{zscalar}
Z[J] = \int\!{\cal{D}}\varphi \:\: \exp\{-\half\varphi \cdot \Delta^{-1}\cdot
\varphi - S_{\Lam_0}^{\mathrm{int}}[\varphi] + J\cdot \varphi \},
\ee
with the propagator denoted by $\Delta$, the (bare) interactions contained within $S_{\Lam_0}^{\mathrm{int}}$ and we have included a source $J$ for the
field. An effective ultra-violet cutoff is introduced via a modification of
the propagator
\be
\Delta = {1\over q^2} \hspace{1cm} \rightarrow  \hspace{1cm} \Delta_{UV} =
{C_{UV}(q^2/\Lam^2) \over q^2}. 
\ee
$C_{UV}(q^2/\Lam^2)$ is an as yet unspecified function of its argument
(the argument has to be $q^2/\Lam^2$ from Lorentz invariance and
dimensions) with the properties $C_{UV}(0)=1$ and $C_{UV}\rightarrow 0$
sufficiently fast as $q \rightarrow \infty$. In a similar fashion, we
define an IR modified propagator $\Delta_{IR} = C_{IR}(q^2/\Lam^2)\,/q^2$
(with $C_{UV}(p^2/\Lam^2) + C_{IR}(p^2/\Lam^2) = 1$).
Figure \ref{cutoffs} shows the properties of these cutoff functions.

Although it is not immediately obvious, we are able to rewrite
(\ref{zscalar}) (up to an uninteresting constant of proportionality which
we drop) as  
\bea\label{zscalar2} 
\lefteqn{
Z[J] = \int\!{\cal{D}}\varphi_>{\cal{D}}\varphi_< \:\: \exp\{-\half\,\varphi_> \cdot 
\Delta_{IR}^{-1}\cdot \varphi_> -\half\,\varphi_< \cdot
\Delta_{UV}^{-1}\cdot \varphi_< } \hspace{4cm}
\nonumber \\
& & - S_{\Lam_0}^{\mathrm{int}}[\varphi_> + \varphi_<] + J\cdot (\varphi_> +
\varphi_<) \},
\eea
The equivalence of \eq{zscalar2} to (\ref{zscalar}) is evident once the substitutions
\be
\varphi_> = \varphi - \varphi_<, \hspace{5mm} \mathrm{followed \  by}
\hspace{5mm} \varphi_< =
\varphi^\prime + C_{UV}\cdot\varphi
\ee
are made in \eq{zscalar2}.  This leaves \eq{zscalar2} in the form
\be
Z[J] = \int\!{\cal{D}}\varphi{\cal{D}}\varphi^\prime \:\: \exp\{-\half
\,\varphi  \cdot 
\Delta^{-1}\cdot \varphi -\half\,\varphi^\prime \cdot
(\Delta_{UV}^{-1} + \Delta_{IR}^{-1})\cdot \varphi^\prime 
- S_{\Lam_0}^{\mathrm{int}}[\varphi] + J\cdot \varphi \}, 
\ee
at which point the Gaussian integral over $\varphi^\prime$ can be performed,
resulting in  \eq{zscalar} (up to the aforementioned constant of
proportionality). 

From the manner in which it propagates, the $\varphi_>$ ($\varphi_<$) can be
interpreted as the momentum modes higher (lower) than $\Lam$, with the modes
lower (higher) than $\Lam$ damped.  If the integral over the higher modes
is isolated in \eq{zscalar2}, it can be expressed as
\be
Z[J] = \int\!{\cal{D}}\varphi_< \, \exp \{-\half \varphi_< \cdot
\Delta^{-1}_{UV} \cdot \varphi_< \}\,Z_{\Lam}[\varphi_<,J],
\ee
where $Z_{\Lam}[\varphi_<,J]$ is defined as
\be\label{zoflam}
Z_{\Lam}[\varphi_<,J] = \int\!{\cal{D}}\varphi_> \exp\{-\half \,\varphi_> \cdot \Delta_{IR}^{-1}\cdot \varphi_> -
S_{\Lam_0}^{\mathrm{int}}[\varphi_> + \varphi_<] + J\cdot (\varphi_> + 
\varphi_<) \}. 
\ee
However, $Z_{\Lam}[\varphi_<,J]$ does not depend upon  $\varphi_<$ and $J$
separately but rather on the sum $\Delta_{IR}\cdot J + \varphi_<$.  Upon the
substitution $\varphi_> = \varphi - 
\varphi_<$, \eq{zoflam} becomes
\be
Z_{\Lam}[\varphi_<,J] = \exp \{-\half\varphi_< \cdot
\Delta_{IR}^{-1}\cdot \varphi_< \}
\int\!{\cal{D}}\varphi \exp\{- \half
\,\varphi \cdot \Delta_{IR}^{-1}\cdot \varphi
-S_{\Lam_0}^{\mathrm{int}}[\varphi] + \varphi\cdot (J + \Delta_{IR}^{-1}
\cdot \varphi_<)
\}
\ee
We proceed by integrating over the $\varphi$  variable to obtain
\bea
\lefteqn{Z_{\Lam}[\varphi_<,J] = 
\exp \{-\half\varphi_< \cdot \Delta_{IR}^{-1}\cdot \varphi_< \} \, \times} 
\nonumber \\
& & \hspace{1cm} \times \exp \{-S_{\Lam_0}^{\mathrm{int}} [{\textstyle{\del
\over \del J}}] \} \, \,
\exp \{ \half (J + \Delta_{IR}^{-1} \cdot \varphi_<) \cdot \Delta_{IR}
\cdot (J + \Delta_{IR}^{-1} \cdot \varphi_<) \} \nonumber \\
& & = \exp \{ \half J \cdot \Delta_{IR} \cdot J + J \cdot \varphi_< \} \,\,
 \exp \{ -\half (J + \Delta_{IR}^{-1} \cdot \varphi_<) \cdot \Delta_{IR}
\cdot (J + \Delta_{IR}^{-1} \cdot \varphi_<) \} \times  \nonumber \\
& & \hspace{1cm} \times \exp \{-S_{\Lam_0}^{\mathrm{int}} [{\textstyle{\del
\over \del J}}] \} \, \,
\exp \{ \half (J + \Delta_{IR}^{-1} \cdot \varphi_<) \cdot \Delta_{IR}
\cdot (J + \Delta_{IR}^{-1} \cdot \varphi_<) \} .
\eea
When the derivatives in $S_{\Lam_0}^{\mathrm{int}} [{\textstyle{\del
\over \del J}}]$ are performed, they are replaced by either $\Delta_{IR}
\cdot J + \varphi_<$ or by $\Delta_{IR}$, a fact which can be expressed as
\be\label{zwiths}
Z_{\Lam}[\varphi_<,J] = \exp \{ \half J\cdot \Delta_{IR} \cdot J + J \cdot
\varphi_< - S_{\Lam}[\Delta_{IR} \cdot J + \varphi_<] \},
\ee
for some functional $S_{\Lam}$,  confirming the
statement given below \eq{zoflam}.  We refer to
$S_{\Lam}$ as the Wilsonian effective action.

\begin{figure}[tbh]
\begin{picture}(300,100)(-30,0)
 \psfrag{part}{\Large ${\partial \over \partial \Lam}$}
 \psfrag{sum}{\Large $\ds = \sum_{ \{I_1,I_2\} } $}
 \psfrag{half}{\Large $- \, {1 \over 2}$}
 \psfrag{I1}{\large $I_1$}
 \psfrag{I2}{\large $I_2$}
 \includegraphics[scale=0.5]{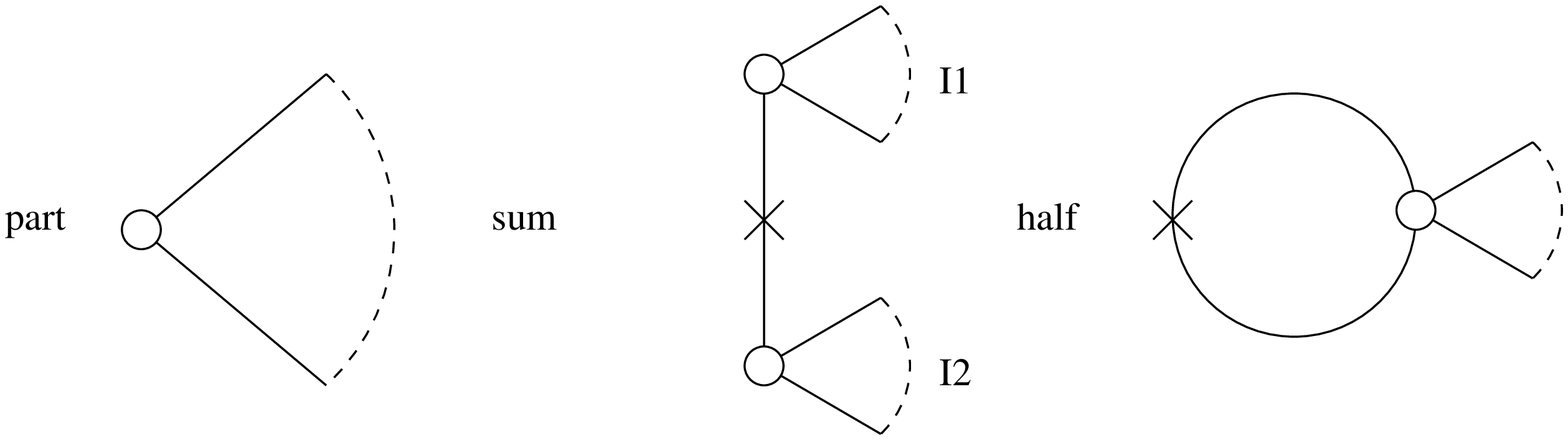}
\end{picture}
\caption{Graphical representation of the momentum expanded
Wilson/Polchinski flow equation.  Crosses denote differentiation with
respect to $\Lam$.}  
\label{fig:wpeqn}
\end{figure}

The exact RG flow equations follow from the observation that \eq{zoflam}
carries its dependence upon $\Lam$ entirely in the $ \varphi_> \cdot
\Delta_{IR}^{-1}\cdot \varphi_>$ term. Consequently,  when $Z_{\Lam}$ is
differentiated with respect to $\Lam$, the flow equation for $Z_{\Lam}$ is
found to be
\be\label{flowofz}
{\partial \over \partial \Lam} Z_{\Lam}[\varphi_<,J]
=
-{1 \over 2} \left( {\delta \over \delta J} - \varphi_< \right)
\cdot
\left( {\partial \over \partial \Lam} \Delta_{IR}^{-1} \right)
\cdot
\left( {\delta \over \delta J} - \varphi_< \right)
Z_{\Lam}[\varphi_<,J] \, .
\ee
By explicitly performing these functional derivatives using \eq{zwiths},
Polchinski's version of the Wilson flow 
equation\footnote{Wilson's version is recovered from that of Polchinski via
the substitution  
$\cal{H} = -S$ and the change of variables $\varphi \rightarrow
\varphi\sqrt{C_{UV}}$.} is obtained\footnote{$\Tr$ stands for a spacetime
trace \ie $\Tr (a \cdot b) = \int \! d^D\!x \!\int \! d^D\!y \, a(x,y) \,
b(y,x)$.} (we shall refer to it as the Wilson/Polchinski flow equation):
\be\label{wp}
{\partial \over \partial \Lam} S_{\Lam}[\varphi] =
{1 \over 2} {\delta S_{\Lam} \over \delta \Lam} \cdot 
{\partial \Delta_{UV} \over \partial  \Lam} \cdot 
{\delta S_{\Lam} \over \delta \Lam}
-
{1 \over 2} \Tr \, {\partial \Delta_{UV} \over \partial  \Lam} \cdot 
{\delta^2 S_{\Lam} \over \delta \varphi \delta \varphi} \,. 
\ee
Furthermore, with the momentum expanded action given by
\be
S(\p_1,\ldots,\p_n;\Lam) \equiv {\delta^n S_{\Lam}[\varphi]
\over \delta 
\varphi(\p_1) \cdots \delta \varphi(\p_n)} ,
\ee
we can obtain the momentum expanded Wilson/Polchinski flow equation which
will be extensively used in chapter \ref{beta}:
\bea\label{wpfl}
\lefteqn{{\partial\over\partial\Lam}S(\p_1,\ldots,\p_n;\Lam)=
\sum_{\left\{I_1,I_2\right\}}
S(-\Pone,I_1;\Lam)K_{\Lam}(P_1)S(\Pone,I_2;\Lam)} \nonumber \\
& & \hspace{5cm} -{1\over 2}\int\!\!{d^4q\over(2\pi)^4}K_{\Lam}(q)
S(\q,-\q,\p_1,\ldots,\p_n;\Lam),\hspace{1cm}
\eea
where $I_1$ and $I_2$ are disjoint subsets of external momenta such that
$I_1\cap I_2=\emptyset$ and $I_1\cup I_2=\{\p_1,\ldots,\p_n\}$, and we
define $K_{\Lam}(p) \equiv {\partial \over \partial \Lam}
\duv(p^2/\Lam^2)$. The sum over ${\left\{I_1,I_2\right\}}$  utilises the
Bose symmetry so pairs are counted only once \ie 
${\left\{I_1,I_2\right\}}={\left\{I_2,I_1\right\}}$. The momentum $\Pone$
is defined to be $\Pone={\sum_{\p_i\epsilon I_1}}\p_i$.  The equation can
be represented graphically as in figure \ref{fig:wpeqn}, which manifestly
displays how the Wilson/Polchinski flow equation is composed of a tree
structure contribution and a one-particle irreducible (1PI) part.

\normalsize

\section{Legendre flow equation}\label{sec:leg}

We need not consider a Wilson inspired RG flow equation only for the action.
A flow equation for the Legendre effective action may also be constructed
which has many additional beneficial properties.  We start by observing that 
the cutoff $\Lam $ can also be regarded as an infrared cutoff for the modes
that have already been integrated out. This can most easily be seen in
\eq{zoflam} 
which we can reinterpret as the partition function of an infrared cutoff
theory with $\varphi_<$ regarded as an external field. 

We start by introducing 
the Wilsonian generating functional for connected Green's functions
$W_{\Lam}[\varphi_<,J] \equiv \ln Z_{\Lam}[\varphi_<,J]$. Furthermore, with
the classical field $\clas$ defined via $\clas \equiv \del  W_{\Lam}
/ \del J$, we can construct $\Gam_\Lam[\clas]$, the interaction part of the
Legendre effective action:  
\be\label{defleg}
\half(\clas - \varphi_<) \cdot \dir^{-1} \cdot (\clas - \varphi_<)
+ \Gam_\Lam[\clas] = -W_{\Lam}[\varphi_<,J] + J \cdot \clas.
\ee
Using \eq{flowofz} we obtain the flow equation for $W_{\Lam}$
\be
{\partial \over \partial \Lam} W_{\Lam}
= - {1 \over 2} \left\{ \left({\del  W_{\Lam} \over  \del J} -
\varphi_<\right) 
\cdot \left({\partial \over \partial \Lam} \dir^{-1} \right) 
\cdot \left({\del  W_{\Lam} \over  \del J} - \varphi_<\right) 
+ \Tr \left(  {\del^2 W_{\Lam} \over \del J \,
\del J}  \cdot {\partial \over \partial \Lam} \dir^{-1}  
\right) \right\},
\ee
where $\Tr$ again denotes a spacetime trace. From \eq{defleg} we
can derive the relation  
\be
{\del^2 W_{\Lam} \over \del J \, \del J} =
\left( \dir^{-1} + {\del^2 \Gam_{\Lam} \over \del \clas \, \del \clas}
\right)^{-1},
\ee
which, after exploiting  \eq{defleg} once more, results in the following flow
equation for the effective action:
\be
{\partial \over \partial \Lam} \Gam_{\Lam}[\clas] = 
{1 \over 2} \Tr \left\{ {\partial \dir^{-1} \over \partial \Lam} \cdot 
\left( \dir^{-1} + {\del^2 \Gam_{\Lam} \over \del \clas \, \del \clas}
\right)^{-1} \right\} .
\ee
This equation is most usefully expressed when we separate off the
uninteresting vacuum energy by splitting the two-point function into its
field dependent and field independent (effective self energy) parts: 
\be
{\del^2 \Gam_{\Lam} \over \del \clas_x \, \del \clas_y}
= \hat{\Gam}_{x y}[\clas] + \Sigma_{x y}. 
\ee
This leads to the equation we shall refer to as the Legendre flow equation:
\be\label{legfl}
{\partial\over\partial\Lam}\Gam[\varphi^c]=-{1\over 2}\Tr
\left\{{K_{\Lam}\over
(1+\Delta_{IR}\Sigma)^2}.\hat{\Gam}.(1+[\Delta_{IR}^{-1} +
\Sigma]^{-1}.\hat{\Gam})^{-1}\right\}.
\ee 
In the work contained in chapter \ref{beta}, the most useful form of this
expression is in terms of a flow equation for the 1PI vertices
\be\label{legflsm}
{\partial\over\partial\Lam} \Gam(\p_1,\ldots,\p_n;\Lam)
=\int\!\!{d^4q\over(2\pi)^4}
{q^2{\partial\over \partial\Lam}C_{UV}({q^2}/{\Lam^2})\over 
\left[q^2+C_{IR}(q^2/\Lambda^2)\Sigma(q;\Lam)\right]^2}
E(\q,\p_1,\ldots,\p_n;\Lam),
\ee
where
\bea\label{legflE}
\lefteqn{E(\q,\p_1,\ldots,\p_n;\Lam) = - {1\over 2} 
\Gam(\q,-\q,\p_1,\ldots,\p_n;\Lam)} \nonumber \\ 
& &  \hspace{0.1in}+\sum_{\left\{I_1,I_2\right\}}\Gam(\q,-\q-\Pone,I_1;\Lam)
G(|\q+\Pone|;\Lam)\Gam(\q-\Ptwo,-\q,I_2;\Lam) \nonumber\\ 
& &  -\sum_{\left\{I_1,I_2\right\},I_3}\Gam(\q,-\q-\Pone,I_1;\Lam)
G(|\q+\Pone|;\Lam) \times \nonumber\\ 
& & \hspace{0.4in} \Gam(\q+\Pone,-\q+\Ptwo,I_3;\Lam)
G(|\q-\Ptwo|;\Lam)\Gam(\q-\Ptwo,-\q,I_2;\Lam) \nonumber \\
& & \hspace{4in}+\cdots \hspace{0.1in}.
\eea
Similarly to before, ${\bf P}_i=\sum_{\p_j\in I_i}\p_j$ and
$\sum_{\{I_1,I_2\},I_3,\ldots,I_m}$ is a sum over disjoint subsets
$I_i\cap I_j=\emptyset$ $(\forall i,j)$ with
${\bigcup_{i=1}^{m}}I_i=\{\p_1,\ldots,\p_n\}$.  Again, the
symmetrization $\{I_1,I_2\}$ means this pair is counted only
once. $G(p;\Lam)$ is defined by
\be\label{Gsm}
G(p;\Lam)\equiv
{C_{IR}(p^2/\Lambda^2)\over
p^2+C_{IR}(p^2/\Lambda^2)\Sigma(p;\Lam)} \nonumber
\ee
where $\Sigma$ is again the (field independent) self energy.

All the equations following \eq{legfl} apply to smooth cutoff profiles
only,  as care needs to be taken with regard to
sharp cutoffs.  If we denote the width over which the cutoff effectively
varies as $2\epsilon$, \ie $C_{UV}(q^2/\Lam^2) \approx 1$ for $q < \Lam -
\epsilon$ and  $C_{UV}(q^2/\Lam^2) \approx 0$ for $q > \Lam + \epsilon$, we
can investigate the effect of letting $\epsilon \rightarrow 0$. First we
need to establish the following lemma: 
\be\label{timslemma}
-\left( {\partial \over \partial \Lam} C_{IR}(p^2/\Lam^2) \right) 
\, f(C_{IR}(p^2/\Lam^2), \Lam) \rightarrow 
\del(\Lam -p) \int^1_0\!dt \, f(t,p)  \hspace{0.7cm} \mathrm{as\ } \epsilon
\rightarrow 0,
\ee
in which we require $f(C_{IR},\Lam)$ to be a function whose dependence upon
$\Lam$ is continuous at $\Lam = p$  as $\epsilon \rightarrow 0$.  The proof
of \eq{timslemma} lies with the identity
\be\label{sharpid}
\left. \left[ {\partial \over \partial \Lam} \int^1_{C_{IR}(p^2/\Lam^2)} \!
dt \, f(t,\Lam_1) \right] \, \right|_{\Lam_1 = \Lam} = \, - \left(
{\partial \over \partial \Lam} C_{IR}(p^2/\Lam^2) \right) \,
f(C_{IR}(p^2/\Lam^2),\Lam).  
\ee
We now note that the  integral on the left hand side (LHS) is a
representation of a (smoothed) step function of height $\int_0^1 dt \,
f(t,\Lam_1)$.  On taking the limit $\epsilon \rightarrow 0$,  the LHS of
\eq{sharpid} becomes the right hand side (RHS) of \eq{timslemma}.  Thus we
find, for example   
\be\label{funlim}
\begin{array}{cl}
- \left( {\partial \over \partial \Lam} C_{IR}(p^2/\Lam^2) \right)
\C_{IR}(p^2/\Lam^2) \rightarrow {1 \over 2} \delta(\Lam - p),
\\
- \left( {\partial \over \partial \Lam} C_{IR}(p^2/\Lam^2) \right)
\C_{IR}^2(p^2/\Lam^2) \rightarrow {1 \over 3} \delta(\Lam - p), &
\mathrm{etc.,} 
\end{array} 
\ee
in the sharp cutoff limit.

Thus returning to \eq{legfl} we now have the mathematical tools  to allow
the cutoff to become sharp.  We find 
\be\label{legflsh}
{\partial\over\partial\Lam} \Gam(\p_1,\ldots,\p_n;\Lam)
= \int\!\!{d^4q\over(2\pi)^4}{{\delta(q-\Lam)}\over
{q^2+\Sigma(q;\Lam)}}E(\q,\p_1,\ldots,\p_n;\Lam),
\ee
where $E(\q,\p_1,\ldots,\p_n;\Lam)$ is as given  in (\ref{legflE})
except now  $G(p;\Lam)$ is defined by
\be\label{Gsh}
G(p;\Lam)\equiv
{\theta(p-\Lam)\over{p^2+\Sigma(p;\Lam)}}.
\ee

\section{Renormalisability}

Perhaps the greatest success to date of the Wilsonian approach has been the
elegance with which the issue of renormalisability is addressed.  The
standard cumbersome and complicated method involving skeleton expansions is
replaced by a much more physically intuitive argument. As mentioned in chapter
\ref{introduction}, it is now a question of whether the theory retains any
predictive power with an infinite number of couplings.  One manner in which
this could be demonstrated is via the introduction of an overall cutoff
$\Lam_0$, and to check the $\Lam_0 \rightarrow \infty$ limit exists.
However, the exact RG does not require such artificial constructions and
allows us to deal directly in the continuum using renormalised quantities,
an approach which we will follow here.  In this section we will use only
dimensionless quantities constructed from the dimensionful ones using
appropriate powers of $\Lam$.

A fixed point of the flow of the action\footnote{These arguments also be
extended to the case of the Legendre flow equation.} in the space of all
possible (\ie an infinite number of) interactions, $S^*$, is defined by 
\be
\Lam {\partial \over \partial \Lam} S^*[\varphi] = 0.
\ee
Since the flow equation is written in terms of dimensionless quantities,
the independence of $\Lam$ exhibited by $S^*$ 
means that the action at the fixed point must have no scale dependence at all.
Since a massive theory has a mass to set a scale and a non-continuum
theory has an upper cutoff to perform the same r\^{o}le, we conclude that
the physics of massless continuum theories must be described entirely by 
fixed points.

Near a fixed point, we introduce new couplings 
\be
\eta_i = g_i - g^*_i,
\ee
where $g^*_i$ is the value of the coupling at the fixed point. We can then
approximate the flow of these couplings as 
\be\label{betaineta}
\Lam {\partial \eta_i \over \partial \Lam} = \sum_j Y_{ij} \, \eta_j,
\ee
where we have neglected contributions of $O(\eta^2)$ and higher, and
$Y_{ij}$ is a matrix of constants.  The eigenvalues and eigenvectors of $Y$
are defined by 
\be
\sum_j Y_{ij} \, \xi^{(k)}_j = -\lam_k \, \xi^{(k)}_i  \hspace{1cm}
(\mathrm{no\ sum\ on\ }k). 
\ee 
The couplings can be expanded in terms of these eigenvectors
\be
\eta_i = \sum_k \alpha_k \,\, \xi^{(k)}_i,
\ee
and \eq{betaineta} requires that the $\Lam$-dependent coefficients
$\alpha_k$ satisfy 
\be
\Lam {\partial \alpha_i(\Lam) \over \partial \Lam} = - \lam_i \,
\alpha_i(\Lam) \hspace{1cm} (\mathrm{no\ sum\ on\ }i), 
\ee
where again we have neglected terms of $O(\alpha^2)$. This implies that to
linear order, we have
\be
\eta_i(\Lam) = \sum_j \left({\mu \over \Lam}\right)^{\lam_j} \, \alpha_j(\mu)
\, \, \xi^{(j)}_i, 
\ee 
for some arbitrary mass scale $\mu$.  Thus near a fixed point a coupling
can flow away from the fixed point value or towards it depending on whether
$\lam_i$ is positive or negative.\footnote{If $\lam_i=0$ the behaviour has
to be followed to second  order where the power law dependency is replaced
by logarithmic evolution. In the following discussion these
so-called marginal couplings will not be explicitly considered.}  The 
$\eta_i$ (or $g_i$) for which $\lam_i>0$ are referred to as \emph{relevant}
couplings and those where $\lam_i<0$ are known as \emph{irrelevant}.  In
the vicinity of the fixed point we can also write (to linear order)
\be
S_{\Lam}[\varphi] = S^* + \sum_i \alpha^i(\mu) \left( {\mu \over \Lam}
\right)^{\lam_i} {\cal O}_i[\varphi], 
\ee  
which defines the scaling operators ${\cal O}_i$.

This formalism allows us to tackle the  slightly less straight
forward situation of  massive theories.  In the (infinite dimensional)
space of actions through which RG  trajectories flow, we are able to define
the critical manifold.  This manifold contains all the actions that will
yield a  massless continuum limit. The critical manifold is spanned  
by the infinite number of irrelevant operators  with other directions
spanned by (typically only one) relevant operators.   

\begin{figure}[tbh]
\begin{picture}(100,220)(-50,0)
  \psfrag{Critical manifold}{\small Critical manifold}
  \psfrag{Massless}{\small `Massless'}
  \psfrag{Renormalised}{\small Renormalised}
  \psfrag{trajectory}{\small trajectory}
  \psfrag{High temperature}{\small `Infinitely massive'}
  \psfrag{fixed point}{\small fixed point}
  \includegraphics[scale=0.5]{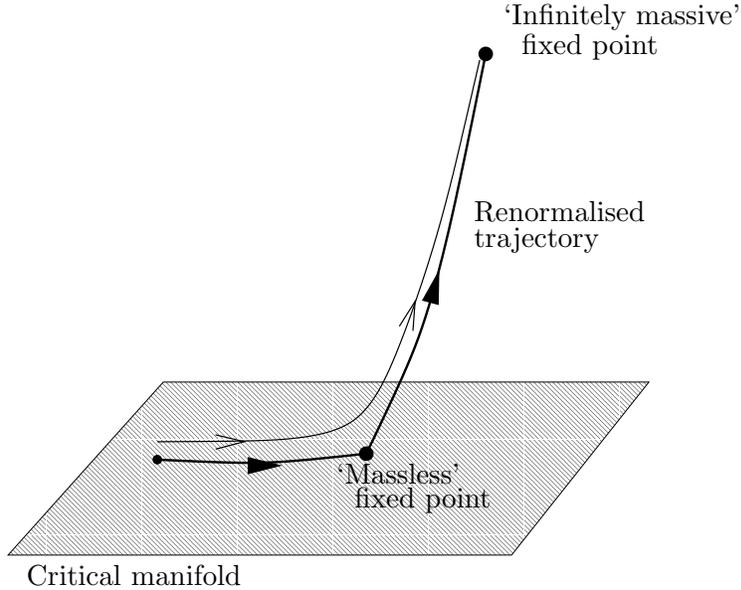}
\end{picture}
\caption{Procedure for tuning to the massive continuum limit} \label{fig:manifold}
\end{figure}

After parameterising the bare action, we are able to move slightly off the
critical manifold (see figure \ref{fig:manifold}).  Initially the RG flow
will still move towards the fixed 
point.  However as the fixed point is approached, the flow will shoot off
along one of the relevant directions to reach a fixed point that describes
an infinitely massive theory.  The continuum limit of a (finite) massive
theory can be extracted by the following procedure. With the bare action
being tuned back towards the critical manifold, physical quantities are
re-expressed in terms of renormalised ones accounting for the diverging
correlation length. When it reaches the critical manifold, the RG
trajectory splits into a part going into the fixed point and a part that
leaves the fixed point in the relevant directions.  This is known as a
renormalised trajectory and the actions that lie upon it are referred to as
`perfect actions'.  The end of this path has a finite limit when
expressed in renormalised quantities.  This trajectory is determined by
\be\label{bdry}
S_{\Lam}[\varphi] = S^*[\varphi] + \sum_{\stackrel{j=1}{\{\mathrm{relevant}\}}}^n
\alpha^j \left( {\mu \over \Lam}
\right)^{\lam_j} {\cal O}_j[\varphi], 
\ee
(the sum over $j$ is restricted to the $n$ relevant directions). 

Thus given the boundary condition \eq{bdry} and the RG flow equation, the
continuum limit is fully specified 
\be\label{sinalpha}
S_{\Lam}[\varphi] \equiv S_{\Lam}[\varphi](\alpha^1, \ldots, \alpha^n).
\ee
The next step is to define the renormalised couplings $g^i(\Lam)$.  We
choose  the renormalisation conditions such that
\be
g^j \sim \alpha^j(\mu/\Lam)^{\lam_j}  
\ee
as $\Lam\rightarrow \infty$ to be consistent with the form of ${\cal
O}_j[\varphi]$. But these relations can also be inverted \ie
\be
\alpha^j = \lim_{\Lam \rightarrow \infty} \left( {\Lam \over \mu}
\right)^{\lam_j} g^j(\Lam),
\ee
and substituting into \eq{sinalpha} returns the continuum action in terms
of the renormalised field ($\varphi$) the  relevant renormalised couplings
($g^1$  
to $g^n$) and the anomalous dimension\footnote{The anomalous dimension
$(\gam)$ is obtained from the wavefunction renormalisation $(Z)$ via $\gam
=\Lam {\partial \over \partial \Lam} Z$.  The wavefunction renormalisation
factor is  introduced to ensure the coefficient of the kinetic term is ${1
\over 2}$.} ($\gam$):  
\be
S_{\Lam}[\varphi] \equiv S_{\Lam}[\varphi](g^1(\Lam), \ldots,
g^n(\Lam),\gam(\Lam)), 
\ee
and $n$ is finite. This  is an equivalent statement to renormalisability
since only a finite number of finite quantities need be considered in
describing the theory.

\section{$\beta$ function}

An important concept in the field of QFTs is that of $\beta$ functions.
This contains information on how the renormalised couplings\footnote{\ie we
restrict ourselves to the relevant directions.} ($g_i$) vary
according to scale and is defined as:   
\be\label{betadefnoz}
\beta_i := \Lam {\partial \over \partial \Lam} g_i(\Lam)
\ee
In the next chapter  we will make extensive use of the $\beta$
function for $\lam \varphi^4$ scalar field theory.  It will prove useful to
include wavefunction renormalisation
separately within these calculations so there we choose to redefine it as 
\be\label{betadef}
\beta(\lam)  = \Lam {\partial \over \partial \Lam} {\lam \over Z^2(\Lam)},
\ee
where $Z$ is the wavefunction renormalisation.  An important property
displayed by the $\beta$ function  is that  
the first two orders in the perturbative expansion are universal, \ie they
are independent of renormalisation scheme. We write the perturbative
expansion as
\be\label{betaexp}
\beta(\lam) = \beta_0\lam^2 + \beta_1 \lam^3 + \cdots.
\ee
If we have another renormalisation scheme with different coupling
$\lam^{\prime}$, we can define a $\beta$ function for this scheme as well:
\be
\beta(\lam^{\prime}) = \beta^{\prime}_0\lam^{\prime 2} + \beta^{\prime}_1
\lam^{\prime 3} + \cdots. 
\ee
The couplings in the two schemes must be related
\be
\lam^{\prime} = \lam + a_1\lam^2 + \cdots,
\ee
which can be re-written as
\be\label{newscheme}
\lam = \lam^\prime - a_1\lam^{\prime 2} + \cdots.
\ee
If we operate with $\Lam {\partial \over \partial \Lam}$ upon
(\ref{newscheme}) and use the definition of (\ref{betadefnoz}) we find
\be\label{twobeta}
\beta_0\lam^2 + \beta_1 \lam^3 + \cdots
= \beta^{\prime}_0\lam^{\prime \gap 2} + \beta^{\prime}_1 \lam^{\prime \gap
3} - 
2a_1\lam^\prime(\beta^{\prime}_0\lam^{\prime \gap 2} + \beta^{\prime}_1
\lam^{\prime \gap 3}) 
+ \cdots,
\ee
and expressing $\lam^\prime$ as a function of $\lam$ on the RHS
of (\ref{twobeta}) shows that
\be
\beta_0\lam^2 + \beta_1 \lam^3 + \cdots
=
\beta^{\prime}_0\lam^2 + \beta^{\prime}_1 \lam^3 + \cdots,
\ee
from which we can see that, as promised, $\beta^{\prime}_0 = \beta_0$ and
$\beta^{\prime}_1= \beta_1$.  The $\beta$ function for massless $\lam
\varphi^4$ theory in four dimensions can be calculated using  standard
perturbation theory \cite{zinn}  to be 
\be\label{truebeta}
\beta(\lam)=3{{\lam^2}\over {(4\pi)^2}} - {17\over 3}{{\lam^3}\over 
{(4\pi)^4}} + O(\lam^4).
\ee

\section{Approximations}\label{sec:approx}

The complexity of the flow equations has prevented the formulation of
general solutions.  This has resulted in a number of approximation
techniques being developed and investigated.  In this section two of the
most widely used methods are discussed.

\subsection{Truncation}

The most obvious method of approximation that can be employed is to
truncate the number of operators that appear in the effective action
$S_{\Lam}$.   We can then construct a number of flow equations for the
coefficients of these operators by equating the terms on the two sides of
the original flow equation \eq{wp} [or \eq{legfl}].  The approximation lies
in neglecting terms from the RHS of the equation which are not
members of the chosen set of operators.

The main problem with using such an approximation scheme is its restricted
area of applicability.  Since this approach corresponds to a truncation in
the powers of the field about a selected point,  sensible answers can only
be obtained if the field $\varphi$ does not fluctuate much.  This amounts
to stating that $\varphi$ is always close to the mean field,  a regime
in which weak coupling theory is valid anyhow.  In non-perturbative
settings it is 
found that the expansion fails to converge and spurious fixed points are
also generated \cite{mor:trunc}.

\subsection{Derivative/momentum scale expansion}

Within statistical mechanics a successfully applied approximation has been 
truncations in real space spin systems. The analogue in QFT is to
perform a short distance expansion of the effective action.  If the cutoff
utilised has a smooth profile, this corresponds to a derivative expansion
\be
S_{\Lam} \sim \int\!\!d^Dx \left\{ V(\varphi,\Lam) + \half
(\partial_{\mu}\varphi)^2 K(\varphi,\Lam) + O(\partial^4) \right\},
\ee
Such an expansion seems a particularly natural one, amounting to an
expansion in external momenta around $\p = {\bf 0}$.  If the higher
derivative terms are not `small' the expansion will fail, but this is
probably also an indication that the description of the theory in terms of
the field content is not appropriate and that other degrees of freedom need
to be considered. 

When a sharp cutoff is imposed, care needs to be taken when taking a short
distance expansion.  Due to the
non-analyticity that is introduced, we are no longer able to expand in
powers of momentum.  The solutions to the flow equation \eq{legflsh} depend
upon the angles between the $\p_i$ even when any  $\p_i \rightarrow 0$; \ie
the solutions are not analytic in this regime. This behaviour is displayed by
terms such as  $\theta(|\p+\q| - \Lam) \sim \theta(\q.\p)$ for $p<<\Lam$
since $|\q|=\Lam$, which could appear in the second term of \eq{legflE}. As
a consequence, expansions have to be made in momentum scale $|\p|$.

It is evident that if a sharp cutoff (\ie $C_{UV}(p^2/\Lam^2) =
\theta(\Lam - p)$) is employed then the momentum scale expansion of the
Wilson/Polchinski equation runs into additional problems.  The expansion
corresponds to expanding $S_{\Lam}$ in the scale of external
momenta, regarding this as small compared to $\Lam$. The differentiation of
the internal propagator of the tree term of (\ref{wpfl}) (\cf figure
\ref{fig:wpeqn})  results in a delta function restricting 
momentum flow to be $\Lam$.  However momentum conservation requires 
the flow should be of order the external momenta which is typically much
lower.  Consequently, the tree term of the Wilson/Polchinski equation
gets discarded along with loop diagrams with more than one vertex (since
these arise from the substitution of the tree parts of
$S_{\Lam}$ into the second term of the equation).  Since
this is such a great mutilation of the theory we apply the such an
expansion only to the 1PI parts of the action, \ie only to the Legendre
flow equation.

The momentum scale expansion can be incorporated via the introduction of a
parameter, $\rho$, which can be set equal to one at a later time.  The 1PI
vertices of \eq{legflE} are expanded in terms of homogeneous functions of
non-negative integer degree
\be
\Gam(\p_1,\ldots, \p_n;\Lam) = \sum_{m=0}^{\infty} \Gam^{(m)}(\p_1,\ldots,
\p_n;\Lam )
\ee
where we define $\Gam^{(m)}(\p_1,\ldots, \p_n;\Lam)$ via
\be
\Gam^{(m)}(\rho\p_1,\ldots, \rho\p_n;\Lam) = \rho^m \, \Gam^{(m)}(\p_1,\ldots,
\p_n;\Lam ) 
\ee
Other external momentum dependence in the flow equation can also be
expanded in integer degree homogeneous functions.

The approximation lies in restricting these sums over an infinite number of
terms to some designated order.  If the derivative/momentum scale expansion
is truncated at $O(p^0)$, we obtain the well established  
local potential approximation which has proved to be both reliable and
accurate \cite{lpa}. 

\vspace{1cm}

Of the approximation methods mentioned the most promising appears the
derivative/momentum scale expansion and this is the one which we will
investigate in the following chapter.
 
\chapter{Convergence of derivative expansion}\label{beta}

As discussed in section \ref{sec:approx}, the difficulty in
dealing with the functional differential equation that expresses the exact
RG flow usually results in one of a variety of
analytic non-perturbative approximation methods  being employed.
Of the methods available, the derivative expansion (or momentum scale
expansion if 
a sharp cutoff is utilised) appears the most promising.  However, the
question that must be addressed when using this  approximation
scheme  is whether the expansion  converges and, if so, whether it
converges to the correct answer.

Obviously it is an extremely challenging task to settle this issue
non-perturbatively and in all generality.  It must be stressed that this is
not a controlled expansion in some small parameter.  Rather, the
approximation we make in using the derivative expansion lies in neglecting
powers of $(p^2/\Lam^2 )$ where $p$ is some typical momentum of the system
and $\Lam$ the effective cutoff.  Consideration of the flow equations \eq{wp}
and \eq{legfl} leads to the conclusion that the typical  momentum that
contribute are of order the effective cutoff, \ie $p \sim \Lam$. Hence the
issue is a numerical one.

In this chapter
we investigate some aspects of the applicability of the derivative
expansion in the weak coupling regime of
massless scalar $\lam \varphi^4$ theory\footnote{with
$\varphi\leftrightarrow - \varphi$ symmetry} in four Euclidean spacetime
dimensions. The derivative expanded 
$\beta$ function at one- and two-loop order is calculated and convergence
(or otherwise) is shown for a variety of different cutoff functions and
flow equations \cite{beta:pap, beta:rome}.

\section{Wilson/Polchinski flow equation}

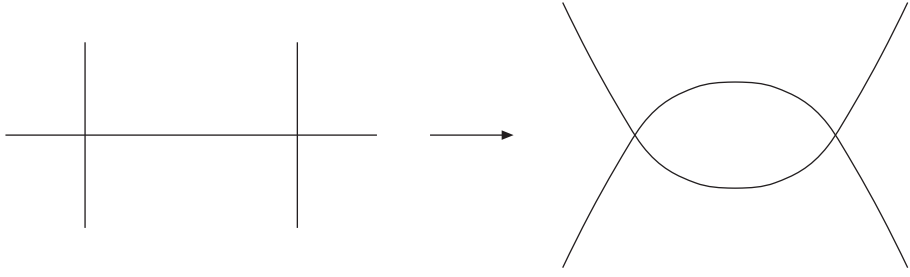
\begin{figure}[tbh]
\begin{picture}(220,100)(-70,0)
\Line(-40,50)(100,50)
\Line(-10,85)(-10,15)
\Line(70,85)(70,15)
\LongArrow(120,50)(150,50)
\Curve{(170,100)(180,80)(190,62)(200,46)(220,32)(234,30)
(236,30)(250,32)(270,46)(280,62)(290,80)(300,100)}
\Curve{(170,0)(180,20)(190,38)(200,54)(220,68)(234,70)
(236,70)(250,68)(270,54)(280,38)(290,20)(300,0)}
\end{picture}
\caption{Feynman diagram contributing to the four-point function 
at one loop, constructed from the  tree-level six-point function with two legs
joined} 
\label{fig:wpone}
\end{figure}

We start from the expanded Wilson/Polchinski flow equation \eq{wpfl}
\bea\label{wpfl2}
\lefteqn{{\partial\over\partial\Lam}S(\p_1,\ldots,\p_n;\Lam)=
\sum_{\left\{I_1,I_2\right\}}
S(-\Pone,I_1;\Lam)K_{\Lam}(P_1)S(\Pone,I_2;\Lam)} \nonumber \\
& & \hspace{2in} -{1\over 2}\int\!\!{d^4q\over(2\pi)^4}K_{\Lam}(q)
S(\q,-\q,\p_1,\ldots,\p_n;\Lam),
\eea
utilising the same notation as before.  We impose the renormalisation
condition  
\be\label{wprencon}
S(0,0,0,0;\Lam) \equiv \lam.
\ee
If the four-point vertex is considered exactly (\ie without a derivative
expansion or similar approximation), the exact one-loop $\beta$ function
can be obtained.   The sole contribution to the flow equation at this order
comes from the tree-level six-point function that has two of its legs joined
together to give figure \ref{fig:wpone}.

\SetScale{1.0}
The tree-level six-point function is found by setting $n=6$ in  \eq{wpfl2} and
substituting $S(\p_1,\p_2,\p_3,\p_4) = \lam$ (\ie to lowest order) in the
tree-level part of the RHS: 
\bea\label{wp6tree}
\lefteqn{ \hspace{-0.5cm}
S(\p_1,\p_2,\p_3,\p_4,\p_5,\p_6;\Lam)
= -\lam^2 \int_{\Lam}^{\infty} d\Lam_1 
\Bigg[ \sum_{i=3}^{6} K_{\Lam_1}({p_1 + p_2 + p_i})} 
\nonumber \\
& & \hspace{-0.5cm} +
\sum_{j=4}^{6} K_{\Lam_1}( {p_1 + p_3 + p_j}) +
\sum_{k=5}^{6} K_{\Lam_1}( {p_1 + p_4 + p_k}) +
K_{\Lam_1}( {p_1 + p_5 + p_6}) \Bigg].
\eea
Note that because the integral over $\Lam_1$ is UV convergent we can
proceed directly to the continuum without introducing an overall
cutoff. This is a reflection of the ability of the exact RG to deal directly
in the continuum using renormalised quantities as we will see later.
Substituting \eq{wp6tree} into the quantum correction part of (\ref{wpfl2})
for the flow of the four-point vertex (\ie fix $n=4$), we
set all external momenta to zero and obtain
\bea
{\partial\over\partial\Lam}\lam & = &
-{1 \over 2}\int\!\!{d^4q\over(2\pi)^4}K_{\Lam}(q)
\bigg[ -6{\lam}^2\int_{\Lam}^{\infty}
{d\Lam_1}K_{\Lam_1}(q)\bigg]
\label{wp6pt} \\[2mm]
& = &
{3\lam^2}\int\!\!{d^4q\over(2\pi)^4}K_{\Lam}(q)
\Delta_{IR}(q^2/\Lambda^2) \label{wptder}
\\
& = &
{6{\lam}^2\over(4\pi)^2}{1\over\Lambda}\int^\infty_0\!\!dx\, 
 C_{IR}'(x) C_{IR}(x) \label{wpots} 
\\
& = & 
{3{\lam}^2\over (4\pi)^2} {1\over \Lam}
\left[C_{IR}^2(\infty)-C_{IR}^2(0)\right], \label{wpans}
\eea
where the term in the square bracket of (\ref{wp6pt}) is the six-point
contribution of (\ref{wp6tree}), and  the prime in (\ref{wpots})  means
differentiation with respect to the argument of the function. Using the
definition for the $\beta$ function \eq{betadef}, we find that 
\be
\beta(\lam) = {3 \over (4 \pi)^2} \lam^2,
\ee
the  expected result at this order [\cf \eq{truebeta}].

Although we have seen that no approximation is required at this stage, we are
investigating the consequences of using a derivative expansion.  As such,
we expand the six-point function in terms of its external momentum.  In
effect, $\Delta_{IR}(q^2/\Lam^2)$ of (\ref{wptder}) is expanded in $q^2$,
and so, recalling that $C_{IR} = 1 -C_{UV}$, we find that (\ref{wpots}) is
replaced by  
\be\label{betexp}
\beta={6\lambda^2\over(4\pi)^2}\sum_{n=1}^\infty{C_{UV}^{(n)}(0)\over n!}
\int_0^\infty\!\!\!\!\!\!dx\, x^n C'_{UV}(x), \label{watsit}
\ee
with the $n$-th derivative with respect to $x$ denoted by
$C^{(n)}_{UV}(x)$.  If one na\"{\i}vely allows the cutoff to be sharp, \ie
$C_{UV} = \theta(1-x)$, we see immediately that this converges to the wrong
answer.  Since, $C^{(n)}_{UV}(0) = 0$ for all $n \ge 1$, (\ref{watsit})
will yield a zero $\beta$ function at this order.
However, as discussed earlier, the sharp cutoff should not be applied to
the Wilson/Polchinski flow equation, so for the remainder
of this section, we shall only consider smooth cutoffs.

If we impose a power law cutoff, then there is a finite value of $n$ larger
than which the integrals in (\ref{watsit}) diverge.  Choosing a cutoff
which falls faster than a power is also not sufficient to obtain a convergent
series.  Consider a cutoff of the form $C_{UV}(q^2/\Lam^2) =
\exp(-q^2/\Lam^2)$.  The $\beta$ function is found to be
\be 
\beta = {6 \lam^2 \over (4 \pi)^2} \sum_{n=1}^\infty (-1)^{n+1} .
\ee
Clearly this is an oscillating series that fails to converge. 

However convergence can be found with certain UV cutoff profiles if the
chosen function falls fast enough as $x \rightarrow \infty$.  Two
such examples are
\bea
C_{UV}(x) &=& \exp(1 - e^x), \label{cuveg1} \\
C_{UV}(x) &=& \exp[e - \exp(e^x)\,]. \label{cuveg2}
\eea
We can (numerically) calculate the one-loop $\beta$ function for these
cutoffs using \eq{betexp}.  From the first choice of cutoff \eq{cuveg1} we
find 
\be\label{wp28}
\beta = {3 \lam^2 \over (4 \pi)^2} \left( 1.193 + 0 - 0.194 -0.060 + 0.032
+ \cdots \right).
\ee
(The second term in this series vanishes since $\left.{d^2 \over dx^2} \,
\exp(1 - e^x)\right|_{x=0} = 0$.)

\begin{figure}[tbh]
  \begin{picture}(100,230)(-60,0)
    \psfrag{Partial sum contribution to beta function}{\small
    \hspace{-0.5cm}Partial sum contribution to $\beta_0 \times (4 \pi)^2$}  
    \psfrag{Number of terms in expansion, n}{\small \hspace{-0.5cm}Number
    of terms in expansion} 
    \includegraphics[scale=0.6]{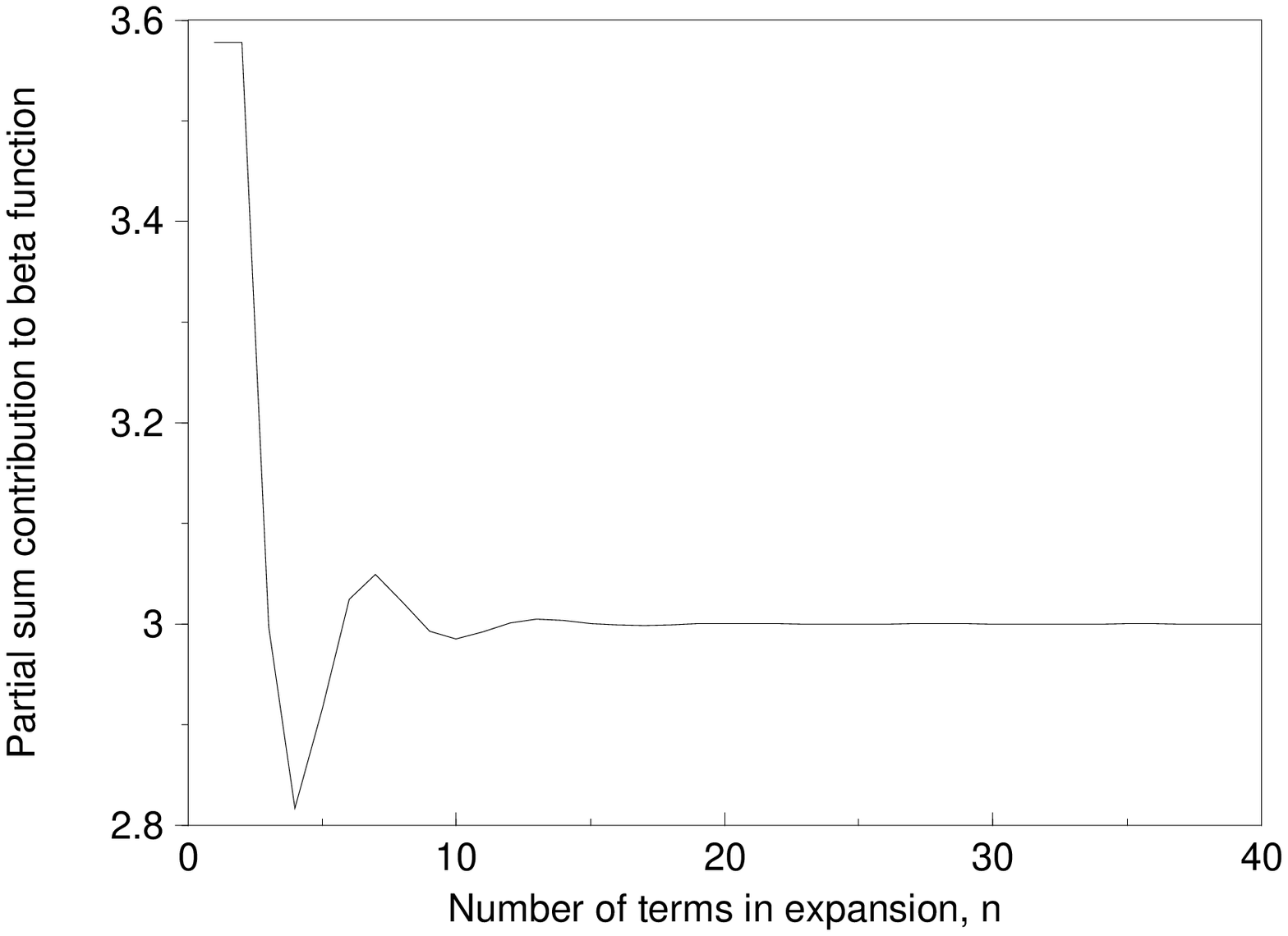}
   \end{picture}
\caption{Graph of partial sum contribution to $\beta_0$ coefficient against
    number of terms in expansion for the series \eq{wp28}}\label{fig:wp28} 
\end{figure}
\begin{figure}[tbh]
  \begin{picture}(100,230)(-60,0)
    \psfrag{Partial sum contribution to beta function}{\small
    \hspace{-0.5cm}Partial sum contribution to $\beta_0 \times (4 \pi)^2$}  
    \psfrag{Number of terms in expansion, n}{\small \hspace{-0.5cm}Number
    of terms in expansion}   
     \includegraphics[scale=0.6]{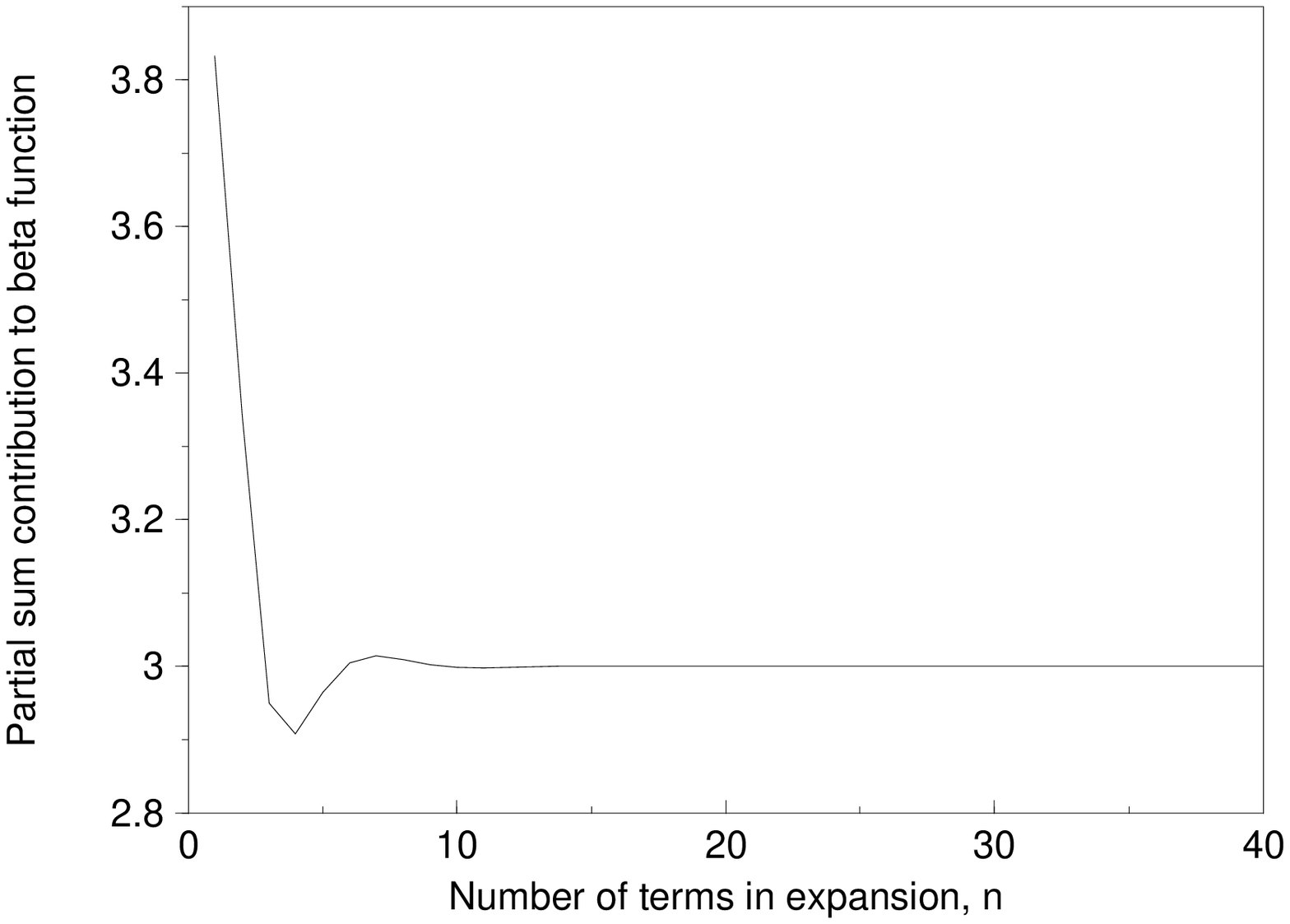}
  \end{picture}
\caption{Graph of partial sum contribution to $\beta_0$ coefficient against
    number of terms in expansion for the series \eq{wp29}} \label{fig:wp29}
\end{figure}

If we calculate the partial sum contribution to the $\beta_0$ coefficient
[\cf \eq{betaexp}] at each order of the expansion 
in \eq{wp28}, we obtain the graph contained in
figure \ref{fig:wp28}.  With the second choice of cutoff function
\eq{cuveg2}, the following expansion is obtained for the one-loop $\beta$
function, with the graph of the partial sums of the series displayed in 
figure \ref{fig:wp29}: 
\be\label{wp29}
\beta = {3 \lam^2 \over (4 \pi)^2} \left(1.278 -0.164 -0.130 - 0.014 +
0.019 +\cdots \right). 
\ee 
In both these cases convergence towards the correct value of the one-loop
$\beta$ 
function is manifest.  Although such convergence is encouraging, we now
leave the Wilson/Polchinski flow equation to  concentrate on the rather
more promising Legendre flow equation. 
This has inherently better convergence properties, not least because as we
are dealing with 1PI functions, hence  there are no tree-level corrections
and so a numerical series arising from a derivative expansion cannot arise
until at least two-loop order.

\section{Legendre flow equation at one loop}

The momentum expanded Legendre flow equation was given in
\eq{legflsh}, \eq{legflE} and \eq{Gsh} for sharp cutoff or \eq{legflsm},
\eq{legflE} and \eq{Gsm} if the
cutoff profile is smooth. Irrespective of  whether the cutoff is sharp
or smooth, the flow of the four-point function at one-loop is given by
\bea\label{legfl1lp}
\lefteqn{{\partial\over\partial\Lam} \Gam(\p_1,\p_2,\p_3,\p_4;\Lam)
= \int\!\!{d^4q\over(2\pi)^4}K_{\Lam}(q) \, \,\times}
\nonumber \\
& & \times\sum_{\left\{I_1,I_2\right\}}\Gam(\q,-\q-\Pone,I_1;\Lam)
\Delta_{IR}(|\q+\Pone|)\Gam(\q-\Ptwo,-\q,I_2;\Lam),
\eea
Imposing the renormalisation condition (\cf condition \eq{wprencon})
\be
\Gam(0,0,0,0;\Lam) = \lam,
\ee
and substituting the tree-level value of the four-point 1PI vertex
($\,\Gam(\p_1,\p_2,\p_4,\p_4;\Lam) = \lam$)  in the RHS of
(\ref{legfl1lp}), the $\beta$ function is found to be 
\be
\begin{array}{rl} \ds
\beta(\lam) & \ds = -3\lam^2\Lam\int\!\!{d^4q\over(2\pi)^4}{1 \over
{q^4}}\left({d \over d\Lam} C_{IR}(q^2/\Lam^2) \right) C_{IR}(q^2/\Lam^2)
\\[3mm] & \ds
= {6 \lam^2 \over {(4\pi)^2}} \int^\infty_0\!\!dx\, 
 C_{IR}'(x) C_{IR}(x)
\\[3mm] & \ds
= {3 \lam^2 \over {(4\pi)^2}}.
\end{array}
\ee
Note that no derivative expansion has been (or indeed can be) performed.
Unlike  the previous situation with the Wilson/Polchinski equation, there is
nothing to expand in. At one-loop, the
Wilson/Polchinski equation had 
the external momentum of the tree-level six-point function in which to
expand; in the case of the Legendre flow equation 
the property of being 1PI means that the only object is that of figure
\ref{fig:wpone} which (within the calculation of the $\beta$ function) has no
external momentum.  Hence the exact one-loop $\beta$ function is obtained
irrespective of the exact form of cutoff function.

\section{Legendre flow equation at two loops} 
 
To iterate the flow equation to two-loop order, care must be taken in using
renormalised quantities.  The four-point function is split into two parts,
momentum free $[\lam(\Lam)]$ and momentum dependent
$[\gam(\p_1,\p_2,\p_3,\p_4;\Lam)]$ \cite{bonini}: 
\bea\label{ren4pt}
\Gam(\p_1,\p_2,\p_3,\p_4;\Lam) &=& \lam(\Lam) +
\gam(\p_1,\p_2,\p_3,\p_4;\Lam), \\
{\rm where}\hskip2cm\gam(0,0,0,0;\Lam) &=&0. \label{ga0}
\eea
It is $\gam(\p_1,\p_2,\p_3,\p_4;\Lam)$ which must be iterated as the
momentum dependent four-point function.

The three topological variants allowed for two-loop diagrams with four
external legs are shown in figure \ref{fig:2loop}.  Actually only
topologies $(b)$ and $(c)$ contribute to the $\beta$ function. Upon setting
external momenta to zero, \eq{ga0} ensures that the iterand of diagram (a)
vanishes.  In fact diagram (a) is already incorporated in the one-loop
running $\lam(\Lam)$ since renormalised quantities are being calculated
directly.  If this calculation was to be performed in the more traditional
manner using bare parameters, topology (a) would only contribute a
divergent part which would be removed upon renormalisation.
\begin{center}
\SetScale{0.7}
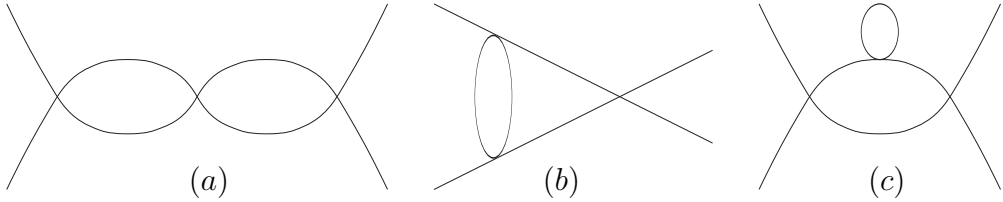
\begin{figure}[tbh]
\begin{picture}(410,80)(20,0)
\Curve{(70,100)(80,80)(90,62)(100,46)(120,32)(134,30)
(136,30)(150,32)(170,46)(175,54)(195,68)(209,70)
(211,70)(225,68)(245,54)(255,38)(265,20)(275,0)}
\Curve{(70,0)(80,20)(90,38)(100,54)(120,68)(134,70)
(136,70)(150,68)(170,54)(175,46)(195,32)(209,30)
(211,30)(225,32)(245,46)(255,62)(265,80)(275,100)}
\put(118,0) {$(a)$}
\Line(300,0)(450,75)
\Line(300,100)(450,25)
\Oval(332,50)(33,10)(0)
\put(252,0) {$(b)$}
\Curve{(475,100)(485,80)(495,62)(505,46)(525,32)(539,30)
(541,30)(555,32)(575,46)(585,62)(595,80)(605,100)}
\Curve{(475,0)(485,20)(495,38)(505,54)(525,68)(539,70)
(541,70)(555,68)(575,54)(585,38)(595,20)(605,0)}
\Oval(540,85)(15,10)(0)
\put(375,0) {$(c)$}
\end{picture}
\caption{Feynman diagrams contributing to four-point function 
at two loops.}
\label{fig:2loop}
\end{figure}
\end{center}
\vspace{-1cm}
Topology (b) of figure \ref{fig:2loop} can be formed in the flow equation
in one of two ways: by joining two legs from different vertices of 
the one-loop six-point 1PI function (shown in figure \ref{fig:6ptprop}
(a)), or by attaching two legs from different vertices of the one-loop
four-point function to the tree-level vertex. Topology (c) can also be
formed 
two ways: from the one-loop six-point 1PI function by joining two legs from
the same vertex, or by inserting the one-loop correction to the propagator
(shown in figure \ref{fig:6ptprop} (b)) into the one-loop four-point
function. As we shall see in the next section, the two contributions to the
$\beta$ function of the form of topology \ref{fig:2loop} (c) cancel  one
another, irrespective of the exact form of the cutoff profile employed.

\begin{center}
\begin{figure}[tbh]
\begin{picture}(200,100)(-80,0)
\Line(0,20)(100,20)
\Line(6,8)(56,94)
\Line(94,8)(44,94)
\put(50,0) {$(a)$}
\Line(154,40)(246,40)
\Oval(200,60)(20,10)(0)
\put(200,0) {$(b)$}
\end{picture}
\caption{Diagrams used in forming two loop four-point functions}
\label{fig:6ptprop}
\end{figure}
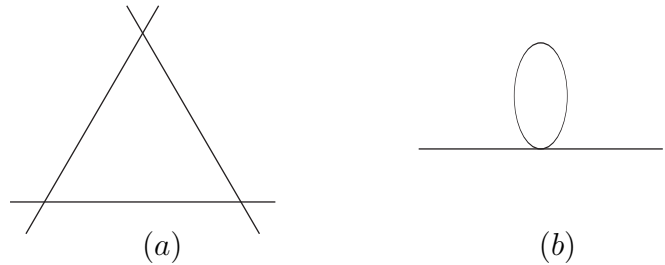
\end{center}
\vspace{-1cm}
At this order of the $\beta$ function, wavefunction renormalisation also
needs to be taken into account:  
\be
\begin{array}{rl} \ds
\beta(\lam) & \ds = \Lam {\partial \over \partial \Lam} {\lam \over Z^2}
\\[3mm] 
& \ds = {\Lam \over Z^2} {\partial \over \partial \Lam} \lam
-2\Lam \lam {\partial \over \partial \Lam} Z,
\end{array}
\ee  
where $Z$ is the wavefunction renormalisation and up until now use has been
made of the fact that $Z(\Lam) = 1 + O(\lam^2)$.  At two-loop order its
contribution to the $\beta$ function arises from the diagram of figure
\ref{fig:wfn2lp}. 
\begin{center}
\SetScale{1.0}
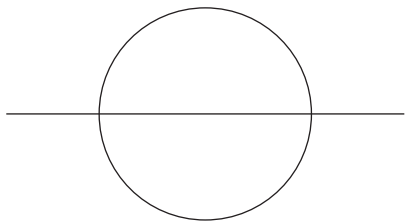
\begin{figure}[tbh]
\begin{picture}(100,100)(0,0)
\Oval(200,50)(40,40)(0)
\Line(125,50)(275,50)
\end{picture}
\caption{Feynman diagram contributing to wave function renormalization
at two loops.}
\label{fig:wfn2lp}
\end{figure}
\end{center}

\section{Cancellation of self energy diagrams}\label{secan}

We shall now demonstrate that the diagrams of the form of figure
\ref{fig:2loop} (c) do not contribute to the $\beta$ function irrespective
of the specific shape of the  cutoff. We will start by looking at a general
smooth cutoff and then consider the special case of the sharp cutoff. 

The self energy correction to the propagator is obtained directly from the
flow equation.  To $O(\lam)$ it is:
\be\label{sefl}
{\partial \over \partial \Lam}\Sigma(p;\Lam) = -{\lam \over 2}
\int\!\!{d^4q\over(2\pi)^4} {\partial\over
\partial\Lam}\Delta_{UV}({q^2}/{\Lam^2}) 
\ee
Integrating up \eq{sefl} we must not introduce a mass scale as we are
dealing with a massless theory. Consequently, the uniquely determined self
energy has to be 
\be\label{se}
\Sigma(p;\Lam) = -{\lam \over 2}
\int\!\!{d^4q\over(2\pi)^4} \Delta_{UV}({q^2}/{\Lam^2}).
\ee  
For example if an exponential cutoff, $C_{UV}(q^2/\Lam^2)=e^{-q^2/\Lam^2}$,
is utilised, the self energy is $ - 
{\lam\over (4\pi)^2} {\Lam^2 \over 2}$, while a power law cutoff such as
$C_{UV}(q^2/\Lam^2)=(1+(q^2/\Lam^2)^3)^{-1}$ gives rise to a self energy of
$ - {\lam\over (4\pi)} {\Lam^2 \over 16\sqrt{2}}$.
The general contribution to the flow of $\lam$ by the insertion of the self
energy correction into the one-loop four-point function is found to be
\bea
\lefteqn{\left. -9  \Lam\lam^2
\int\!\!{d^4q\over(2\pi)^4} 
\Delta_{IR}^2({q^2}/{\Lam^2})\left({\partial\over
\partial\Lam}\Delta_{UV}({q^2}/{\Lam^2}) \right) \Sigma(q;\Lam) 
\right|_{O(\lam)} } \label{secont1}\\
& & \hspace{3cm}
= - {3\over 2} \Lam\lam^3 \int\!\!{d^4q\over(2\pi)^4}{\partial\over
\partial\Lam}\Delta_{IR}^3({q^2}/{\Lam^2}) \int\!\!{d^4p\over(2\pi)^4}
\Delta_{UV}({p^2}/{\Lam^2})\hspace{0.5cm}\label{secont2}
\eea
with the numerical factor 9 in \eq{secont1} arising from various
combinatorics.  Since there are disjoint integrals over $\p$ and $\q$, it
is evident that no derivative expansion is possible.

The flow of the one-loop six-point 1PI function is 
\be\label{sixpt1lp}
\begin{array}{ll}
\ds{\partial \over \partial \Lam}  \Gam(\p_1,\p_2,\p_3,\p_4,\p_5,\p_6;\Lam) 
=\int\!\!{d^4q\over(2\pi)^4} \left({\partial\over
\partial\Lam}\Delta_{UV}({q^2}/{\Lam^2}) \right) \times 
\\
\ds \hspace{2cm} \times
\sum_{i=1}^3 \sum_{j=i+1}^6 \dir(\q+\p_i+\p_j)
\sum_{k=i+1}^6
\sum_{l=j+1}^6
\dir(\q-\p_k-\p_l).\\[-3mm]
\hspace{7.3cm}{\scriptstyle k \neq j}\hspace{0.5cm}  {\scriptstyle l \neq j}
\end{array}
\ee
This provides contributions to the $\beta$ function of both topology (b) and
(c) of figure \ref{fig:2loop}.  Extracting just the part relating to 
topology (c) we find that its contribution to the $\beta$ function is:
\bea
-{9 \over 2}\Lam\lam^3 \int\!\!{d^4q\over(2\pi)^4} 
{\partial \over \partial\Lam} \Delta_{UV}(q^2/\Lam^2)
\int^{\infty}_{\Lam}\!\!d\Lam_1
\int\!\!{d^4p\over(2\pi)^4} \Delta_{IR}^2({p^2}/{\Lam^2}){\partial\over
\partial\Lam_1}\Delta_{UV}({p^2}/{\Lam_1^2})
\\
=-{3 \over 2}\Lam\lam^3 \int\!\!{d^4q\over(2\pi)^4} 
{\partial \over \partial\Lam} \Delta_{UV}(q^2/\Lam^2)
\int\!\!{d^4p\over(2\pi)^4} \Delta_{IR}^3({p^2}/{\Lam^2}) \hspace{3cm}
\label{sixptcont}
\eea
Combining \eq{secont2} and \eq{sixptcont}, the total input to the $\beta$
function from figure \ref{fig:2loop} (c) is 
\bea 
-{3 \over 2}\Lam\lam^3 
{\partial \over \partial\Lam}
&\ph{\!}& \hspace{-7mm}
\int\!\!{d^4q\over(2\pi)^4} 
\Delta_{UV}(q^2/\Lam^2)
\int\!\!{d^4p\over(2\pi)^4} \Delta_{IR}^3({p^2}/{\Lam^2})
\\[1mm]
&&=-{6 \over (4\pi)^2}\Lam\lam^3 
{\partial \over \partial\Lam}
\int_0^{\infty}\!\! dx \: C_{UV}(x)
\int_0^{\infty}\!\! dy {C_{IR}^3(y) \over y^2}
\label{secanc}
\\[1mm]
&&=\, 0
\eea 
The situation with  regard to the sharp cutoff is very similar, with the
self energy \eq{se} replaced by
\be\label{sesh}
\Sigma(p;\Lam) = -{\lam \over 2}
\int\!\!{d^4q\over(2\pi)^4} {\theta(\Lam-q) \over q^2} = -{1 \over
(4\pi^2)} {\Lam^2 \over 2}.
\ee  
The different form of the flow equation for sharp cutoffs means that the
part of the $\beta$ function arising from self energy insertion is
\be
3\Lam\lam^3 \int\!\!{d^4q\over(2\pi)^4} \:\:
{\delta(q-\Lam) \: \theta(q-\Lam) \over q^6}
\int\!\!{d^4p\over(2\pi)^4} 
{\theta(p-\Lam) \over p^2} 
={3 \over (4\pi)^2} \lam^3. \label{shseins}
\ee
The $\beta$ function contribution of the topology of figure \ref{fig:2loop}
coming from the six-point function is
\be
-{9 \over 2} \Lam\lam^3 \int\!\!{d^4q\over(2\pi)^4}\: 
{\delta(q-\Lam) \over q^2}
\int^{\infty}_{\Lam}\!\! d\Lam_1
\int\!\!{d^4p\over(2\pi)^4 } \: \:
{\delta(p-\Lam_1) \over p^2} \: {\theta^2(p-\Lam_1) \over p^4}
=-{3 \over (4\pi)^2} \lam^3, \label{shsesix}
\ee
where we use the fact that here\footnote{See section \ref{sec:leg}.}
$\theta^2(0) = 1/3$. Obviously \eq{shseins} and 
\eq{shsesix} combine to provide the desired cancellation.

\section{Sharp cutoff}\label{sec:sharp}

We use the momentum expanded Legendre flow equation for a sharp cutoff
contained in (\ref{legflsh}), (\ref{legflE}) and (\ref{Gsh}).
The momentum dependent part of the four-point function at one loop is
calculated to be
\bea\label{shfl1lp}
\lefteqn{\gam(\p_1,\p_2,\p_3,\p_4;\Lam)}  \nonumber \\
& & = - \lam^2 \int_{\Lam}^{\infty}d\Lam_1
\int\!\!{d^4q\over(2\pi)^4}{{\delta(q-\Lam_1)}\over q^2}\sum_{i=2}^{4} 
\left\{{\theta(|\q+{\bf\cal P}_i|-\Lam_1)
\over (\q+{\bf\cal P}_i)^2}-{\theta(q-\Lam_1)\over q^2} 
\right\} \label{sh1lp} \\
& & =  - {\lam}^2\sum_{i=2}^{4}\int_{\Lam}^{\infty}
\!\!{d^4q\over(2\pi)^4} \left({\theta({{\cal P}_i\over 2q} + x)\over q^{\,4}} 
\left\{ 
1+2x{{\cal P}_i\over q}+ \left({{\cal P}_i\over q}\right)^2 
\right\}^{-1} -{1\over 2}\right)\label{sh1lpex}\\
& & = +{\lam^2\over4\pi^3}\sum_{i=2}^{4} 
\left\{{1\over6}{{\cal P}_i\over \Lam}
+{1\over720}\left({{\cal P}_i\over\Lam}\right)^3+{3\over44800}\left({{\cal
P}_i\over \Lam}\right)^5
+\cdots \right\}, \label{sh1lpans}
\eea
where ${\bf{\cal P}_i}=\p_1+\p_i$ and $x={\bf{\cal P}_i}\cdot\q\, /\, {\cal
P}_iq$.  Note that it is the subtraction in (\ref{sh1lp}) of the part
independent of external momentum that allows the upper limit of the
$\Lam_1$ integral to be set as $\infty$ as the integral is now convergent.
In (\ref{sh1lp}) the integral over $\Lam_1$ is performed by noting that
$\theta(0)$ can be treated as being equal to ${1 \over 2}$ [\cf
\eq{funlim}].  After changing variables, the step function in \eq{sh1lpex}
can be accounted for in the limits  the integral over $x$ allowing  the
term in braces to be expanded in momentum-scale  
${\cal P}_i=|{\bf{\cal P}_i}|$ \cite{mor:approx,mor:momexp}. Alternatively
the step function may be 
expanded directly \cite{mor:approx}  
\bea
\theta\left({{\cal P}_i\over 2q} + x\right)=\theta(x)
+\sum_{n=1}^{\infty}{1\over {n!}}\left({\cal P}_i\over2q\right)^n
\delta^{(n-1)}(x),
\eea
where $\delta^{(n-1)}(x)$ is the $(n-1)$th derivative of $\delta(x)$
with respect to x.  The same result as (\ref{sh1lpans}) was calculated in
ref.\ \cite{mor:approx} but using bare parameters instead of renormalised
ones as here.

Dropping the terms related to the self energy diagram of figure
\ref{fig:2loop} $(c)$, we find the flow equation for $\lam$ to $O(\lam^3)$
is 
\bea\label{shmain2lp}
\lefteqn{{\partial\over\partial\Lam}\lam(\Lam) = {1\over \Lam}
{{3{\lam}^2}\over (4\pi)^2}} \nonumber\\
& & - \, { 3{\lam}^3 \over2}
\int\!\!{d^4q\over(2\pi)^4}{{\delta(q-\Lam)}\over q^2}  
\int_{\Lam}^{\infty}\!\!d\Lam_1
\int\!\!{d^4p\over(2\pi)^4}{{\delta(p-\Lam_1)}\over p^2} 
 \nonumber \\  
& &  \hspace{0.6in} \left\{ {4\,\theta^2(|\p+\q|-\Lam_1)\over {|\p+\q|^4}} + 
{8\,\theta(|\p+\q|-\Lam_1)\,\theta(p-\Lam_1)\over {p^2|\p+\q|^2}} \right.
\nonumber\\
& & \hspace{1in} \left.
+{8\theta(q-\Lam)\over q^2}\left( {\theta(|\p+\q|-\Lam_1)\over |\p+\q|^2}
- {\theta(p-\Lam_1)\over p^2}\right)
\right\}.
\eea

\begin{figure}[tbh]
\begin{picture}(100,240)(-60,0)
   \psfrag{Partial sum contribution to beta function/stuffstuffstuff}{\small
   \hspace{-0.5cm} Partial sum contribution to $\beta$ function / ${\lam^3
   \over (4\pi)^4}$}  
   \psfrag{Number of terms in expansion}{\small \hspace{-0.5cm}Number
   of terms in expansion}  
  \includegraphics[scale=0.6]{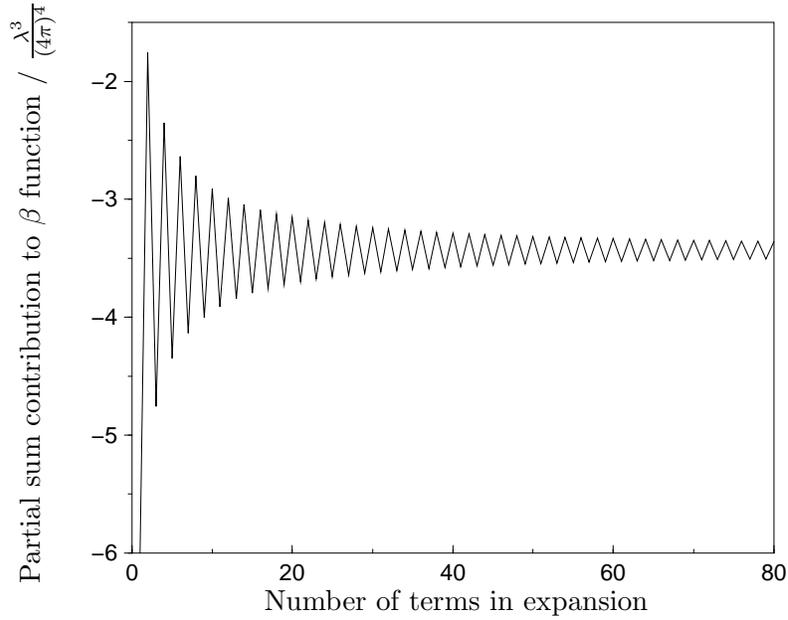}
\end{picture}
\caption{Partial sum contribution to $\beta$ function  against number of
   terms in expansion  for the series \eq{oscil}} \label{fig:twolpconv}
\end{figure}

\begin{figure}[tbh]
\begin{picture}(100,210)(-60,0)
    \psfrag{Average of partial sum contributions to beta function}
    {\scriptsize \hspace{-0.5cm}Average of partial sum contributions to $\beta$
    function /$ {\lam^3 \over(4\pi)^4}$}  
    \psfrag{Number of terms in expansion}{\small \hspace{-0.5cm}Number
    of terms in expansion}  
  \includegraphics[scale=0.6]{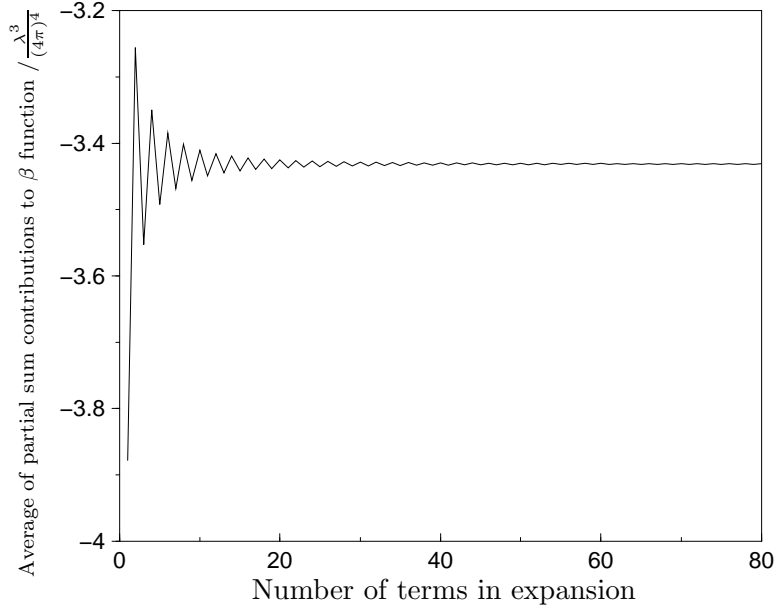}
\end{picture}
\caption{Average of successive partial sum contributions to $\beta$
    function  against number of terms in expansion for series 
    \eq{oscil}} 
\label{fig:twolpav} 
\end{figure}

The first two terms arise from the one-loop six-point 1PI function
\eq{sixpt1lp} with legs joined so as to be of the form of figure
\ref{fig:2loop} (b). The final line of \eq{shmain2lp} arises from iterating
the 
one-loop four-point function of \eq{sh1lpans} through the flow equation.
These contributions can be calculated using the momentum expansion; in this
case the embedded one-loop terms are expanded in $q/p$. For the first one
we find:
\bea
\lefteqn{- \,  6{\lam}^3
\int\!\!{d^4q\over(2\pi)^4}{{\delta(q-\Lam)}\over q^2}  
\int_{\Lam}^{\infty}\!\!d\Lam_1
\int\!\!{d^4p\over(2\pi)^4}{{\delta(p-\Lam_1)}\over p^2} 
{\theta^2(|\p+\q|-\Lam_1)\over {|\p+\q|^4}}}
\nonumber \\
& & =- \,  {6{\lam}^3 \over 4 \pi^3}
\int\!\!{d^4q\over(2\pi)^4}{{\delta(q-\Lam)}\over q^2}  
\int^{\infty}_{\Lam}\!\! {dp  \over p^3} \left\{ {\pi \over 4} 
- {5 \over 6}\left({q \over p}\right) 
+ {\pi \over 4}\left({q \over p}\right)^2 + \right. 
\nonumber \\
& & \left. \hspace{8cm}
-{63 \over 80}\left({q \over p}\right)^3
+{\pi \over 4} \left({q \over p}\right)^4 
+ \cdots \right\}
\nonumber\\
& & =-12{{{\lam}^3}\over (4\pi)^4}{1\over \Lam}{1\over
\pi}\left({\pi\over2}-{10\over9}+{\pi\over4}-{63\over
100}+{\pi\over6}-{7035\over 15680}+\cdots \right) \label{oscil}
\eea
This oscillating series converges, but only does so very slowly. The partial
sum contributions to the $\beta$ function from  the series of \eq{oscil}
are displayed in figure \ref{fig:twolpconv}.  
The average of successive partial sums is shown in
figure \ref{fig:twolpav} allowing an estimate of $-3.430{\lam^3
\over(4\pi)^4\Lam}$ to be made for the convergence of \eq{oscil}. 

\begin{figure}[hbt]
\begin{picture}(100,240)(-50,0)
  \psfrag{Partial sum contribution to beta function/stuffstuffstuff}{\small
  \hspace{-0.5cm} Partial sum contribution to $\beta$ function / ${\lam^3
  \over (4\pi)^4}$} 
  \psfrag{Number of terms in expansion, n}{\small Number of terms in
  expansion} 
  \psfrag{Series \eq{sharp_b}}{\small Series \eq{sharp_b}}
  \psfrag{Series \eq{sharp_c}}{\small Series \eq{sharp_c}}
  \psfrag{Series \eq{shwfn}}{\small Series \eq{shwfn}}
  \includegraphics[scale=0.6]{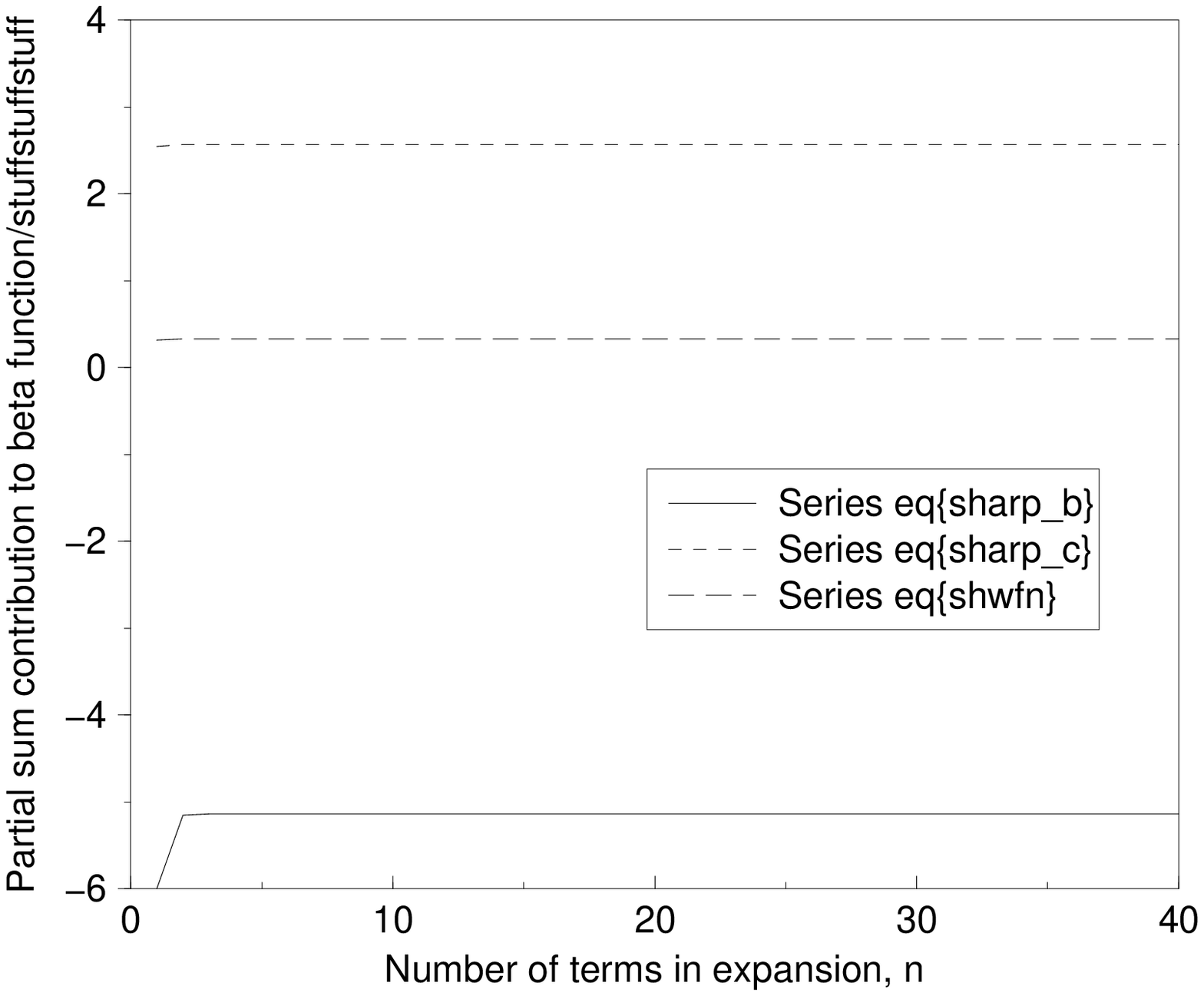}
\end{picture}
\caption{Partial sum contributions to the $\beta$ function against number
  of terms in the expansion of the series of \eq{sharp_b}, \eq{sharp_c} and
  \eq{shwfn}} 
\label{fig:sh_con}
\end{figure}

The next contribution in \eq{shmain2lp} provides a series which
converges rapidly to a value of $-5.13764 {\lam^3 \over(4\pi)^4\Lam}$:
\bea
\lefteqn{- \,  12{\lam}^3
\int\!\!{d^4q\over(2\pi)^4}{{\delta(q-\Lam)}\over q^2}  
\int_{\Lam}^{\infty}\!\!d\Lam_1
\int\!\!{d^4p\over(2\pi)^4}{{\delta(p-\Lam_1)}\over p^2}
{\theta(|\p+\q|-\Lam_1)\,\theta(p-\Lam_1)\over {p^2|\p+\q|^2}} }
\nonumber \\ 
& &=- \,  {6{\lam}^3 \over 4 \pi^3}
\int\!\!{d^4q\over(2\pi)^4}{{\delta(q-\Lam)}\over q^2}  
\int^{\infty}_{\Lam} {dp  \over p^3} \left\{ {\pi \over 4} 
-{1 \over 6}\left({q \over p}\right) 
-{1 \over 240}\left({q \over p}\right)^3
-{3 \over 8960}\left({q \over p}\right)^5 +
\cdots \right\}
\nonumber \\
& & = -12{{{\lam}^3}\over (4\pi)^4}{1\over \Lam}{1\over
\pi}\left({\pi\over2}-{2\over9}-{1\over300}- {3\over15680}+\cdots
\right). \label{sharp_b}
\eea
Iterating the one-loop result of \eq{sh1lpans}, the final part of
\eq{shmain2lp} returns the previously published value\footnote{Calculated
using bare parameters \cite{mor:approx}.} 
\bea
\lefteqn{- \,  6{\lam}^3
\int\!\!{d^4q\over(2\pi)^4}{{\delta(q-\Lam)}\over q^2}  
\int_{\Lam}^{\infty}\!\!d\Lam_1
\int\!\!{d^4p\over(2\pi)^4}{{\delta(p-\Lam_1)}\over p^2} 
{\theta(q-\Lam)\over q^2} \times} \nonumber \\ 
& & \hspace{8cm}\times
\left( {\theta(|\p+\q|-\Lam_1)\over |\p+\q|^2}
- {\theta(p-\Lam_1)\over p^2}\right) 
\nonumber \\
& &  =- \,  {6{\lam}^3 \over 4 \pi^3}
\int\!\!{d^4q\over(2\pi)^4}{{\delta(q-\Lam)}\over q^4}
\int^{\infty}_{\Lam} {dp  \over p} \left\{
{1 \over 6}\left({q \over p}\right) 
+{1 \over 240}\left({q \over p}\right)^3
+{3 \over 8960}\left({q \over p}\right)^5 +
\cdots \right\}
\nonumber \\ 
& & ={{\lam}^3\over (4\pi)^4}{1\over \Lam} {1\over\pi}
\left(8+{1\over15}+{9\over2800}+\cdots \right), \label{sharp_c}
\eea 
which converges to $2.56882 {\lam^3 \over (4\pi)^4\Lam}$.
As previously discussed, wavefunction renormalisation must be accounted for
through $\Sigma(k;\Lam)|_{O(k^2)}=[Z(\Lam)-1]k^2$ arising from
figure \ref{fig:wfn2lp}. 
\bea
\lefteqn{ k^2{\partial\over\partial\Lam}Z(\Lam) } \nonumber
\\
& & \left. =\lam^2\int\!\!{d^4q\over(2\pi)^4}{
{\delta(q-\Lam)}\over q^2}
\int_{\Lam}^{\infty}d\Lam_1\int\!\!{d^4p\over(2\pi)^4} 
{\delta(p-\Lam_1)\over p^2} 
{\theta(|\p+\q+\k|-\Lam_1)\over |\p+\q+\k|^2}
\right|_{O(k^2)} \nonumber
\\
& & \left. =-{\lam^2\over 24\pi^3}\int\!\!{d^4q\over(2\pi)^4}{
{\delta(q-\Lam)}\over q^2}\left\{{|\q+\k|\over
\Lam}+{1\over 120}{|\q+\k|^3\over \Lam^3}+{1\over 22400}{|\q+\k|^5\over
\Lam^5}+\cdots\right\} \right|_{O(k^2)} \nonumber
\\
& & = -{\lam^2k^2\over (4\pi)^4} {1\over \Lam} {1\over \pi}
\left({1\over 2}+{1\over 48}+{3\over 1280}+\cdots\right). \label{shwfn}
\eea
The second line of \eq{shwfn} is obtained using the expanded one-loop
four-vertex of \eq{sh1lpans}. The final line then follows upon the
realisation that the net effect of expanding to second order in $k$ and
then averaging over the angles is to convert $|\q+\k|^n$ into
${1\over8}n(n+2)q^{n-2}k^2$. We find that \eq{shwfn} converges to $-0.16667
{\lam^3 \over (4\pi)^4\Lam}$. In figure \ref{fig:sh_con} we display the
partial sum contributions to the $\beta$ function from each  of the series in
\eq{sharp_b}, \eq{sharp_c} and \eq{shwfn}.

\begin{figure}[htb]
\begin{picture}(100,250)(-60,0)
  \psfrag{Value of beta function / stuff}{\small
  Value of $\beta_1 \times (4\pi)^4$} 
  \psfrag{Number of terms in expansion, n}{\small Number of terms in
  expansion} 
  \includegraphics[scale=0.6]{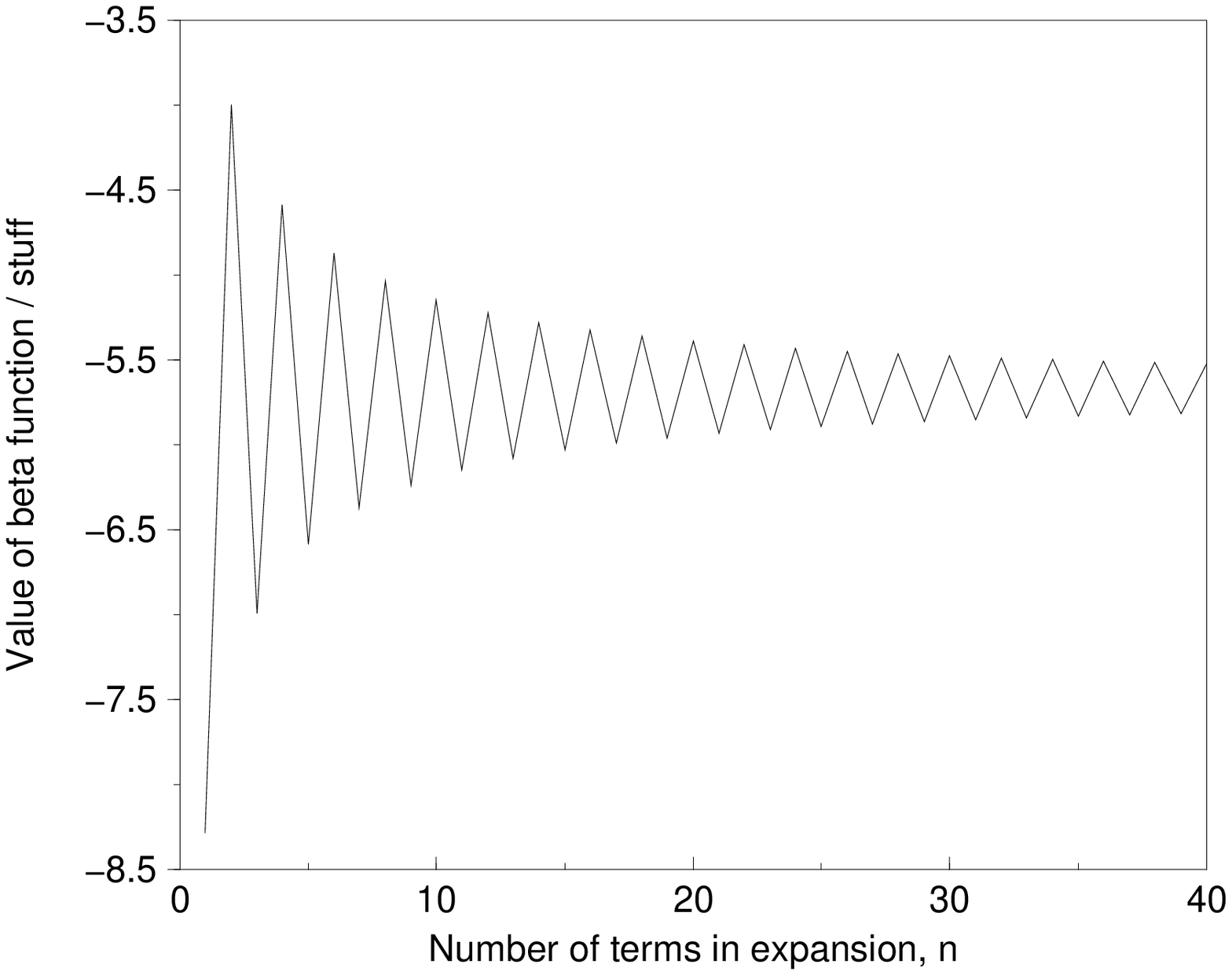}
\end{picture}
\caption{Value of $\beta_1$ coefficient against number of terms in expansion}
\label{fig:sh_beta}
\end{figure}

Thus the momentum expanded $\beta$ function  at two loops using a sharp
cutoff is
\bea\label{betash}
\lefteqn{\beta(\lam)  = 3{{\lam^2}\over {(4\pi)^2}} -{{{\lam}^3}\over
(4\pi)^4}{1\over\pi} \left\{
12\left({\pi\over2}-{10\over9}+{\pi\over4}-{63\over
100}+{\pi\over6}-{7035\over 15680}+\cdots \right) \right.}  
\nonumber \\
& & \hspace{0.5in}  + 12\left({\pi\over2}-{2\over9}-
{1\over300}-{3\over15680}+\cdots \right) 
- \left(8+{1\over15}+{9\over2800}+\cdots
\right) 
\nonumber \\
& & \hspace{2.0in} \left. -\left(1+{1\over 24}+{3\over
640}+\cdots\right) 
\right\},
\eea
which converges (albeit slowly) towards the exact expression \eq{truebeta}.
In figure \ref{fig:sh_beta} we show the value of the $\beta_1$ coefficient
[\cf \eq{betaexp}]  if we
just consider the specified number of terms from each of the series.

\section{Exponential cutoff}\label{sec:smooth}

The momentum expanded Legendre flow equation for smooth cutoffs is given
in (\ref{legflsm}) and (\ref{Gsm}).  Using an exponential cutoff
$C_{UV}=e^{-q^2/\Lam^2}$, the renormalised four-point function is given
by\footnote{In agreement with previous calculations using bare parameters
\cite{mor:momexp}.} 
\bea
\lefteqn{\gam(\p_1,\p_2,\p_3,\p_4;\Lam)}
\nonumber \\
& & =  -{\lam}^2\sum_{i=2}^{4}\int_{\Lam}^{\infty}
\!\!d\Lam_1\int\!\!{d^4q\over(2\pi)^4}\left(
{\partial\over{\partial\Lam_1}}{e^{-{q^2}/{{\Lam_1}^2}}\over q^2}
\right)\left\{{\left(1-e^{-{|\q+{\bf{\cal
P}_i}|}^2/{{\Lam_1}^2}}\right)\over
{{|\q+{\bf{\cal P}_i}|}^2}} \right. \label{sm1lp}
\\
& & \hspace{4.0in} \left. -\,
{\left(1-e^{-q^2/{{\Lam_1}^2}}\right)\over{q^2}} \right\}
\nonumber \\
& & = -2{{{\lam}^2}\over (4\pi)^2}\sum_{i=2}^{4}
\int_{\Lam}^{\infty}\!\!{d\Lam_1\over \Lam_1}
\left\{\left({{\Lam_1}\over {\cal P}_i}\right)^2
\left(1-e^{-{{\cal P}_i}^2/2{\Lam_1}^2}\right)-{1\over 2}\right\}
\label{sm1lpex} 
\\
& & = -{{{\lam}^2}\over 2(4\pi)^2}\sum_{i=2}^{4}
\sum_{n=1}^{\infty}{(-1)^n\over (n+1)!\,n}\left({{{\cal P}_i}^2\over
2{\Lam_1}^2}\right)^{n}. \label{sm1lpans}
\eea
The expression in (\ref{sm1lpex}) can be obtained from (\ref{sm1lp})
either by expanding the exponential of $-|\q+{\bf{\cal
P}_i}|^2/{{\Lam_1}^2}$, performing the integration over
momentum space and then resumming, or by using
\be
{\left(1-e^{-q^2/{{\Lam}^2}}\right)\over{q^2}}
= {1\over \Lam^2}\int_{0}^{1}\!\!\!da\ e^{-aq^2/\Lam^2}\nonumber
\ee 
and interchanging the order of integration.

Using the results of section \ref{secan} and dropping the self energy
diagrams, the flow of the coupling to $O(\lam^3)$ is 
\bea\label{smmain2lp}
\lefteqn{{\partial\over\partial\Lam}\lam(\Lam) = {1\over \Lam}
{{3{\lam}^2}\over (4\pi)^2}} \nonumber
\\
& & - \, {3{\lam}^3 \over 2}\int\!\!{d^4q\over(2\pi)^4}
{{\partial\over{\partial\Lam}}\left(e^{-{q^2}/{{\Lam}^2}}
\right)\over q^2} \int_{\Lam}^{\infty}d\Lam_1
\int\!\!{d^4p\over(2\pi)^4}
{{\partial\over{\partial\Lam_1}}\left(e^{-{p^2}/{{\Lam_1}^2}}\right)
\over p^2}
\nonumber
\\
& & \left\{{4\left(1-e^{-{|\p+\q|^2}/{\Lam_1^2}}\right)^2\over |\p+\q|^4}
+{8\left(1-e^{-{|\p+\q|^2}/{\Lam_1^2}}\right)
\left(1-e^{-{p^2}/{\Lam_1^2}}\right)\over p^2|\p+\q|^2} \right.\nonumber
\\
& & \left.+{8\left(1-e^{-{q^2}/{\Lam^2}}\right)\over q^2}\left(
{\left(1-e^{-{|\p+\q|}^2/{{\Lam_1}^2}}\right)\over {{|\p+\q|}^2}} 
-{\left(1-e^{-p^2/{\Lam_1^2}}\right)\over p^2}\right)\right\}. 
\eea
To perform these integrals, the inner ($p$) integral must be expanded in
terms of the momentum external to it (\ie $q$ momentum).

The first two contributions come from the 1PI one-loop six-point diagram
with two of its legs joined to give figure \ref{fig:2loop} $(b)$.  The
first of these gives the convergent series
\be\label{sm6pt1}
12{{{\lam}^3}\over (4\pi)^4}{1\over
\Lam}\left(\ln{3\over 4} +
\sum_{n=2}^{\infty}{(-1)^n} 
\left[\,\ln{4\over 3} -{1\over n}
\sum_{s=2}^{n}{n\choose s}{(-1)^s\over s-1}
\left\{1 - {1\over 2^{s-2}} + {1\over
3^{s-1}}\right\} \right]\right),
\ee
(when expanded) which numerically sums to ${{\lam^3}\over (4\pi)^4}{1\over
\Lam}(-2.45411725)$.  The second is
\be\label{sm6pt2}
-24{{{\lam}^3}\over (4\pi)^4}{1\over
\Lam}\left(\ln{4\over 3}
+ \sum_{n=1}^{\infty}{(-1)^{n}\over n(n+1)}
\left\{\left({2\over 3}\right)^n -\left({1\over 2}\right)^n  \right\}
\right).
\ee
Using the fact that
\be\label{lnexp}
\ln(1+x) = \sum_{n=1}^{\infty} {(-1)^{n+1} \over n} x^n \hspace{1cm}
(-1 < x \le 1), 
\ee
we integrate to find 
\be\label{smmainsum}
\sum_{n=1}^{\infty} {(-1)^n \over n(n+1)} x^n = 1 - {(x+1) \over
x}\ln(x+1),
\ee
which is of the same form of the sums of (\ref{sm6pt2}).  Hence
(\ref{sm6pt2}) sums exactly to $12{{\lam^3}\over (4\pi)^4}{1\over \Lam}
[9\ln3 -2\ln2 -5\ln5]$.  The final line of (\ref{smmain2lp}) comes from the
iterated value of $\gam(\p_1,\p_2,\p_3,\p_4;\Lam)$ of \eq{sm1lpans} and
gives \cite{mor:momexp} 
\be\label{smmain}
-12{{{\lam}^3}\over (4\pi)^4}{1\over
\Lam}\sum_{n=1}^{\infty}{{(-1)^{n}}\over 
{n(n+1)}} {1\over 2^n} \left(1 - {1\over
2^{n+1}} \right).
\ee
Using (\ref{smmainsum}), this sums to $6{{{\lam}^3}\over (4\pi)^4}{1\over
\Lam}[6\ln3+4\ln2-5\ln5-1]$.  

\begin{figure}[p]
 \begin{picture}(100,225)(-60,0)
   \psfrag{Series \eq{sm6pt1}}{\small Series \eq{sm6pt1}}
   \psfrag{Series \eq{sm6pt2}}{\small Series \eq{sm6pt2}}
   \psfrag{Series \eq{smmain}}{\small Series \eq{smmain}}
   \psfrag{Series \eq{smwfnans}}{\small Series \eq{smwfnans}}
   \psfrag{Partial sum contributions to beta function}{\small \hspace{-1.2cm}
   Partial sum contributions to $\beta$ function \ ${\lam^3 \over (4
   \pi)^4}$} 
   \psfrag{Number of terms in expansion}{\small Number of terms in
   expansion}   
   \includegraphics[scale=0.6]{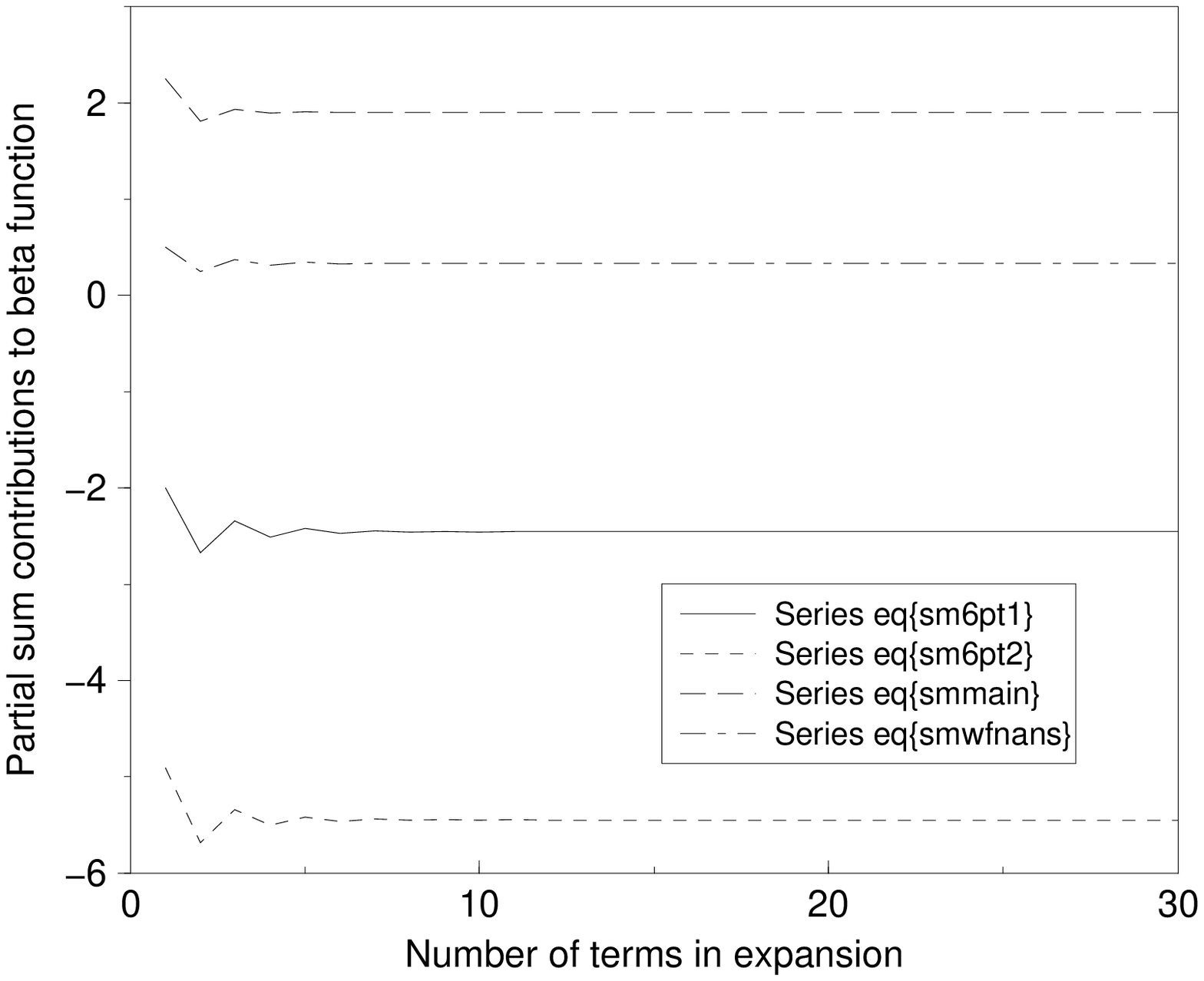}
 \end{picture}
\caption{Partial sum contributions to the $\beta$ function against number
  of terms in the expansion of the series of \eq{sm6pt1}, \eq{sm6pt2},
  \eq{smmain} and \eq{smwfnans}}
\label{fig:exp_con}
\end{figure}

\begin{figure}[p]
 \begin{picture}(100,230)(-60,0)
   \psfrag{Value of beta function / stuff}{\small Value of $\beta_1 \times
   (4 \pi)^4$} 
   \psfrag{Number of terms in expansion}{\small Number of tersm in
   expansion}
   \includegraphics[scale=0.6]{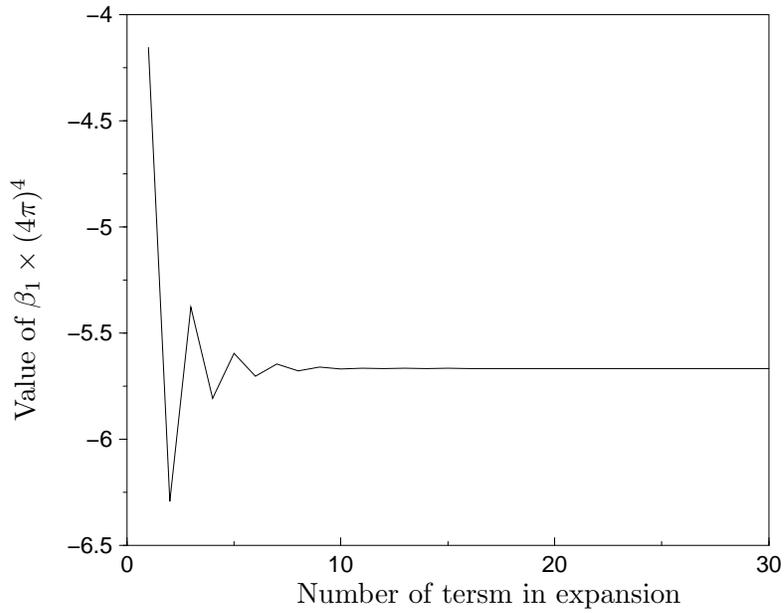}
 \end{picture}
\caption{Value of $\beta_1$ coefficient against number of terms in expansion}
\label{fig:exp_beta}
\end{figure}

As for the case of the sharp cutoff we need figure \ref{fig:wfn2lp} at
second order in external momentum to calculate wavefunction
renormalisation.  We have
\bea
\lefteqn{k^2{\partial\over\partial\Lam}Z(\Lam) =
-{\lam^2}\int\!\!{d^4q\over(2\pi)^4} 
\left({\partial\over\partial\Lam}{e^{-q^2/\Lam^2}\over
q^2}\right)} \nonumber 
\\ 
& & \left. \times \int_{\Lam}^{\infty}d\Lam_1\int\!\!{d^4p\over(2\pi)^4}
\left({\partial\over\partial\Lam_1}{e^{-p^2/{\Lam_1}^2}\over p^2}\right)
\left({1-e^{-|\p+\q+\k|^2/{\Lam_1}^2}\over |\p+\q+\k|^2}\right)
\right|_{O(k^2)} \nonumber
\\
& & = {{\lam^2}k^2\over (4\pi)^4}{1\over
\Lam}\sum_{n=2}^{\infty}{(-1)^{n+1}\over {2^n}}. \label{smwfnans}
\eea
Using  the binomial expansion 
\be
{1 \over 1+x} = \sum_{n=0}^{\infty}(-1)^nx^n,
\ee
and setting $x = {1 \over 2}$ enables the sum of (\ref{smwfnans}) to be
computed, and so 
\be
{\partial\over\partial\Lam}Z(\Lam) = -
{1\over 6}{{\lam^2}\over (4\pi)^4}{1\over\Lam}. 
\ee
In figure \ref{fig:exp_con} we display the partial sum contributions to the
$\beta$ function against the number of terms in the expansion for the
series coming from \eq{sm6pt1}, \eq{sm6pt2},  \eq{smmain} and \eq{smwfnans}. 
The $\beta$ function to two loops using an exponential cutoff is
found by adding together these series (each of which are separately 
convergent): 
\bea\label{betasm}
\lefteqn{\beta(\lam) = 3{\lam^2\over (4\pi)^2}
-{12\lam^3\over (4\pi)^4}\left\{\ln{4\over3} +
\right.}\nonumber
\\
& & \hspace{0.4in}
+\sum_{n=1}^{\infty}{(-1)^{n}}
\left[\ln{4\over 3}+{1\over12}\left(1\over2\right)^n
+{1\over n(n+1)}
\left(2\left(2\over 3\right)^n -{1\over 2^n}  
- {1\over 2^{2n+1}}\right)
\right. \nonumber
\\
& & \left. \left. \hspace{2in}
-{1\over n+1}\sum_{s=2}^{n+1}{{n+1}\choose s}{(-1)^s\over s-1}
\left\{1 - {1\over 2^{s-2}} + {1\over3^{s-1}}\right\}
\right]  \right\} \nonumber
\\
& & = 3{\lam^2\over (4\pi)^2} -{\lam^3\over (4\pi)^4}
\left[72\ln3-48\ln2-30\ln5+2.45411725+6-{1\over 3}\right],
\eea
which gives the expected form of (\ref{truebeta}).   The quick convergence
to the correct value of the $\beta_1$ coefficient is displayed in figure
\ref{fig:exp_beta}.

\section{Power law cutoff}

The final form of cutoff under consideration is that of a power law, \ie
$C_{UV}= 1/(1+(q^2/\Lam^2)^{\kappa + 1})$ where $\kappa$ is a non-negative 
integer. For a derivative expansion of the (non-perturbative) flow equation
\eq{legfl}, it would appear that if $\kappa$ is chosen such that $\kappa >
D/2 -1$ (with $D$ the spacetime dimension) all momentum integrals will
converge.  However within the method we have utilised in this chapter,
major problems arise.  If we consider  the integral
pertaining to figure \ref{fig:2loop} (b) obtained by iterating the one-loop
four-point function, we have
\bea\label{pow2lp}
\lefteqn{\sim {\lam^3\over \Lam^{2\kappa +3}} 
\int\!\!{d^4q}{q^{4\kappa}
\over[1+(q/\Lam)^{2\kappa+2}]^3} 
\int_{\Lam}^{\infty}\!\!{d\Lam_1\over \Lam_1^{2\kappa+3}} 
\int\!\!{d^4p}}
\nonumber \\
& & \hspace{0.25in} \times {p^{2\kappa}\over[1+(p/\Lam_1)^{2\kappa+2}]^2}
\left[1- {1\over 1+(|\q+\p|/\Lam_1)^{2\kappa+2}}\right]
{1\over {|\q+\p|^2}}. 
\eea
A derivative expansion requires an expansion in the external momentum of
the one-loop four-point function which translates as expanding the inner
integral in powers of the momentum $q$.  This expansion can be performed to
all orders.  However, once the power $q^{2m}$ is such that $m \ge \kappa+1$,
the outer integral over $q$ will cease to converge.  Hence with a power law
cutoff even the coefficients of the derivative expansion are infinite and
hence such an expansion ceases to make sense.

This problem arises purely because we are also working within a
perturbation expansion.  As stated earlier, non-perturbatively all
integrals will converge if $\kappa > D/2 -1$.  The improved behaviour here
can be traced to the $(q^2 + C_{IR}\Sigma)^{-3}$ factor contained in
\eq{legflsm} and \eq{Gsm}.  For small $\lambda$, $\Sigma\sim
\lambda^2q^{2m}$  at $O(\partial^{2m})$, from figure \ref{fig:wfn2lp}. The
extra powers of $q$ in the denominator always stabilise the integral
providing $\kappa>D/2-1$, but clearly the integral will then diverge as
$\lambda\to0$.

\section{Operators of higher powers}

The results presented in this chapter can, to a certain extent, also
address  the issue of convergence for operators of higher momentum powers
than the (zeroth order) $\beta$ function.  Let us first consider the sharp
cutoff of section \ref{sec:sharp}.

The $n$th term in the momentum scale expansion of an $O(\partial^{2r})$
operator is given by that of the derivative-free operator (with the same
number of fields) but with the $q^n$ in the expanded terms of
\eq{shmain2lp}  replaced by $ \sim {n/2 \choose
2r}q^{n-2r} k^{2r}$, with $k$ some external momentum.  For large $n$ this
will yield a multiplier $\sim n^{2r}$.  Thus, if convergence is to occur
for all operators, the coefficients of the expansion must fall faster than a
power of $n$. We saw that the series \eq{oscil} barely managed to converge
and the coefficients certainly did not fall faster than any power of $n$.
Without the need for further calculation, it is evident that the
contribution of figure  \ref{fig:2loop} (b) will have a momentum scale
expansion that ceases to converge at second or higher order in its external 
momenta.   In particular, the $O(k^{2r})$ $r \ge 1$ coefficients of   
\be
- \,  6{\lam}^3
\int\!\!{d^4q\over(2\pi)^4}{{\delta(q-\Lam)}\over q^2}  
\int_{\Lam}^{\infty}\!\!d\Lam_1
\int\!\!{d^4p\over(2\pi)^4}{{\delta(p-\Lam_1)}\over p^2} 
{\theta^2(|\p+\q|-\Lam_1)\over {|\p+\q|^4}}
\ee
will not provide a convergent momentum scale expansion.

The situation is much more promising with the  smooth exponential cutoff of
section \ref{sec:smooth}.  The series \eq{sm6pt2}, \eq{smmain} and
\eq{smwfnans} all fall faster than $1/ R^n$, with $R>1$ (\ie faster than a
power of $n$).  The equivalent of the troublesome diagram for the sharp
cutoff is series \eq{sm6pt1} which can best be analysed by recasting
the original integral
\bea
\lefteqn{-6 \lam^3 
\int\!\!{d^4q\over(2\pi)^4}
{{\partial\over{\partial\Lam}}\left(e^{-{q^2}/{{\Lam}^2}}
\right)\over q^2} \int_{\Lam}^{\infty}d\Lam_1
\int\!\!{d^4p\over(2\pi)^4}
{{\partial\over{\partial\Lam_1}}\left(e^{-{p^2}/{{\Lam_1}^2}}\right)
\over p^2}
{\left(1-e^{-{|\p+\q|^2}/{\Lam_1^2}}\right)^2\over |\p+\q|^4} }
\nonumber
\\
& & =
{6 \lam^3 \over \Lam^3 (4 \pi)^2}
\sum_{n=0}^{\infty}
\int\!\!{d^4q\over(2\pi)^4} {e^{-q^2/\Lam^2} \over q^2}
\int_{\Lam}^{\infty}d\Lam_1
{(-1)^n \over \Lam_1n!}
\left( {q \over \Lam_1} \right)^{2n} \hspace{3cm}
\nonumber \\
& & \hspace{6cm} \times
\int_0^1 da \left\{ \left( {1+a \over 2+a} \right)^n - \left( {a \over 1+a}
\right)^n \right\}
\eea
It is evident that the integral over $a$ is bounded from above by
$(2/3)^n$ and from below by zero.  Together with the $1/n!$ factor, this
provides a sufficiently fast rate of convergence. 

We can take this analysis further by repeating that used with the sharp
cutoff. For an $O(\partial^2)$ operator, the power of $q^{2n}$ of the
expanded terms of \eq{smmain2lp}, is replaced by $\sim
n^{2r}k^{2r}q^{2n-2r}$ (with $k$ again being some external momentum).
With the coefficients going like $\sim 1/ n!$ or better, the $q$ integral
will not completely cancel, leaving a remainder $\sim 1/n^{2r}$ for
large $n$.  Hence  we  see that  the derivative expansions of
higher derivative operators will converge.

\section{Summary and conclusions}

In summary, the derivative expansion for the $\beta$ function was
calculated at one-loop order for the Wilson/Polchinski flow equation and
was found to converge for certain fast falling profiles.  The equivalent
for the Legendre flow equation trivially converged as no expansion was
possible.  With a sharp cutoff used within the Legendre flow equation, slow
convergence was found for the two-loop $\beta$ function and it was demonstrated that
higher momentum 
operators have divergent momentum scale expansions.  While a power law
cutoff proved not to provide meaningful results, the most promising profile
was an exponential which, when used in the Legendre flow equation, has
rapid convergence of the derivative expansion of the $\beta$ function and
higher momentum operators.  The properties exhibited by the exponential
cutoff has resulted in it being favoured by many authors \cite{berv,
wett}.  

The technique of approximation using derivative expansions within exact RG
flow equations has been shown to be applicable in the calculation of
perturbative quantities.  However, it has been demonstrated that scalar field
theory is largely perturbative in nature both in $D=4$ \cite{luscher} and
$D=3$ \cite{zinn}, so the successes
recorded here also go some way to explaining the accurate results found
using derivative expansions in a non-perturbative setting (see \eg
\cite{mor:der, mor:momexp,lpa,denoper}).
 			
\chapter{Towards a gauge invariant exact RG}\label{super}

In this chapter we identify the manner of regularisation as one of the
stumbling blocks in the establishment of a gauge invariant RG. One approach
suggests a regularisation scheme using the ideas of supersymmetry might
exist and we introduce some of the necessary concepts before formulating
the regularisation scheme in the next chapter.  Unless otherwise stated,
this chapter is based upon material from sources\footnote{The mathematics
of supergroups is also discussed in \cite{dewitt}} \cite{mor:manerg, zinn,
itz, cornwell, bars}.

\section{Consequences of gauge invariance}

Problems arise when we attempt to carry over the concepts of the Wilsonian
RG to QFTs which have local internal symmetry groups.  To see this we need
only 
consider a theory consisting of just a gauge field  $A_{\mu}(x) \equiv
A_{\mu}^a(x)\, T_a$, where the $T_a$ are generators of the gauge group.
We define the covariant derivative $\nabla_{\mu}$ with coupling $g$ to be
$\nabla_{\mu} := \partial_{\mu} -ig A_{\mu}$ with the 
gauge field acting by commutation. We can take the bare Lagrangian to be  
\be\label{lgauge}
{\cal{L}}_{gauge} = \half \tr (F_{\mu \nu} F_{\mu \nu}),
\ee
where the field strength is given by $F_{\mu \nu} := {i \over g}[\nabla_\mu,
\nabla_\nu]$ and the trace is over group indices. The Lagrangian is
invariant under a gauge transformation of the 
type 
\be\label{gtran}
A_{\mu} \rightarrow \partial_{\mu} \omega -ig[A_{\mu}, \omega],
\ee
where $\omega$ is the gauge parameter. When we transform \eq{gtran} to
momentum space, the second term leads to a convolution over all
momentum. Thus the restrictions placed upon allowed values of
momentum that we employed in the scalar field theory, do not respect this
invariance.    

This restriction upon momentum space is known as a {\sl regularisation
scheme}. We therefore have to make a choice: either break gauge
invariance and hope to restore at a later stage \cite{gbreak} or find an
alternative regularisation scheme.  In the next section we consider whether
a regularisation scheme can be found that will enable Wilsonian ideas to be
applied,  yet allow us to retain gauge invariance. 

\section{Regularisation techniques}

Throughout the subject of QFT, one is confronted with physical calculations
that involve divergent integrals.  The need to manipulate these integrals
and to rigorously define the theory, provide the motivation for
regularisation - the process of making the integrals finite at intermediary
steps of the calculation.  Regularisation is achieved via a modification of
the theory  at high energies which renders all integrals finite in  
a manner that is determined by a single parameter.  This parameter can also
be tuned to a specified limit  in such a way 
as to regain the original (divergent) theory. It must be stressed that the
procedure of regularisation is completely separate from that of
renormalisation (in old-fashioned parlance the process by which divergences
are removed via redefinitions of the couplings). The final renormalised
theory is independent of regularisation technique utilised and the control
parameter will not appear in calculated physical quantities. 

There are a wide variety of regularisation techniques available in the
literature,  each with its own advantages and disadvantages and hence its
own area off applicability.  In this section, some of
the most important  methods of regularisation are introduced and the
reasons for their unsuitability as a regulating scheme for a gauge
invariant exact RG are discussed.

\subsection{Dimensional regularisation}

In the case of gauge invariant theories, the most widely utilised
regularisation is dimensional regularisation as this has the attractive 
property of maintaining gauge invariance at all stages.  The central idea
is to generalise the spacetime dimension from $D$ to an arbitrary (not
necessarily integer) dimension $d$.  For sufficiently small $d$, all the
Feynman diagrams are finite.  The UV 
divergences of the theory appear as simple or multiple poles at  $d=D$.
All symmetries, including gauge symmetries (although problems arise with
chiral fields), that are independent of the dimension of spacetime are
preserved.  

While this regularisation scheme is the most widely used and practical
technique for gauge theories, it suffers from a number of drawbacks if it
is to be incorporated into an exact RG framework.  Firstly, it does not sit
easily with the Wilsonian approach of suppression of high energy modes since
dimensional regularisation has no such physical interpretation.  Secondly,
it is not clear whether this scheme has any meaning non-perturbatively
since it is applied directly to (perturbative) Feynman diagrams and the 
non-perturbative capability of the Wilsonian approach is one which we would
like to preserve. 

\subsection{Pauli-Villars regularisation}

This technique \cite{pauli} (like momentum space cutoffs)  modifies the
behaviour of the propagator at high momenta.  It achieves this by
introducing fictitious particles with the same interactions but which have
no effect at low energies. However,  at high energies 
the propagator of the new field exactly cancels that of the
original one.  This is achieved by giving these Pauli-Villars particles large
masses, $M_k$.  Thus we have
\be
\Delta^{reg}_{\mu\nu}(p) = \Delta_{\mu\nu}(p) + \sum_k C_k \,
\Delta^k_{\mu\nu}(p,M_k) 
\ee
When the limit $M_k \rightarrow \infty$ is taken, the
regularisation is removed.

\subsection{Higher covariant derivatives}

This method attempts to extend the idea of a momentum space cutoff to the
realm of gauge theories \cite{cov}.  If we view the scalar
theory in Euclidean space, 
we have cutoff functions appearing as $C_{UV}(-\partial^2/\Lam^2)$.  Thus to
proceed in a gauge invariant manner we replace all ordinary derivatives by
covariant derivatives.  In this way we may hope to regularise the action of
\eq{lgauge} by replacing it with the following.
\be\label{hvgauge}
{\cal{L}}_{gauge} = \half \tr (F_{\mu \nu}\, C_{UV}(-\nabla^2/\Lam^2) \cdot F^{\mu
\nu}), 
\ee
with the dot signifying that the covariant derivatives act via
commutation. Unfortunately, this is not enough to completely regularise the
theory;  the insertion of covariant derivatives introduces fresh
interactions. If  the cutoff function $C_{UV}(-\nabla^2/\Lam^2)$ is a
polynomial in its argument of rank $n$, then the superficial degree of
divergence of a (1PI) Feynman diagram in $D$ dimensions is (ignoring
gauge-fixing terms which do not affect the argument)
\be\label{dodsimp}
{\cal D}_{\Gam} =  DL - (2n+2)I + \sum_{i=3}^{2n+4}(2n+4-i)V_{A^i}
\ee
where $L$ is the number of loops of the diagram, $I$ the number of internal
propagators it possesses and $V_{A^i}$ the number of vertices at which $i$
$A$ fields are present. Using the relations
\bea
L &=& 1 + I - \sum_i  V_{A^i}, \\
E  &=& -2I + \sum_i  i\, V_{A^i},
\eea
(with $E$ the number of external lines of the Feynman graph), \eq{dodsimp}
becomes 
\be
{\cal D}_{\Gam} = (D -4) L  - 2n (L-1) - E +4.
\ee
In $D=4$ (the case of most physical relevance), the rank $n$ can always be
chosen such that ${\cal D}_{\Gam}$ is always 
negative (and hence all diagrams are superficially convergent)
\emph{except} when $L=1$ and $E \le 4$, where this regularisation
fails. 

\subsection{Hybrid regularisation}\label{hd&pv}

This takes the methods of Pauli-Villars and higher covariant derivatives
and combines them to produce a gauge-invariant regularisation
\cite{hybrid, hybrid2}.  The higher covariant derivatives takes care of all
the divergences appearing in diagrams with more than one loop, and all but
a small subset at one loop which are taken care of by the Pauli-Villars 
regularisation. Unfortunately further one-loop divergences then typically
arise when the  
Pauli-Villars fields are external.  It may be argued that this could be
ignored on the grounds that the Pauli-Villars particles cannot be regarded
as physical.  However, these divergences will reappear in internal
subdiagrams embedded at higher loops.  This is referred to as the
problem of overlapping divergences.  It can be cured by adding yet more
Pauli-Villars fields and by carefully choosing their actions \cite{hybrid2}.
The Pauli-Villars fields need to appear
bilinearly in the action so that upon integrating out, they provide missing
one-loop counterterms.  Unfortunately this is not a property that can be
preserved by the exact RG framework.

In a series of papers \cite{mor:manerg, mor:erg1, mor:erg2} a manner of
constructing a $SU(N)$ gauge invariant exact RG was suggested.  Using the
insight that the freedom in the construction of exact RG equations amounts
to a general field redefinition \cite{latmor}, a flow equation is
formulated. The necessary regularisation is provided by a form of the hybrid 
regularisation.  However, for the
regularisation scheme to prove effective, many requirements were placed upon
the properties of the Pauli-Villars fields including the presence of a
wrong-sign gauge field, fermionic gauge partners and scalar fields.  The
required cancellations forced the flow equation itself to be  of
a complicated form \cite{mor:erg2}.  Other shortcomings of the scheme were
that it could only be applied at $N=\infty$ and that it suffered from the
problem of overlapping divergences. It was realised however that the
plethora of particles might be more elegantly described (except for a few
minor discrepancies) by embedding the $SU(N)$ gauge group within the larger
\SUNN supergroup and allowing this larger group to be spontaneously
broken.  Unlike the bilinear Pauli-Villars regularisation, the
spontaneously broken \SUNN structure can be preserved under exact RG flows.
The aim of chapter \ref{sunn} is to show that spontaneously broken \SUNN
with covariant derivatives provides a regularisation scheme to all loop
orders  at finite $N$.

\section{Supersymmetric groups}

An important concept that has attracted wide attention in the field of
theoretical physics since the 1970s is that of supersymmetry, a
symmetry which mixes bosons and fermions.  Usually it is considered as a
symmetry of the space-time background \cite{susy}. However, work where the
supersymmetry exists in the internal symmetry of the QFT in an ordinary
background actually predates this \cite{sunm} (although largely ignored at
the time)  and is what we shall concern ourselves with here.  In this
section we shall introduce some of the properties of these supergroups
within the context of the supergroups \SUNM and \SUNN.  

\subsection{Grading}

The set of integers provide the simplest example of a graded structure.  They
have the property of being either even or odd. With ordinary addition
denoted by the symbol `${\mathbf \cdot}$', the additive group of integers
has the following behaviour. 
\be\label{gradint}
\begin{array}{ccccc}
\mathrm{even} & {\mathbf \cdot} & \mathrm{even} & = & \mathrm{even,} \\
\mathrm{even} & {\mathbf \cdot} & \mathrm{odd}  & = & \mathrm{odd,}  \\
\mathrm{odd}  & {\mathbf \cdot} & \mathrm{odd}  & = & \mathrm{even.} 
\end{array}
\ee
This structure is the same as that of $Z_2$, the cyclic group of order 2:
\be
e{\mathbf \cdot}e = e, \hspace{5mm} e{\mathbf \cdot}a = a{\mathbf \cdot}e =
a, \hspace{5mm} a{\mathbf\cdot}a = e,
\ee
where $e$ is the identity and $a$ the other element of the group being
identified with even and odd integers respectively.  Hence,
the grading structure of (\ref{gradint}) is known as  $Z_2$ grading and
appears in the Lie superalgebras which are of interest in this work, with the
characteristic of being odd or even replaced by the property of being
fermionic or bosonic.

\subsection{\bSUNM}

An even supermatrix ${\bf M}$ is a $(p+q) \times (r+s)$ matrix partitioned
such that
\be\label{superM}
{\bf M} = \left(\begin{array}{cc} {\bf A} & {\bf B} \\
                                  {\bf C} & {\bf D} \end{array} \right),
\ee
where ${\bf A}$ (${\bf D}$) is a $p \times r$ ($q \times s$) submatrix whose
elements are even under the grading structure and ${\bf B}$ (${\bf C}$) is
a $q \times r$ ($p \times s$) submatrix constructed from odd elements.

The set of $(N + M) \times(N + M)$ even supermatrices define the Lie 
supergroup $U(N|M)$ (with $N \neq M$) if any element ${\bf G}$ of the set
satisfies the condition
\be\label{unmcond}
{\bf G^{\ddagger}\,G}=1,
\ee
where ${\bf G}^{\ddagger}$ denotes the adjoint of ${\bf G}$.  The adjoint
of a supermatrix is defined such that the adjoint of the matrix ${\bf M}$
of \eq{superM} is 
\be
{\bf M}^{\ddagger} = \left(\begin{array}{cc} 
\tilde{\bf A}^{\sharp} & \tilde{\bf C}^{\sharp} \\
\tilde{\bf B}^{\sharp} & \tilde{\bf D}^{\sharp} 
\end{array} \right),
\ee
where the tilde means that we take the transpose of the submatrix. The hash
operator represents the Grassmann adjoint of the submatrix and is defined
as follows.  In the vector space in which the submatrix ${\bf N}$ lies,
it can be written as
\be
{\bf N} = \sum_{\mu} (u_{\mu} + iv_{\mu}) \epsilon_{\mu},
\ee
where each $u_{\mu}$ and $v_{\mu}$ are real numbers and the
$\epsilon_{\mu}$ form a particular basis. Then we define
\be
{\bf N}^{\sharp} = \sum_{\mu} (u_{\mu} - iv_{\mu}) \epsilon^{\sharp}_{\mu},
\ee
where for this basis
\be
\epsilon^{\sharp}_{\mu} = \left\{ \begin{array}{l} 
\ph{-i}\epsilon_{\mu} \: \:  \mathrm{if} \, \epsilon_{\mu} \, \mathrm{is\
bosonic},\\ 
-i\epsilon_{\mu} \: \: \mathrm{if} \, \epsilon_{\mu} \, \mathrm{is\
fermionic}. 
\end{array} \right.
\ee
This implies for a general bosonic element written as $X + i Y$ ($X$ and
$Y$ real) that 
\be\label{adjbos}
(X + i Y)^{\sharp} = X - iY,
\ee
while for a general fermionic element $\Theta + i \Psi$ ($\Theta$ and
$\Psi$ real Grassmann numbers) we find
\be\label{adjferm}
(\Theta + i \Psi)^{\sharp} = - i\Theta- \Psi.
\ee

In turn, \SUNM is the subgroup of $U(N|M)$ whose supermatrices have the
additional property that 
\be\label{sunmcond}
\sdet \,{\bf G} = 1,
\ee
with the superdeterminant defined [again with regard to the supermatrix
of \eq{superM}] as 
\be\label{sdetdef}
\sdet\, {\bf M} = {\det \, ({\bf A} - {\bf B}\,{\bf D}^{-1}\,{\bf
C}) \over \det{\bf D}}.
\ee

Let $\H$ be a member of the Lie superalgebra of $U(N|M)$, partitioned as
an even supermatrix:
\be
\H = \left( \begin{array}{cc} {\bf P} & {\bf Q} \\
                              {\bf R} & {\bf S}	
            \end{array}
       \right).
\ee
Condition \eq{unmcond} implies that
\be
\H + \H^{\ddagger} = 0,
\ee
which in turn implies that
\bea
{\bf P} &=& - \tilde{{\bf P}}^{\sharp}, \\
{\bf S} &=& - \tilde{{\bf S}}^{\sharp}, \\
{\bf Q} &=& - \tilde{{\bf R}}^{\sharp},
\eea
where, for example $(\tilde{{\bf P}}^{\sharp})^i_{\gap j} = (P^j_{\gap
i})^{\sharp}$.  Using \eq{adjbos} and \eq{adjferm}, we find we are able 
to write the algebra of $U(N|M)$ in the $(N + M)$
dimensional fundamental representation  is of the form of $(N + M) \times(N
+ M)$ even supermatrices
\be 
{\cal H} = \left( \begin{array}{cc} H_N & \theta \\
                               \theta^\dagger & H_M	
               \end{array}
       \right).			
\ee
$H_N$ ($H_M$) is an $N\times N$  ($M \times M$) Hermitian  matrix. $\theta$
is a $N\times M$ matrix composed of  complex Grassmann numbers and
$\theta^\dagger$ is its Hermitian conjugate. Together $\theta$ and
$\theta^\dagger$ contain $2NM$ real anti-commuting Grassmann numbers.

Using the supermatrix of \eq{superM} as an example once more, the supertrace
is defined to be 
\bea
\str \,{\bf M} &=& \tr {\bf A} - \tr {\bf D} \\
             &=& \tr \,( \sig3 \,{\bf M}),
\eea
where we have taken the opportunity to introduce the $(N + M) \times(N +
M)$ version of the third Pauli matrix.\footnote{\ie $\sig3=
{\left(\!\begin{array} {cc}  \one_N & 0 \\ 0 & -\one_M \end{array}
\!\!\right)}$ with $\one_{N(M)}$ being the $N\times N \,(M\times M)$
identity matrix} The supertrace of supermatrices is cyclically invariant.

If we require that ${\cal H}$ is a member of the Lie superalgebra of
$SU(N|M)$, we take account of the
condition placed on the elements of the group \eq{sunmcond},  by noting
that (see Appendix \ref{app:superid})  
\be\label{superid}
\sdet \big (\exp({\bf M}) \big) = \exp (\str \,{\bf M}),
\ee
and hence require
\be\label{strless}
\str \,{\cal H} = 0.
\ee
Imposing \eq{strless} has the effect that the traceless
part of $H_N$ ($H_M$) can be identified with an $SU(N)$ ($SU(M)$) subgroup
with the traceful part giving rise to a $U(1)$. Hence the bosonic sector of 
\SUNM forms a $SU(N) \times SU(M) \times U(1)$ subgroup.  

As a concrete example of an \SUNM group we can consider $SU(2|1)$; a
supergroup that has been  studied  within the context of the
Standard Model \cite{su21}. A general element of the algebra may be written
as  
\bea
{\cal H} &=& {1\over 2} \left(
\begin{tabular}{c|c}
& $ \theta^1 - i \theta^2$ \\[-5mm]
$\ds \sum_{i=1}^3 \eta^i \sigma _i + \eta^4 \one_2$ & \\[-5mm] 
& $\theta^3 - i\, \theta^4$ \\ \hline
$\theta^1 + i \theta^2 \hspace{1cm} \theta^3 + i \theta^4$ & $2\, \eta^4$
\end{tabular}
\right) \\
&=& \sum_{m=1}^4 \eta^m U_m + \sum_{n=1}^4 \theta^{\mu} V_{\mu},
\eea
where $\sigma_i$ are the $(2 \times 2)$ Pauli matrices, $\eta^m$ are real
bosonic parameters and $\theta^{\mu}$ real fermionic variables. The bosonic
generators $U_M$ are
\be
\begin{array}{cc} \ds
U_i = {1 \over 2} \left( \begin{tabular}{c|c}
& $0$ \\[-4mm]
$\sigma_i$ & \\[-4mm]
& $0$ \\ \hline
$0 \hspace{6mm} 0$ & $0$ \end{tabular} \right),
&
U_4 = \left( \begin{tabular}{c|c}
$\half \hspace{6mm} 0$ & $0$ \\ 
$0 \hspace{6mm} \half$ & $0$ \\ \hline
$0 \hspace{6mm} 0$ & $1$ \end{tabular} \right), 
\end{array}
\ee
and the fermionic sector  generators $V_{\mu}$ are given by
\be
\begin{array}{cc} \ds
V_1 = {1 \over 2}\left( \begin{tabular}{c|c}
$0 \hspace{6mm} 0$ & $1$ \\
$0 \hspace{6mm} 0$ & $0$ \\ \hline
$1 \hspace{6mm} 0$ & $0$ \end{tabular} \right),
& \ds
V_2 = {1 \over 2}\left( \begin{tabular}{c|c}
$0 \hspace{6mm} 0$ & $-i$ \\
$0 \hspace{6mm} 0$ & $0$ \\ \hline
$i \hspace{6mm} 0$ & $0$ \end{tabular} \right),
\\[10mm] \ds 
V_3 = {1 \over 2}\left( \begin{tabular}{c|c}
$0 \hspace{6mm} 0$ & $0$ \\
$0 \hspace{6mm} 0$ & $1$ \\ \hline
$0 \hspace{6mm} 1$ & $0$ \end{tabular} \right),
& \ds
V_4 = {1 \over 2}\left( \begin{tabular}{c|c}
$0 \hspace{6mm} 0$ & $0$ \\
$0 \hspace{6mm} 0$ & $-i$ \\ \hline
$0 \hspace{6mm} i$ & $0$ \end{tabular} \right).
\end{array}
\ee
While these generators bear many similarities to those of $SU(3)$, note
that the fermionic generators $V_n$ close onto the bosonic generators by
\emph{anti}commutation, \ie $\{V_{\mu},V_{\nu}\} = C^m_{\mu\nu} \, T_m$.
However, by including the parameters in these
relations, we can retain the usual Lie commutation rule:
\be
[{\cal H}, {\cal H}^\prime] = i\, {\cal H}^{\prime \prime}.
\ee
For general \SUNM we have (in the adjoint representation) 
\be
{\cal H}^i_{\gap j} = \omega^A \, (T_A)^i_{\gap j},
\ee
where $\omega^A$ is a bosonic or fermionic parameter depending on the index
$A$. The first $N^2 + M^2 -1$ are chosen to be bosonic and the remaining
$2NM$ are fermionic.  

\begin{table}[!h]
\begin{tabular}{||l|c|c||} \hline \hline
& & \\[-3mm]
$\ds
A=0$ & $\sqrt{{NM \over 2 \, |N-M|}} \left( \begin{tabular}{c|c} 
${\one_N \over N}$ & $0$ \\ \hline
$0$ & ${\one_M \over M}$ \end{tabular} \right)$
& $U(1)$ \\[8mm] \hline 
& & \\[-3mm]
$A= 1, \ldots, N^2-1$ & $\left( \begin{tabular}{c|c}  
$\tau^{(N)}_A$ & $0$ \\ \hline
$0$ & $0$ \end{tabular} \right)$ & $SU(N)$ \\[8mm] \hline
& & \\[-3mm]
$A= N^2, \ldots, N^2 + M^2 - 2$ & $\left( \begin{tabular}{c|c}  
$0$ & $0$ \\ \hline
$0$ & $\tau^{(M)}_A$ \end{tabular} \right)$ & $SU(M)$ \\[8mm] \hline
& & \\[-3mm]
$
\begin{array}{l}
A=N^2+M^2 -1, \ldots, \\
\hspace{8mm}\ldots (N + M)^2 -1 
\end{array} 
$
&  
{\tiny $
\begin{array}{l}
\left( 
\begin{tabular}{ccc|ccc}
$0$ & $0$ & $\cdots$ & $1$ & $0$ & $\cdots$ \\
$0$ & $0$ & $\cdots$ & $0$ & $0$ & $\cdots$ \\
$\vdots$ & $\vdots$ & & $\vdots$ & $\vdots$ & \\ \hline
$1$ & $0$ & $\cdots$ & $0$ & $0$ & $\cdots$ \\
$0$ & $0$ & $\cdots$ & $0$ & $0$ & $\cdots$ \\
$\vdots$ & $\vdots$ & & $\vdots$ & $\vdots$ & 
\end{tabular}
\right) \\[15mm]
\left(
\begin{tabular}{ccc|ccc}
$0$ & $0$ & $\cdots$ & $-i$ & $0$ & $\cdots$ \\
$0$ & $0$ & $\cdots$ & $0$ & $0$ & $\cdots$ \\
$\vdots$ & $\vdots$ & & $\vdots$ & $\vdots$ & \\ \hline
$i$ & $0$ & $\cdots$ & $0$ & $0$ & $\cdots$ \\
$0$ & $0$ & $\cdots$ & $0$ & $0$ & $\cdots$ \\
$\vdots$ & $\vdots$ & & $\vdots$ & $\vdots$ & 
\end{tabular}
\right)
\end{array} $} \etc 
& Super \\[30mm] \hline \hline
\end{tabular}
\caption{Table of generators of \SUNM}
\label{tab:sunmgen}
\end{table}

The generators thus contain only ordinary numbers.  They are chosen to be
those of  table 
\ref{tab:sunmgen}, where $\tau^{(N)}_A$ are the traceless generators of
$SU(N)$ normalised such that 
\be
\tr \, \big(\tau^{(N)}_A \,\tau^{(N)}_B \big) = \half \del_{AB} \hspace{2cm} \left\{
\begin{array}{l} 1\leq A \leq N^2 -1  \\ 1\leq B \leq N^2 -1 \end{array}
\right. 
\ee
and similarly for $\tau^{(M)}$.

This enables us to define the super Killing metric of the group as 
\be\label{metric}
g_{AB} = 2\, \str \,(T_A \, T_B), 
\ee
and with the  generators normalised as in table \ref{tab:sunmgen}, this
results in \\
\begin{minipage}[!h]{6in} 
\bea\label{sunmmet}
g_{AB} = 
\left( \mbox{ {\small
\begin{tabular}{c|ccc|ccc|ccccc} 
$\pm 1$& &&& &&& &&&& \\ \hline
& $1$&&& &&& &&&& \\ 
& &$1$&& &&& &&&& \\ 
& &&$\ddots$& &&& &&&& \\ \hline
& &&& $-1$&&& &&&& \\
& &&& &$-1$&& &&&& \\ 
& &&& &&$\ddots$& &&&& \\ \hline
& &&& &&& $0$& $i$& && \\ 
& &&& &&& $-i$& $0$& && \\ 
& &&& &&& && $0$& $i$&  \\ 
& &&& &&& && $-i$& $0$&  \\ 
& &&& &&& &&&& $\ddots$ \\
\end{tabular} } }
\right) \\
\underbrace{\hspace{0.8cm}}_{U(1)}\,
\underbrace{\hspace{2cm}}_{SU_1(N)}\,
\underbrace{\hspace{2.6cm}}_{SU_2(N)}\,
\underbrace{\hspace{4cm}}_{\mathrm{Fermionic}}
\hspace{0.2cm} \nonumber \\ \nonumber
\eea \end{minipage}
with the sign of the $U(1)$ sector determined as positive for $N>M$ and
negative for $N<M$.  Note that while the metric is symmetric in the bosonic
part, it is antisymmetric in the fermionic sector, a fact that we express
as 
\be\label{sunmasym}
g_{AB} = g_{BA} \, (-1)^{\f(A)\,\f(B)},
\ee
by introducing $\f(A)$,  the grade of the index $A$ defined such that 
\be\label{grade}
\f(A) = \left\{ 
\begin{array}{l} 
0 \hspace{1cm} \mathrm{if \ } A \mathrm{\ is\ a\ bosonic\ index} \\
1 \hspace{1cm} \mathrm{if \ } A \mathrm{\ is\ a\ fermionic\ index}
\end{array}
\right.
\ee
We are also able to define another metric $g^{AB}$ which is the inverse of
that of  \eq{sunmmet}
\be
g^{AB}g_{BC} = g_{CB}g^{BA} =  \del^A_{\gap C},
\ee
with a sum over $B$. This  enables us to lower indices on the parameters:
\be\label{parlower}
X_A := g_{AB}\,X^B,
\ee
(note that it is the second index of the metric that is summed over; from 
\eq{sunmasym} it is clear that the ordering of indices is important), and
raise indices on the generators
\be\label{genraise}
T^A :=  g^{AB} \, T_B .
\ee

Since the generators of \SUNM form a complete set of $(N+M) \times (N+M)$
supertraceless matrices, we can derive (see Appendix \ref{app:nm}) a
completeness relation: 
\be\label{sunmcomp}
(T^A)^i_{\gap j}\,(T_A)^k_{\gap l} = {1 \over 2} \, (\sig3)^i_{\gap l}
\,\del^k_{\gap j} - {1 \over 2(N-M)}\,\del^i_{\gap j}\,\del^k_{\gap l} \, .
\ee

\subsection {\bSUNN}\label{sec:sunn}

It is evident from consideration of the denominator of the $U(1)$ generator
in table \ref{tab:sunmgen} and that of the last term in the completeness
relation \eq{sunmcomp} that a na\"{\i}ve setting of $N=M$ will not be
sufficient to define $SU(N|N)$.  All of the problems that arise can be traced
back to the $U(1)$ subgroup of the bosonic sector.  In the case of $SU(N|N)$, this
generator becomes proportional to the identity in $2N$ dimensions, 
$\one_{2N}$, and commutes with every other generator in the Lie algebra.
This will give rise to a number of interesting properties when \SUNN is
employed as a gauge group, a discussion of which we delay until chapter
\ref{sunn}.  In fact the $U(1)$ part has proved to be so unpalatable, that
some authors have dropped it completely \cite{bars}.  We will not take such
a drastic step (a deeper discussion of such subtleties is contained in
chapter \ref{sunn}).  Instead we 
note that the identity matrix does indeed have a special r\^{o}le to play
and so separate it from the other generators.

We split the generators $T_A = \{\one, S_{\alpha}\}$, \ie $S_{\alpha}$ are
the traceless generators of $SU(N|N)$, $A\equiv\{0,\alpha\}$ and $\alpha$ runs
from 1 to $4N^2 -1$, with the first $2(N^2 -1)$ of these being bosonic
indices. 
Once again we can define a super Killing metric as in \eq{metric}.  The
normalisation of the generators means that the metric is 
\bea\label{sunnmet} 
g_{AB} =
\left( \mbox{ \small
\begin{tabular}{c|ccc|ccc|ccccc}
$0$& &&& &&& &&&& \\ \hline
& $1$&&& &&& &&&& \\ 
& &$1$&& &&& &&&& \\ 
& &&$\ddots$& &&& &&&& \\ \hline
& &&& $-1$&&& &&&& \\
& &&& &$-1$&& &&&& \\ 
& &&& &&$\ddots$& &&&& \\ \hline
& &&& &&& $0$& $i$& && \\ 
& &&& &&& $-i$& $0$& && \\ 
& &&& &&& && $0$& $i$&  \\ 
& &&& &&& && $-i$& $0$&  \\ 
& &&& &&& &&&& $\ddots$ \\
\end{tabular} }
\right)\\
\underbrace{\hspace{0.8cm}}_{U(1)}\,
\underbrace{\hspace{2cm}}_{SU_1(N)}\,
\underbrace{\hspace{2.7cm}}_{SU_2(N)}\,
\underbrace{\hspace{4cm}}_{\mathrm{Fermionic}}
\hspace{0.2cm} \nonumber
\eea
Obviously we cannot define an inverse to this metric.  However if we
restrict ourselves to just the traceless $S_{\alpha}$ generators, we are
able to define
\be
h_{\alpha \beta} = h_{\beta \alpha} \,(-1)^{\f(\alpha)\,\f(\beta)}
= 2\, \str \,(S_{\alpha} S_{\beta}),
\ee
with the inverse $h^{\alpha \beta}$ determined by 
\be
h^{\alpha \beta}h_{\beta \gamma} = h_{\gamma \beta}h^{\beta \alpha} =
\del^{\alpha}_{\gap \gamma}. 
\ee
This then allows us to raise indices as in \eq{parlower} and \eq{genraise}.
Since the 
$S_{\alpha}$ generators form a complete set of supertraceless and traceless
matrices, a completeness relation can be constructed for them (see
Appendix \ref{app:nn}):
\be\label{sunncomp}
(S^{\alpha})^i_{\gap j}\,(S_{\alpha})^k_{\gap l} = {1 \over 2} \,
(\sig3)^i_{\gap l} \,\del^k_{\gap j} - {1 \over 4N}\left[ (\sig3)^i_{\gap
j} \, \del^k_{\gap l} + \del^i_{\gap j}\,(\sig3)^k_{\gap l} \right]\, .
\ee
This is most usefully cast in the following forms
\bea
\str(X S_{\alpha}) \, \str(S^{\alpha}  Y) &=& {1 \over 2} \,\str(X Y) 
- {1 \over 4N} \left[ \tr X \, \str Y + \str X \, \tr Y \right],
\label{comptree} \\[2mm]
\str(S_{\alpha}XS^{\alpha} Y) &=& {1 \over 2}\, \str X \, \str Y - {1 \over
4N} \tr (XY +YX), \label{comploop}
\eea
for arbitrary supermatrices $X$ and $Y$. In chapter \ref{sunn} we will use
\SUNN as a gauge group and demonstrate how it can act as a regulator.

\chapter{Regularisation via \bSUNN}\label{sunn}

As we saw in subsection \ref{hd&pv} it is possible to construct a gauge
invariant regularisation scheme by combining the techniques of
regularisation via covariant higher derivatives and Pauli-Villars fields.
It could also be noted that such a technique appears cumbersome and unsuited
to the exact RG approach.  In this
chapter we introduce an extension of these ideas in which the combination
of these methods appears more natural and also more promising as regards
the exact RG 
\cite{su:tver}--\cite{su:pap}.

\section{The action of the  regulating scheme}

In the this section we describe the action for the regulating scheme using
covariant derivatives in spontaneously broken \SUNN gauge theory. 

\subsection{The gauge field sector}

We start by introducing $\Amu$, the gauge field of \SUNN:
\bea\label{defa}
\Amu &\equiv& \Amu^{A} \, T_{A} \nonumber \\
&=& 
\Amu^{\alpha}\,  S_{\alpha} + \Amu^0\,  \one 
\nonumber \\[2mm]
&=& \left( \begin{array}{cc} A^1_\mu & B_{\mu} \\
                               \Bbar_{\mu} & A^2_\mu	
               \end{array}
       \right)
+ \left( \begin{array}{cc} \A^0_\mu & 0 \\
                           0 & \A^{0}_\mu \\
          \end{array}
       \right),
\eea
where $S_{\alpha}$ are  traceless and supertraceless generators of
$SU(N|N)$.  Note we have also included the unity 
generator and its associated bosonic  field in \eq{defa}.  The $A_1$ field
is the usual $SU(N)$ gauge boson which we wish to regulate, with  the $A_2$
field being a $SU(N)$ copy  which, as we 
shall see, will enter the Lagrangian with the wrong sign.  The $B$ field is
fermionic and will eventually play the r\^{o}le of the fermionic
Pauli-Villars 
regulating particles.

The Lagrangian we require will be ultra-violet regulated.  The first  step in
achieving  this is to utilise the supergroup variant of higher covariant
derivatives.  The covariant derivative is chosen to be 
\be \label{covder}
\nabla_{\mu} :=\partial_{\mu} - ig\Lam^{2-D/2}\Amu,
\ee
where we have chosen to make the coupling dimensionless by explicitly
including the appropriate powers of $\Lam$. The field strength is then given
by 
\be\label{field}
{\cal F}_{\mu\nu}:= \Lam^{D/2 -2}{i \over g}[\nabla_{\mu},\nabla_{\nu}]. 
\ee 
Using the wine notation explained
in Appendix \ref{sec:wine} we can then write the pure Yang-Mills part of the
action as
\be\label{YMact}
S_{YM} = {1\over {2}}{\cal F}_{\mu \nu} \{\c\}{\cal
F}_{\mu \nu}. 
\ee 
The function $\c$ that appears in the wine is chosen to be a polynomial in
its argument [in this case $(\nabla^2/\Lam^2)$] of rank $r$.  The action is
invariant under the gauge transformations
\be\label{agauge}
\del \Amu = {1 \over g}\Lam^{D/2 -2}  [\nabla_{\mu}, \omega].
\ee

There are two features of \eq{YMact} which must be commented upon.
Firstly, the $\A^0$ field plays no part in it. We note that all $\A$ field
interactions occur via commutators and since the  $\A^0$ field (uniquely)
commutes with everything, it cannot interact. Furthermore, because
$\str(\one \, T_A) = 0$, we  see that it cannot propagate and is
non-dynamical.  The effect of 
integrating over the $\A^0$ field in the partition function is therefore just to introduce an
(infinite) constant which can be factored out.  However, we are not allowed
to simply exclude $\A^0$ as gauge transformations do appear in the
$\one$ direction since the identity is generated by fermionic elements of
the superalgebra, \eg
\be
{1 \over 2} \left\{ 
\left( \begin{array}{cc} 0 & \one_N \\ \one_N & 0 \end{array} \right) , 
\left( \begin{array}{cc} 0 & \one_N \\ \one_N & 0 \end{array} \right)
\right\} = \one_{2N}.
\ee

An alternative procedure for tackling the troublesome $U(1)$ sector is that
favoured by ref.\ \cite{bars}.  This redefines the Lie bracket to ensure
that $\one_{2N}$ does not appear.  Thus the *bracket is given by
\be
[\; , \;]_{\pm}^* = [\; , \;]_{\pm} - {\one \over 2N} \tr [\; , \;]_{\pm} 
\ee
where $[\; , \;]_{\pm}$ is a graded commutator.\footnote{\ie it is a
commutator if at least one of its two arguments is bosonic or an
anticommutator otherwise.} The super Jacobi identity is still satisfied
since 
\bea
[\H_1, [\H_2, \H_3]^*]^* &=& [\H_1, [\H_2, \H_3]]^*  \label{jacstep} \\
&=&[\H_1, [\H_2, \H_3]] - {\one \over 2N} \tr [\H_1, [\H_2, \H_3]].
\eea
The equality in \eq{jacstep} follows upon the realisation that $\tr  [\H_2,
\H_3]$ is always bosonic.  Thus we can conclude that the *bracket is a
perfectly acceptable representation of the super Lie product.
Hence, a member of the Lie algebra may be written as $\omega^{\alpha}
S_{\alpha}$ and the gauge field as $\A \equiv \A^{\alpha} S_{\alpha}$, with
the commutators of \eq{field} and \eq{agauge} being replaced by the
*bracket.

These two alternatives actually amount to the same thing.  The r\^{o}le of
the *bracket is to set to zero all the structure constants that generated
$\one$. However, since the Killing supermetric that appears in the
`$\A^0$-free' representation vanishes in the $\one$ direction [\cf
\eq{sunnmet}], the interactions in the two choices are the same and hence
are physically equivalent.  We shall concentrate  on the `$\A^0$-free'
representation as it is more elegant.

The second aspect of \eq{YMact} worthy of comment is with regard to the
$A_2$ field.  Due to the properties of the supertrace and its position
within the $\A$ supermatrix, the $A_2$ propagator comes from 
\be
- \tr \left\{ (\partial_{\mu} A^2_{\nu} - \partial_{\nu} A^2_{\mu}) \,\,
c^{-1}\!\left(-{\partial^2 / \Lambda^2}\right) \,\partial_{\mu} A^2_{\nu}
\, ,
\right\}
\ee
\ie it has the wrong sign.  This has been interpreted as a sign of
instability and deemed physically unacceptable \cite{spoon}, but we argue
on the basis of the consideration of a quantum mechanical analogue
described in subsection \ref{sec:qm} that rather it is a loss of unitarity.
However, we expect this not to be problematic since such a loss of unitarity
is confined to terms that will disappear when the regularisation cutoff
($\Lam$) is removed.  

Of course more  has to be added to this scheme if we are to have a 
satisfactory regularisation technique for $SU(N)$ gauge theory.  The
problem is we have also altered the low energy physics of the embedded
$SU(N)$ Yang-Mills theory by the introducion of new fields.  To 
redress this shortcoming, we must ensure that the fermionic and $A_2$
fields only 
have a influence on the $A_1$ sector at high energies, and this can be
achieved by giving the fermionic fields large masses. The $A_2$ field can
only interact with the physically important $A_1$ gauge boson via the $B$
fields. 

It could be asked whether in giving the fermionic fields mass, we really
need to maintain the full \SUNN invariance 
as ultimately the only physically relevant group is one of the $SU(N)$
subgroups;
\ie could the $B$ fields be given mass by introducing explicit  mass terms.
Unfortunately if the action was of the form
\be
S= {1\over {2}}{\cal F}_{\mu \nu} \{\c\}{\cal F}^{\mu \nu}  
+ {1\over {2}} m B \bar{B},
\ee
the $B$ propagator would not be transverse (as such a property  is only
guaranteed by gauge invariance) and divergences would
appear in the longitudinal direction.  These can be regulated  by the
introduction of a scalar field \cite{mor:manerg}.  Since the appearance of
this scalar field seems to be essential, we incorporate it in the most
elegant method available, keeping  the full \SUNN invariance and
introducing spontaneous symmetry breaking. 

\subsection{Spontaneous symmetry breaking sector}

To this end we introduce a superscalar field
\bea
\C = \left( \begin{array}{cc} C^1 & D \\
                              \Dbar & C^2	
               \end{array}
       \right).
\eea
We require that the fermionic parts of the $\A$ field obtain masses so we
must spontaneously break in these (and only these) directions.  This is
achieved by introducing a non-zero vacuum expectation value along a
direction 
\be
\sig3 + a \one
\ee
($a$ real) in the Lie superalgebra.  Thus $\C$ must lie in the adjoint
of $U(N|N)$ but transform locally under $SUN(N|N)$.  Under gauge
transformations \eq{agauge}, $\C$ transforms as 
\be\label{ctrans}
\del \C = - i \, [\C , \omega].
\ee
It is \emph{not} possible to replace this commutator by the *bracket as the
result would not be gauge invariant in general.  This can be seen by
considering an example such as the supertrace of an $n^{\mathrm{th}}$ order
monomial that could arise in a potential term.  With the gauge
transformation given by $\del \C = - i [\C ,\omega]^*$, we find 
\be
\del \, \str \, \C^n = {i  n \over 2N} \, \str \, \C^{n-1} \, \,
\tr[\C,\omega],
\ee
\ie non-vanishing in general.  Thus the identity cannot be
excluded\footnote{We could still dispense with the $\A^0$ in a consistent
manner by using the *bracket for all pure gauge interactions but using the
usual commutator for interactions concerning $\C$ fields.} from the $\C$
field which can be expanded as   
\be\label{Cexp}
\C= \C^{0}\one_{2N} \, + \, \C^{\sigma}\sig3 \, + \, \C^{\alpha} \,
S_{\alpha}.  
\ee
In the unbroken action we introduce a kinetic term for
the $\C$ field and the usual form for the Higgs' potential
\be\label{unbCact}
S^{unbroken}_{\C} =
{1 \over 2} \nabla_{\mu}\cdot \C \{\ctil \}\nabla_{\mu}\cdot \C  + 
{\lam \over {4}} \Lam^{4-D}\, \str \dint (\C^2 - \Lam^{D-2})^2.
\ee
We have introduced another cutoff function, $\ctil$, which is chosen to be a 
polynomial of rank $\rtil$.  The combined action \eq{YMact} and
\eq{unbCact}, is invariant under the transformations of the fields
\eq{agauge} and \eq{ctrans}.  In contrast to the gauge field, the $\sig3$
and $\one$ components of $\C$ are dynamical.  They propagate into one
another through the term
\be
2N  \partial_{\mu}\C^0 \,\ctil(-\partial^2/\Lam^2) \,
\partial_{\mu}\C^{\sigma}. 
\ee

When we shift to the stationary point of the $\C$ field, the \SUNN gauge
group will be
spontaneously broken to $SU(N) \times SU(N)$ (\ie the
symmetries of the bosonic sector).  Upon expanding about the stationary
point (\ie $\C \rightarrow \Lam^{D/2-1} \sig3 + \C$), \eq{unbCact} becomes
\bea\label{brkCact}
\lefteqn {S^{broken}_{\C} = 
-{1 \over 2} g^2\Lam^2[\Amu,\sig3]\{\ctil\}[\A_{\mu},\sig3] 
-ig\Lam [\Amu,\sig3]\{\ctil\}\nabla_{\mu}\cdot \C }
\nonumber \\
& & \hspace{1cm} + {1 \over 2}\nabla_{\mu}\cdot \C \{\ctil
\}\nabla_{\mu}\cdot \C 
+{\lam \over 4} \Lam^{4-D} \, \str\dint(\Lam^{D/2-1}\{\sig3,\C\} +\C^2)^2.
\,\,
\eea
The first term of \eq{brkCact} gives a mass of order the effective
cutoff, $\Lam$ to the fermionic part of $\A$. The bosonic part of $\C$ also
gains a mass via the last part of \eq{brkCact}. The action given by
\eq{YMact} and \eq{brkCact} is invariant when the fields transform as
\eq{agauge} and 
\be\label{cbrkgauge}
\del \C \rightarrow -i\, [\C, \omega] - i \Lam^{D/2-1} \, [\sig3,  \omega].
\ee

\subsection{Gauge fixing sector}

As with all gauge invariant theories,  the gauge must be
fixed\footnote{This is true in standard perturbation theory as this procedure
is required to properly define propagators; gauge
invariant ERG does not require gauge fixing to calculate certain quantities
\cite{mor:manerg, mor:erg1,mor:erg2}.}  to extract 
physically relevant quantities from the theory.  
Otherwise when Greens' functions are computed, integrating over an infinite
number of copies 
of the same theory occurs, leading to spurious divergences being obtained.
Obviously the manner in which the gauge is fixed should not have an influence
on the physical predictions extracted from the theory.  At
this point we also note  that the second term in \eq{brkCact} gives
rise to  a term linear in both $\A$ and $\C$ which could prove troublesome.
As such, we follow the lead of 't Hooft who faced a similar problem
\cite{thooft} and fix the gauge in a manner  which enables  this contribution to the
action to be  cancelled.  

The following choice of gauge fixing function is made 
\be\label{gfunction}
F = \partial_{\mu}\A_{\mu} - ig{\Lam \over 2\xi} {\ctil \over
{\chat}}[\sig3,\C], 
\ee
where yet another new cutoff function, $\chat$, has been employed.
However, since this  term is not required to be gauge invariant, $\chat$
is not covariantised; \ie it is a polynomial of rank $\rhat$ in 
$(-\partial^2/\Lam^2)$ rather than $(-\nabla^2/\Lam^2)$.  The 
process of 't Hooft averaging results in  the gauge fixing contribution to
the action  being 
\bea\label{gaugeact}
S_{gauge} &=&  
\xi F\cdot \chat \cdot F  \nonumber  \\
&=&  
\xi (\partial_{\mu}\A_{\mu})\cdot\chat\cdot(\partial_{\nu}\A_{\nu})
-ig\Lam(\partial_{\mu}\A_{\mu})\cdot\ctil\cdot[\sig3,\C] 
\nonumber  \\
&\ph{=}& 
- g^2{\Lam^2 \over 4{\xi}}[\sig3,\C]\cdot{{\tilde{c}^{-2}} \over {\chat}}
\cdot[\sig3,\C] , \label{sgauge}
\eea
where we have used the notation ${\ds  u\cdot W\cdot v
\equiv \str \dint \, u(x)\,W(-\partial^2/\Lam^2)\,v(y)}$. When
this is combined with the other parts of the broken action, the second term
provides the required cancellation.  The final part of \eq{sgauge} contains
a mass term for the fermionic subfield of the superscalar $\C$. 

The final contribution to the action comes from the Faddeev-Popov
superghosts which are defined to be   
\be
\eta = \left( \begin{array}{cc} \eta^1 & \phi \\
                                         \psi & \eta^2	
               \end{array}
       \right). 
\ee
In the case of the usual bosonic symmetry groups, the process of gauge
fixing leads to the Faddeev-Popov determinant which can be rewritten in
terms of fermionic ghosts \cite{faddeev}.  Likewise we would na\"{\i}vely
expect the ghosts in our theory to have the opposite grading to that of the
gauge and scalar fields. However, it must be stressed that superfields are
actually of indeterminate grading and the usual requirement of
(anti)commutativity is replaced by (anti)cyclicity of the supertrace.  As
the grading stands however, we find that, as required, $\str \, \eta X =
-\str X \eta$ if $X$ is ghost number odd, but that $\str \, \eta X = \str X
\sig3 \eta \sig3$ if $X$ has even ghost number.

There is an elegant solution to this problem.  Since we are free to choose
whether different fermionic flavours commute or anticommute \cite{wein}, we
take the opportunity to introduce multiple grading.  As well as the usual
supergroup grading $\f$ [\cf \eq{grade}], we also assign a ghost grading $\g$.
All superfields (including ghosts) have supergroup-odd block off-diagonal
elements ($\f=1$) and supergroup-even block diagonal entries ($\f=0$).
$\A$ and $\C$ are both ghost-even ($\g=0$) while $\eta$ and $\bar{\eta}$ are
ghost-odd ($\g=1$).  We therefore require 
that elements commute up to a multiplicative extra sign whenever odd
elements of the same grading are pushed passed one another, \ie for elements
$a$ and $b$
\be
ab= ba (-1)^{\f(a)\f(b) + \g(a)\g(b)}.
\ee
We now find that 
\be
\str \, \eta X= (-1)^{g(X)} \, \str X \eta,
\ee
as required.  
The ghost action arises from the variation of  the gauge fixing
function \eq{gfunction} with gauge transformations \eq{agauge} and
\eq{cbrkgauge}.  We find that  
\be
S_{ghost} = 
- {2 \over g}\Lam^{D/2-2} {\bar\eta}\cdot\partial_{\mu}\nabla_{\mu}\cdot\eta  
-  g{\Lam \over {\xi}} \, {\ctil \over \chat} \,\str \dint \,
[\sig3,{\bar\eta}]  [\Lam^{D/2-1}\sig3 + \C,\eta] .
\ee
The contribution to the bare action can be 
tidied up by shifting the antighost variables $\bar{\eta} \rightarrow
g \Lam^{2-D/2}\chat \tilde{c} \bar{\eta}$.  We shall see in section \ref{sec:power} that
this shift has the added benefit of assigning the correct momentum
behaviour to the different legs of the ghost interaction vertices.   The
ghost sector of the action is then
\be\label{ghostact}
S_{ghost} = 
- 2{\bar\eta} \cdot \chat \tilde{c}\, \partial_{\mu}\nabla_{\mu}\cdot\eta
-  {g^2 \over {\xi}} \Lam^{3-D/2}\, \str \dint \, [\sig3,{\bar\eta}] 
[\Lam^{D/2-1}\sig3 + \C,\eta] .
\ee

\newpage

\subsection{Total action}

Gathering together all the elements of the action  we have 
\bea
S^{broken}&=& {1\over {2}}{\cal F}_{\mu \nu} \{\c\}{\cal
F}^{\mu \nu}
-g^2\Lam^2[\Amu,\sig3]\{\ctil\}[\A_{\mu},\sig3] 
-ig\Lam [\Amu,\sig3]\{\ctil\}\nabla_{\mu}\cdot \C 
\nonumber \\[2mm] &\ph{=}&
 +{1 \over 2} \nabla_{\mu}\cdot \C \{\ctil \}\nabla_{\mu}\cdot \C
+{\lam \over 4}\Lam^{4-D}\str\dint(\Lam^{D/2-1}\{\sig3,\C\} +\C^2)^2
\nonumber  \\[2mm] &\ph{=}&
+\, \xi (\, \partial_{\mu}\A_{\mu})\cdot\chat\cdot(\partial_{\nu}\A_{\nu})
+ ig\Lam(\partial_{\mu}\A_{\mu})\cdot\ctil\cdot[\sig3,\C] 
\nonumber \\[2mm] &\ph{=}&
- g^2{\Lam^2 \over 4 {\xi}}[\sig3,\C]\cdot{{\tilde{c}^{-2}} \over {\chat}}
\cdot[\sig3,\C]
- 2  \, {\bar\eta} \cdot \chat \tilde{c}\,
  \partial_{\mu}\nabla_{\mu}\cdot\eta  
\nonumber \\[2mm] &\ph{=}&
- {g^2 \over {\xi}} \Lam^{3-D/2} \, \str \dint \, [\sig3,{\bar\eta}] 
[\Lam^{D/2-1}\sig3 + \C,\eta] . \label{bigact}
\eea
Some of the Feynman rules for this action are contained in Appendix
\ref{sec:feyn}. To ensure that the high momentum behaviour of the $\A$
propagator is unaffected by the introduction of the scalar field and
gauge-fixing we are forced to bound the ranks of the polynomials:
\be\label{bounds}
\rhat \ge r > \rtil -1.
\ee
 
If we had not  spontaneously broken the symmetry, the action
would  be (in covariant gauge, $F=\partial_{\mu}\Amu$) 
\bea
S^{unbroken}&=& {1\over {2}}{\cal F}_{\mu \nu} \{\c\}{\cal
F}^{\mu \nu} \,
+ \,
{1 \over 2} \nabla_{\mu}\cdot \C \{\ctil \}\nabla_{\mu}\cdot \C
\nonumber \\[2mm]
&\ph{=}&  
+ \,
{\lam \over {4}} \Lam^{4-D}\, \str \dint (\C^2 - \Lam^{D-2})^2 \,
+ \, 
\xi (\partial_{\mu}\A_{\mu})\cdot\chat\cdot(\partial_{\nu}\A_{\nu})
\nonumber \\[2mm]
&\ph{=}& 
- \,
2\,  {\bar\eta} \cdot \chat \tilde{c}\,
\partial_{\mu}\nabla_{\mu}\cdot\eta, 
\label{unbroken}
\eea
a form which will prove to be of use later as many of the important
aspects of the physics (especially as regards to issues of  finiteness) can
be discovered by consideration of just the unbroken sector.

\section{Power counting} \label{sec:power}

Within the theory defined in \eq{bigact} and \eq{bounds}, the superficial
degree of divergence, ${\cal D}_{\Gam}$, of a 1PI diagram in $D$ spacetime
dimensions is calculated using the standard rules \cite{itz} to be
\bea 
{\cal D}_{\Gamma} &=& D L - (2r+2) \, I_{\A} -(2\rtil+2) \, I_{\C}
-(2 \rhat -2\rtil+2) \, I_{\eta}  + \sum_{i=3}^{2r+4} (2r+4-i) \, V_{{\A}^i}
\nonumber \\
&+& \sum_{j=2}^{2\rtil+2} (2\rtil+2-j) \, V_{{\A}^j {\C}}
+ \sum_ {k=1}^{2\rtil+2}(2\rtil+2-k) \,
V_{{\A}^k {\C}^2} + (2 \rhat -2\rtil+1) \, V_{{\eta}^2 \A}, \,\,
\label{initdod} 
\eea
utilising the following nomenclature: $L$ is the number of loops, $I_s$ the
number of internal propagators of type $s$, and $V_t$ the number of
vertices containing the set of fields $t$. We aim to show that for all but
a small sub-class of 1PI diagrams, the ranks of the polynomials can be
chosen so that $\dgam$ is negative.  This sub-class will then be shown to be
finite by other methods developed in sections \ref{sec:strmech} and
\ref{sec:ward}. Since the superficial degree of divergence of any given
diagram and all its connected subdiagrams is thus shown to be negative,
finiteness to all orders of perturbation theory follows from the convergence theorem \cite{itz}.

Unfortunately, \eq{initdod} 
does not adequately take account of 1PI diagrams with external antighost
legs. The formula treats the whole momentum dependence of the associated
$V_{\eta^2\A}$ vertex as if it was loop momentum, whereas it depends only
upon the (external) $\bar{\eta}$ line.  Thus the superficial degree of
divergence calculated via \eq{initdod} is overestimated in these diagrams.
To remedy this we include an extra term: $-(2\rhat-2\rtil
+1)E^{\A}_{\bar{\eta}}$, where $E^{\A}_{\bar{\eta}}$ is the number of
external antighost lines which enter a $V_{\eta^2\A}$ vertex.  The improved
formula for the superficial degree of divergence is \vspace{-3mm}
\bea 
{\cal D}_{\Gamma} &=& D L - (2r+2) \, I_{\A} -(2\rtil+2) \, I_{\C}
-(2 \rhat -2\rtil+2) \, I_{\eta}  + \sum_{i} (2r+4-i) \, V_{{\A}^i}
\nonumber \\
&+& \sum_{j} (2\rtil+2-j) \, V_{{\A}^j {\C}}
+ \sum_ {k}(2\rtil+2-k) \,
V_{{\A}^k {\C}^2} + (2 \rhat -2\rtil+1) \, (V_{{\eta}^2 \A} -
E^{\A}_{\bar{\eta}}). \nonumber   \\[-2mm]
\label{secdod}
\eea

The form of \eq{secdod} is unhelpful as written since diagrams are easier
to classify by external, rather than internal, propagators.  As such, we use
the geometric relations 
\bea
&&\hspace{-3.5cm} L= I_{\A} + I_{\C} + I_{\eta} + 1 - \sum_i V_{\A^i} - \sum_j V_{\A^j\C} -
\sum_k V_{\A^k \C^2} - V_{\eta^2 \A} - V_{\eta^2\C} - V_{\C^3} - V_{\C^4},
\nonumber \\[-1mm]
\hspace{2cm} \label{euler} 
\\ 
&&\hspace{-3.5cm}E_{\A} = -2I_{\A} + \sum_i i V_{\A^j\C} +  \sum_j j V_{\A^j\C} + \sum_k k
V_{\A^k \C^2} + V_{\eta^2 \A},\label{geo:a} \\
&&\hspace{-3.5cm}E_{\C} = -2I_{\C} + \sum_j V_{\A^j\C} + 2 \sum_k V_{\A^k \C^2} +
V_{\eta^2\C} + 3V_{\C^3} +4  V_{\C^4}, \label{geo:c}\\
&&\hspace{-3.5cm}E_{\eta}^{\A} + E_{\eta}^{\C} + E_{\bar{\eta}}^{\A} +
E_{\bar{\eta}}^{\C} = -2 I_{\eta} + 2V_{\eta^2\A} + 2 V_{\eta^2\C}.
\label{geo:gh} 
\eea
The Euler relation \eq{euler} assumes that all diagrams are connected since
the first term on the RHS  (denoting the number of connected components) has
been set to 1.  Note that in the relation \eq{geo:gh}, the external ghost and
antighost lines have been classified according to the vertex to which they
are attached. They satisfy the constraint $E_{\eta}^{\A} + E_{\eta}^{\C} = 
E_{\bar{\eta}}^{\A} + E_{\bar{\eta}}^{\C}$, and so \eq{geo:gh} can be
rewritten as
\be
E_{\bar{\eta}}^{\A} + E_{\bar{\eta}}^{\C} = - I_{\eta} + V_{\eta^2\A} +
V_{\eta^2\C}.  
\ee   

The four relations \eq{euler} -- \eq{geo:gh} are used to rewrite ${\cal
D}_{\Gamma}$ as  
\bea
{\cal D}_{\Gamma} &=& \ph{+}  (D-2r-4)(L-2) - E_{\A}- (r - \rtil - \rhat) E_{\C} -
2(r + \rtil - \rhat +1)E^C_{\bar{\eta}} 
\nonumber \\
&\ph{=}&  - (2r+3) E^{\A}_{\eta} - (r - \rtil +1)\sum_j V_{\A^j\C} + (r -
3\rtil -1) V_{\C^3} + 2\,(r - 2\rtil)V_{\C^4} \nonumber \\
&\ph{=}& + (r + \rtil - 2\rhat -1) V_{\eta^2\C} +
2(D-r-2). \label{findg} 
\eea

While it is straightforward to choose sufficient conditions so that all
diagrams (except certain one-loop cases) are superficially convergent, it
is trickier to ascertain those that are also  necessary.  The strategy we
adopt is to consider  one- and multi-loop diagrams separately and 
establish the sufficient conditions required to make ${\cal D}_{\Gamma}$ as 
negative in as many diagrams as possible.  We will introduce a theorem
which will demonstrate that some of these conditions are not strictly
necessary. We will then show the remaining  conditions are necessary by
considering examples where they are needed.  

We will temporarily relax the condition that $r$, $\rtil$ and $\rhat$ are
integers.  Instead we consider them as general real numbers and re-impose
the restriction to integers at the end.  In this case we then have to
impose the additional constraint  
\be\label{notint}
\rtil > -1,
\ee
which is required to ensure the high momentum behaviour of the $\C$
propagator is unaffected by the spontaneous symmetry mass term in
\eq{bigact}. 

\subsection{Multiloop diagrams}\label{sec:multi}

If we stipulate that $L>1$, all such 1PI diagrams can be made superficially
convergent merely by requiring that all the coefficients  in
\eq{findg} are negative. Hence the following sufficient conditions are 
obtained: 
\bea
r &>& D - 2, \label{rlow}\\
r  &>& -3/2, \label{r32} \\
r &<& 2\rtil,\label{rup} \\
\rhat &<& r + \rtil + 1, \label{rhatup}
\eea
as well as the assumed conditions \eq{bounds} and \eq{notint}. 

Combining the inequalities \eq{rlow}--\eq{rup}, we obtain a
lower bound upon $\rtil$ as well, namely $\rtil > \half \, \max(D-2,-{3\over
2})$. 
The lower bounds on $r$ and $\rtil$ are to be expected since the higher the
number of spacetime dimensions the more divergent the diagrams.  However,
there is no obvious physical reason why upper bounds such as \eq{rhatup}
are found and one is lead to suspect that such conditions are not necessary.
We can prove that these restrictions are not necessary by the use of the
following proposition:
\newtheorem{prop1}{Proposition}
\begin{prop1}
If we denote by $\S$ the collection of triples $(r,\rtil,\rhat)$ s.t.\
$\dgam < 0$, then $\forall \, (r_0,\rtil_0,\rhat_0) \in \S$, the subset
$\{(r,\rtil,\rhat) \, 
s.t.\ r \ge r_0, \, \rtil = \rtil_0, \, \rhat \ge \rhat_0, \, \rtil_0 \le r
\le \rhat\} \subset \S$.
\end{prop1}
{\it Proof}:\\ We note that \eq{findg} depends upon $\rhat$ as
$+2\rhat(E^{\C}_{\bar{\eta}} - V_{\eta^2\C})$.  This term is always non-positive
since it is not possible to have more external antighost lines entering
$V_{\eta^2C}$ vertices than $V_{\eta^2C}$ vertices themselves.  Thus
increasing $\rhat$ above $\rhat_0$ cannot increase $\dgam$.

The dependence upon $r$ is carried by
\bea
\lefteqn{
r\left(-2L +2 - E_C - 2E^{\C}_{\bar{\eta}} - 2E^{\A}_{\bar{\eta}} - \sum_j
V_{\A^j\C} + V_{\C^3} + 2 V_{\C^4} + V_{\eta^2\C}\right) \hspace{5cm} } 
\nonumber \\
& & \hspace{8cm}
= 2r\left(\sum_i
V_{\A^i} - I_{\A} \right),
\eea
with the equality following from \eq{initdod}.  Since every $V_{\A^i}$ must
be attached to  at least two internal $\A$ lines in a 1PI diagram, this
contribution is also non-positive and so increasing $r$ above $r_0$ does
not increase $\dgam$. \hfill $\Box$

\begin{figure}[tbh]
$
\begin{array}{ll}
\begin{picture}(100,120)(-80,-20)
\GlueArc(50,50)(40,0,180){5}{6}
\GlueArc(50,50)(40,180,360){5}{6}
\Gluon(10,50)(90,50){5}{6}
\put(49,-20) {(a)}
\end{picture} &
\begin{picture}(100,120)(-110,-40)
  \includegraphics[scale=0.35]{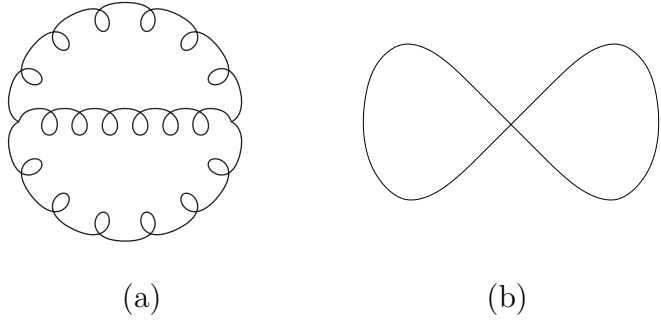}
\end{picture}
\put(57,0) {(b)}
\end{array}
$
\caption{1PI diagrams which by power counting alone require conditions
\eq{r:great} -- \eq{r&rtil} to be finite. (Curly lines represent $\A$ fields
and straight lines $\C$ fields.)}
\label{fig:multi}
\end{figure}

Proposition 1 implies that inequalities \eq{rup} and \eq{rhatup} are
unnecessary so the sufficient relations for convergence of multiloop
diagrams are:
\bea
r &>& \, \max\left( D-2, -{3 \over 2} \right),  \label{r:great}\\
\rtil &>& \, \max \left( {D \over 2} - 1, -{3 \over 4} \right), \\
\rhat \ge &r& > \rtil - 1. \label{r&rtil}
\eea
At first glance these conditions are apparently necessary to regulate the
diagrams of figure \ref{fig:multi} (in $D\ge \half$).  However the
na\"{\i}ve power counting we have employed does not take into account other
considerations  such as the supergroup factors.  It transpires
that these two diagrams are already regulated by the supertrace mechanism
that will be  discussed in section \ref{sec:strmech} and as such, the
necessity of the above conditions is actually unproven.  We leave a
demonstration of why these conditions really are necessary until after the
discussion of the one-loop case. 

\subsection{One-loop diagrams}

While the covariant derivatives are not able to regularise all one-loop
diagrams, they are able to ensure finiteness in a number of sub-classes.
At one-loop the superficial degree of divergence is 
\bea
\dgam^{1-loop} &=& D - E_{\A}- (r - \rtil +1) E_{\C} -
2(r + \rtil - \rhat +1)E^C_{\bar{\eta}} 
- (2r+3) E^{\A}_{\bar{\eta}} \nonumber \\
&\ph{=}&
- (r - \rtil +1)\sum_j V_{\A^j\C} + (r - 3\rtil -1) V_{\C^3} + 2\,(r -
2\rtil)V_{\C^4} \nonumber \\
&\ph{=}& 
 + (r + \rtil - 2\rhat -1) V_{\eta^2\C} \label{dgam1lp}
\\
&=& {\bf \alpha} \cdot {\bf v}, 
\eea
where we define the elements of ${\bf \alpha}$ and ${\bf v}$ to be
\be\label{alp&v}
\begin{tabular}{l@{\hspace{15mm}}l}
$\alpha_1 = D$ & $ v_1 = 1$, \\
$\alpha_2 = -1 $  &  $ v_2  = E_{\A} $, \\
$\alpha_3 = - (r - \rtil +1)$  &  $ v_3  = E_{\C} $, \\
$\alpha_4 = -2(r + \rtil - \rhat +1) $  &  $ v_4  = E^C_{\bar{\eta}} $, \\
$\alpha_5 = - (2r+3) $  &  $ v_5  = E^{\A}_{\bar{\eta}} $, \\
$\alpha_6 =  - (r - \rtil +1)$  &  $ v_6  = \sum_j V_{\A^j\C} $, \\
$\alpha_7 =  (r - 3\rtil -1)$  &  $ v_7  = V_{\C^3} $, \\
$\alpha_8  = 2\,(r - 2\rtil) $  &  $v_8  = V_{\C^4} $, \\
$\alpha_9  = (r + \rtil - 2\rhat -1) $  &  $v_9  = V_{\eta^2\C} $. \\
\end{tabular}
\ee
The general stratergy we shall follow is to consider specific classes of
one-loop diagrams.  With the strictures placed by these classes we  shall
then change some of the $v_i$ to ensure that  all $v_i$ are non-negative.
This is done in such a manner that \eq{dgam1lp} is unchanged so we must
also adapt the corresponding $\alpha_i$s.  To ensure that $\dgam$ is then
negative, we  require that all the $\alpha_i<0$.  This gives us a number
of sufficient conditions, some of which can be shown not to be necessary by
appealing to Proposition 1.  It then remains to show that the final list of
conditions are also necessary.  

The cases we consider are:

\noindent
{\bf (i) $\bf E_{\A} \ge D+1$; any number of $\bf E_{\C}$, $\bf
E^{\A,\C}_{\eta}$, $\bf E^{\A,\C}_{\bar{\eta}}$} \\
The combination $(E_{\A} -D -1)$ is always non-negative, so we make the
following replacements in \eq{alp&v} which leave \eq{dgam1lp} unchanged
\be
\begin{tabular}{l@{\hspace{15mm}}l}
$\alpha_1 \rightarrow \alpha_1=-1$, & $v_1 \rightarrow v_1 = E_{\A}-D-1$,\\
$\alpha_2 \rightarrow \alpha_2=-1$, & $v_2 \rightarrow v_2 = 1$.
\end{tabular}
\ee
All other $\alpha_i$ and $v_i$ remain  unaltered.
Requiring all the coefficients $\alpha_i$ to be negative results, after the
assumption of \eq{bounds} and \eq{notint}, in the following restraints being placed on the
parameters 
\bea
r &<&2\rtil, \label{1lpa}\\
2r &>& -3 , \\
\rhat &<& r + \rtil +1. \label{1lpc}
\eea

\noindent
{\bf (ii) $\bf E^{\A}_{\bar{\eta}} \ge 1$; any number of $\bf E_{\A}$, $\bf
E_{\C}$, $\bf E^{\C}_{\bar{\eta}}$}\\ 
The new variable will be $(E^{\A}_{\bar{\eta}}-1)$ rather than
$E^{\A}_{\bar{\eta}}$ so we need to change the following components in 
\eq{alp&v}
\be
\begin{tabular}{l@{\hspace{15mm}}l}
$\alpha_1 \rightarrow \alpha_1=(D-2r-3)$, & $v_1 \rightarrow v_1 = 1$, \\
$\alpha_5 \rightarrow \alpha_5=-(2r+3)$, & $v_5 \rightarrow v_5 =
(E^{\A}_{\bar{\eta}}-1)$. 
\end{tabular}
\ee
If all $\alpha_i$
coefficients are to be negative, \eq{1lpa}--\eq{1lpc} are regained  along
with the extra condition $D-2r-3<0$.

\noindent
{\bf (iii) $\bf E^{\C}_{\bar{\eta}} \ge 1$; any number of $\bf E_{\A}$,
$\bf E_{\C}$, $\bf E^{\A}_{\bar{\eta}}$}\\ 
The only changes to \eq{alp&v} that must be made are 
\be
\begin{tabular}{l@{\hspace{15mm}}l}
$\alpha_1 \rightarrow \alpha_1=D-2(r+ \rtil - \rhat +1)$, & $v_1 \rightarrow
v_1 = 1$, \\ 
$\alpha_4 \rightarrow \alpha_4=-2(r+ \rtil - \rhat +1)$, & $v_4 \rightarrow
v_4 = (E^{\C}_{\bar{\eta}} -1)$,
\end{tabular}
\ee
and we obtain the new constraint $D-2(r+\rtil-\rhat+1)<0$, which has to
replace the previous weaker bound \eq{1lpc} (for any $D \ge 0$).

\noindent
{\bf (iii) $\bf E_{\C} \ge 2$; any number of $\bf E_{\A}$, $\bf
E^{\A,\C}_{\eta}$, $\bf E^{\A,\C}_{\bar{\eta}}$} \\ 
With the new variable $(E_{\C} -2)$ introduced, we adapt  
\be
\begin{tabular}{l@{\hspace{15mm}}l}
$\alpha_1 \rightarrow \alpha_1=D-2(r- \rtil +1)$, & $v_1 \rightarrow
v_1 = 1$, \\ 
$\alpha_3 \rightarrow \alpha_3=-(r-\rtil)$, & $v_4 \rightarrow
v_4=(E_{\C}-2)$, 
\end{tabular}
\ee
and an additional constraint is found: $r-\rtil > {D \over 2} -1$.

\noindent
{\bf (iv) $\bf E_{\C} = 1$, $\bf
E_{\A}=E^{\A,\C}_{\eta}=E^{\A,\C}_{\bar{\eta}}=0$}\\ 
This gives rise to three possibilities since the internal loop can  be one
of three flavours.

\begin{tabular}{l@{\hspace{15mm}}l}
Internal $\A$ loop: & 
$\dgam^{1-loop} = D - 2r+2\rtil -2$ \\
Internal $\C$ loop: &
$\dgam^{1-loop} = D - 2\rtil-2$ \\
Internal ghost loop: &
$\dgam^{1-loop} = D +2\rtil - 2 \rhat -2$
\end{tabular}

We require $r -\rtil >{D \over 2} -1$ if the first diagram is to be finite
(which will also make the third diagram finite). The second diagram
requires the bound $\rtil > {D \over 2} -1$.  The more
general case with any allowed number of external $\A$ and (anti)ghost lines
does not change these conditions as they both contribute negatively to
\eq{dgam1lp}.  

\subsection{Final list of constraints}

By the use of Proposition 1 it is possible to remove the upper bounds in
these constraints.  The final list of constraints for both multi- and
single loop graphs is therefore:
\bea
r&>& \max \left( D-2, {D-3 \over 2}, -{3 \over 2} \right),
\label{rcon1}\\[1mm] 
\rtil &>& {1 \over 2} \max  \left( D-2, {D-3 \over 2}, -{3 \over 2} \right),
\\[1mm]
r-\rtil &>& {D \over 2} -1, \label{rcon3}
\eea
and, as ever, \eq{bounds} and \eq{notint}. With $D\ge 1$, these mean
\bea
r &>& D-2, \label{finr}\\
\rtil &>& {D \over 2} -1 ,\\
r-\rtil &>& {D \over 2} -1 ,\\
\rhat \,\,\,\, \ge &r& > \rtil -1 >0. \label{fincon} 
\eea
Suitable ranks for polynomials can be found by selecting integers which
satisfy these bounds.

We now address the question of the necessity of these conditions.  We noted
that the diagrams of figure \ref{fig:multi} seemed to demonstrate necessity. 
However, we have ignored supergroup factors and, when these are allowed for,
we find that the unbroken parts of these diagrams will actually disappear
at large loop momenta through the supertrace mechanism which will be
discussed in the next section (the broken parts are finite by power
counting).  Necessity will  
actually arise from the broken sector of the \SUNN gauge theory.  To see
this we need to borrow a result from the next section, namely \eq{twotraces},
which shows that unbroken one-loop corrections take the form of a product
of two supertraces over the external fields.  This carries over to the
broken theory as well except that $\left<\C\right> = \sig3 \Lam^{D/2-1}$
factors may also arise in these supertraces. Now, the condition
$r-\rtil>{D\over 2}-1$ 
arose from power counting the one-loop graph made by attaching an $\A$
propagator to the $\C\A^2$ vertex [\ie by inspection the vertex from 
$-ig\Lam [\Amu,\sig3]\{\ctil\}\nabla_{\mu}\!\cdot \C$ of \eq{bigact}].
Thus $r-\rtil>{D\over 2}-1$ is necessary for the contributions 
with group theory factor
$\str\,\C\,\str\,\sig3$. The condition $\rtil>D/2-1$ is necessary for 
finiteness of $(\str\,\C)^2$ contributions arising from attaching a $\C$ 
propagator to the $\str\,\C^4$ vertex. The final condition for any
$D\ge1$, namely $r>D-2$, already follows from combining these two.

\begin{figure}[tbh]
$
\begin{array}{ll}
\begin{picture}(100,120)(-60,-10)
\Line(-10,60)(50,60)
\GlueArc(80,60)(30,-180,180){5}{12}
\Vertex(49,60){2}
\put(60,0) {(a)}
\end{picture} &
\begin{picture}(100,120)(-110,-10)
\Line(-10,30)(110,30)
\Oval(50,60)(30,20)(0)
\put(50,0) {(b)}
\end{picture}
\end{array}
$
\caption{1PI diagrams from which the necessity of conditions
\eq{finr}--\eq{fincon} are demonstrated.}
\label{fig:necess}
\end{figure}

\newcounter{bean}
By inspection of \eq{dgam1lp} and use of subsection \ref{sec:multi} we
can deduce that 
the only diagrams that remain unregularised after the
imposition of the constraints listed above, have the following properties:
\begin{list}
{(\roman{bean})}{\usecounter{bean}
  \setlength{\rightmargin}{\leftmargin}}
\item One loop
\item Up to $D$  external $\A$ legs 
\item No external $\C$ or ghost legs
\item Do not have $\A^j\C$, $\C^3$, $\C^4$ or $\C\eta\bar{\eta}$ interactions 
\end{list}  

Diagrams with these properties will be known as `One-loop Remainder
Diagrams'.

\section{Supertrace mechanism}\label{sec:strmech}

The power counting arguments of the previous section are a demonstration of
the well established problem that the introduction of higher covariant
derivatives is not sufficient to regularise all one-loop diagrams in gauge
theories \cite{cov}.  The improvement in the high momentum behaviour of the
propagators is not enough to compensate for the number of new interactions
we have been forced to introduce. We obviously need further regularsiation
and this  is the reason the \SUNN gauge group has been used.  The aim is to
demonstrate that the extra fields introduced by  the supergroup provide a
mechanism for cancellation to occur between component fields and hence
regularise some of the remaining troublesome diagrams.  Actually, gauge
invariance arguments mean that  the one-loop diagrams with $3<E_{\A} \le D$
do not diverge in the manner one would expect from the  na\"{\i}ve power
counting, as will be demonstrated in section \ref{sec:ward}, and so we
concentrate on the cases with $E_{\A} \le 3$ in this section.

There are three varieties of One-loop Remainder Diagrams:
those with just $\C$, $\A$ or ghost internal propagators. We will calculate
the group theory factors for the unbroken parts of these diagrams 
and see that at large loop momentum they disappear, while we demonstrate
that the broken parts are finite by power counting arguments.

\subsection{One-loop Remainder Diagrams with $\A$
propagators}\label{sec:a1lprd}

The large momentum behaviour of the $\A$ propagator can be obtained from
\eq{Aprop} and is found to be 
\be\label{Aprophigh}
\big< \,  \A^{\alpha}(p) \, \A^{\beta}(-p) \, \big> 
= {c_p \over p^2} h^{\alpha \beta} \left[ \del_{\mu\nu} + {p_{\mu} p_{\nu} \over p^2}
\left( {\hat{c}_p  \over \xi c_p} - 1 \right) \right] + O(p^{2 \rtil -4r -4})
\ee
The second term on the RHS arises from the symmetry breaking; the
constraints introduced earlier mean that the parts of diagrams using these
symmetry breaking terms are finite.  We need to use pure $\A^i$ vertices
which either come from the unbroken interactions of \eq{YMact} with index
of divergence $2r+4-i$, or from the regularised mass term in \eq{brkCact}
with index $2\rtil +2 -i$: using the symmetry breaking part of the
propagator and/or the symmetry breaking vertices will result in the degree
of divergence of the ensuing integral being bounded by
$\dgam \le D- E_{\A} - 2(r-\rtil+1)<0$, \ie these contributions are
finite. 

Feynman diagrams are constructed by creating propagators using Wick
contraction between different supertraces originating from 
interactions. Concentrating on the group theory dependence only, we find
tree diagrams take the following form:
\be\label{atree}
\str(X \, \A) \, \str(\A \,  Y) \wick{20}{13} = \str(X \, Y) + \ldots,
\ee
where the ellipsis denotes group theory factors arising from broken symmetry
parts, $X$ and $Y$ are products of supermatrices,  and we have used the
completeness relation in the form of \eq{comptree}.  In general, a term of
the following structure should also be included 
\be\label{bummer}
-{1 \over 4N} (\tr X \, \str Y + \str X \, \tr Y).
\ee
If this term was required, it would imply that the propagation of only
$\A^A$ would be inconsistent (\ie $\A^0$ would also be needed) since such
terms, although they arise in the unbroken theory, actually break
$SU(N|N)$.  However we are saved by the fact that all $\A$ interactions
occur via commutators,\footnote{This is true in the $A^0$-free
representation; in the *bracket version, extra interactions play the same
r\^{o}le.} so by rearrangement $X$ and $Y$ can also be expressed as
commutators.  Since the supertrace of a commutator vanishes, so does
\eq{bummer}. 

One-loop diagrams are formed by Wick contracting within a supertrace.  From
the previous arguments, \eq{atree} and \eq{comptree}, we know this must be
of the form  
\bea
\lefteqn{ \hspace{-1cm}\str( \, [ \A, Z_1 ] \, Z_2 \, [ \A, Z_3 ] Z_4 )
\wick{33}{19}  
= {1 \over 2} \Big[ \, \str(Z_1 \, Z_2) \, \str(Z_3 \, Z_4) 
+ \str(Z_1 \, Z_4) \, \str(Z_2 \, Z_3)
} \nonumber \\
&& 
\hspace{1cm}
- \str (Z_1 \, Z_2 \, Z_3) \, \str(Z_4) 
- \str(Z_1 \,  Z_3 \, Z_4) \, \str(Z_2) \Big]
+ \cdots, \label{azero}
\eea
or 
\be\label{anought}
\str( \, \A \, [\A,Z_1] ) \wick{16}{6} = 0,
\ee
where $Z_i$ are products of superfields and again the ellipsis correspond
to suppressed (finite) terms from the broken sector.  The possible
$O\left({1 \over N}\right)$ corrections from \eq{comptree} cancel out for the
same reasons as above.  In the cases we are
interested in $E_{\A} \le 
3$, so the terms in  \eq{azero} yield either $\str \A =
0$ or $\str \one = 0$.  Thus we can conclude that One-loop Remainder
diagrams with $\A$ internal propagators are finite,  because their
contributions from the spontaneous symmetry breaking  sector are finite by
power counting, while the unbroken  part has   group theory factors which
disappear.

\subsection{One-loop Remainder Diagrams with $\C$ propagators}

From \eq{Cprop} we see that the large momentum behaviour of the $\C$
propagator is 
\be\label{Cprophigh}
\big< \,  \C^i_{\gap j}(p) \, \C^k_{\gap l}(-p) \, \big>
= {\tilde{c}_p \over p^2} \,  \del^i_{\gap l} \, (\sig3)^k_{\gap j} +
O(p^{-4-2m}), 
\ee
where $m = \min(2\rtil,\rhat)$.  Again we note that contributions arising
from the broken part of the theory are finite by power counting as the
degree of divergence of a one-loop diagram using the broken part of \eq{Cprophigh}
is bounded by $\dgam \le D-E_{\A} -2 -2 \min(\rtil,\rhat-\rtil) <0$, and we
have already shown  that we need not consider $\C\A^j$  interactions.

Tree diagrams have the form 
\be\label{ctree}
\str(X_1 \, \C) \, \str(\C \, X_2) \wick{20}{11} = \str(X_1 \, X_2) + \cdots,
\ee
where the ellipsis signifies contributions from the broken sector.  Note
that we do not have to address the issue of $O\left( {1 \over N}
\right)$ corrections here.  The one-loop diagram then takes the general
form  
\be\label{twotraces}
\str(\C \, Y_1 \, \C \, Y_2 ) \wick{16}{8}
= \str Y_1 \, \, \str Y_2 + \cdots, 
\ee
with broken sector contributions represented by the ellipsis.
Similarly to the previous situation, $Y_1$ and $Y_2$ are the products of
the remaining superfields, and with $E_{\A} \le 3$, this leaves us with
either $\str \one= 0 $ or $\str \A = 0$, and so One-loop Remainder Diagrams
with $E_{\A} \le 3$ and $\C$ internal propagators are also finite.

\subsection{One-loop Remainder Diagrams with $\eta$ propagators}

The analysis for these diagrams is the same as that for the $\A$
propagators in subsection \ref{sec:a1lprd}, which is unsurprising as $\eta$
is linked to gauge transformations by BRST (see subsection \ref{sec:brst}).
With the  symmetry breaking 
terms once more finite by power counting since their degree of divergence
is bounded from above by the already negative $D- E_{\A}-2(\rhat -\rtil
+1)$, the unbroken sector yields 
one-loop diagrams of the same form as the RHS of \eq{azero} and/or
\eq{anought}, and so we can draw the same conclusions with regard to
finiteness.

\subsection{Example of explicit calculation of supergroup factors}

In this subsection we present the explicit results of a calculation of the
supergroup factors of diagrams using the
Feynman rules of Appendix \ref{sec:feyn}.  This hides much of the simplicity
and elegance of the previous section since individual diagrams do not allow
for the fact that vertices appear only as commutators. 

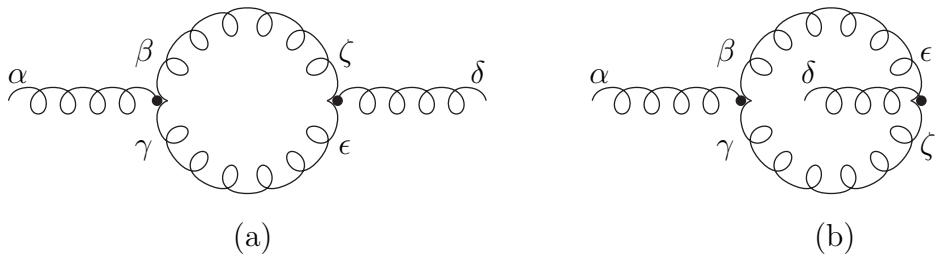
\begin{figure}[tbh]
\begin{picture}(390,115)(30,-10)
\GlueArc(150,50)(30,0,180){5}{6}
\GlueArc(150,50)(30,180,360){5}{6}
\Gluon(60,50)(116,50){5}{4}
\Gluon(184,50)(240,50){5}{4}
\Vertex(116,50){2}
\Vertex(184,50){2}
\put(60,57){$\alpha$}
\put(108,65){$\beta$}
\put(108,30){$\gam$}
\put(235,57){$\del$}
\put(185,30){$\epsilon$}
\put(185,65){$\zeta$}
\put(145,-5){(a)}
\GlueArc(370,50)(30,0,180){5}{6}
\GlueArc(370,50)(30,180,360){5}{6}
\Gluon(280,50)(336,50){5}{4}
\Gluon(360,50)(404,50){5}{3}
\Vertex(336,50){2}
\Vertex(404,50){2}
\put(280,57){$\alpha$}
\put(328,65){$\beta$}
\put(328,30){$\gam$}
\put(360,57){$\del$}
\put(405,65){$\epsilon$}
\put(405,30){$\zeta$}
\put(365,-5) {(b)}
\end{picture}
\caption{One-loop contributions to the {${\cal A}$} propagator}
\label{fig:propcor3}
\end{figure}

Since the Feynman rules have been derived for a strict cycle of fields,
diagrams are calculated by considering all possible topological variants.
For example, figure \ref{fig:propcor3} shows the two possibilities that arise
for the one-loop correction to the $\A$ propagator that uses just $\A^3$
vertices.

We are only interested in the high momentum behaviour
of such graphs; specifically we aim to demonstrate that the leading
contribution  vanishes in this
regime, with subleading terms arising from broken terms already shown to be
finite, and hence the diagram is UV regulated.  We will use only the single
index (\ie adjoint index) notation here as  the example uses only $\A$
fields. Of course the same results are obtained using the double index (\ie
fundamental and complex conjugate indices), which is the notation that
appears more natural if $\C$ fields are involved. 
Inspection of the momentum part of (\ref{Athreesig}) reveals that we need
not consider $\sig3$ insertions; 
such contributions to these diagrams are already finite by power counting.
If we take figure \ref{fig:propcor3} (b) as an example, the 
group theory part of the calculation comes from (suppressing Lorentz
indices and spacetime dependence):
\bea\label{grpb}
\lefteqn{
\A^{\alpha}\A^{\beta}\A^{\gam}\A^{\zeta} \wick{9}{6} 
\A^{\del}\A^{\varepsilon} \bigwick{24}{21} \,
\str(S_{\alpha}S_{\beta}S_{\gam}) \, 
\str(S_{\zeta}S_{\del}S_{\varepsilon}) }
\nonumber \\[3mm]
& &
\hspace{2cm}
\sim \A^{\alpha}\A^{\beta}\A^{\del}\A^{\varepsilon} \wick{14}{11}
\, h^{{\gam}{\zeta}}
\,\str(S_{\alpha}S_{\beta}S_{\gam}) 
\,\str(S_{\zeta}S_{\del}S_{\varepsilon})
\nonumber \\[2mm]
& &
\hspace{2cm}
\sim
\A^{\alpha}\A^{\del}\A^{\beta}\A^{\varepsilon}\wick{9}{6}
\,(-1)^{\f({\beta})\f({\del})}h^{{\gam}{\zeta}}\,\str(S_{\alpha}S_{\beta}S_{\gam})\,\str(S_{\zeta}S_{\del}S_{\varepsilon}) 
\nonumber \\[2mm]
& &
\hspace{2cm}
\sim
\A^{\alpha}\A^{\del}(-1)^{\f({\beta})\f({\del})}
h^{{\beta}{\varepsilon}}h^{{\gam}{\zeta}} \,
\str(S_{\alpha}S_{\beta}S_{\gam})\,\str(S_{\zeta}S_{\del}S_{\varepsilon})   
\nonumber \\[1mm]
& &
\hspace{2cm}
\sim \A^{\alpha}\A^{{\del}(b)}h^{{\beta}{\varepsilon}}h^{{\gam}{\zeta}} \,
\str(S_{\alpha}S_{\beta}S_{\gam})
\,\str(S_{\zeta}S_{{\del}({\beta})}S_{\varepsilon})   
\nonumber \\
& & \hspace{2.8cm}+
\A^{\alpha}\A^{\del(f)}h^{{\varepsilon}{\beta}}h^{{\gam}{\zeta}} \,
\str(S_{\alpha}S_{\beta}S_{\gam}) \,\str(S_{\zeta}S_{\del(f)}S_{\varepsilon}), 
\eea
where we have used the property of cyclicity under the supertrace and taken
the opportunity of Wick contracting two fields (to form an internal
propagator) on the
second line  using the high momentum behaviour exhibited in
(\ref{Aprophigh}). 
The notation $\del(b)$ $(\del(f))$ introduced in the final line means we only
consider the bosonic (fermionic) parts of the $\A^{\del}$ field.
Similarly for figure \ref{fig:propcor3} (a) the group theory part is
calculated to be
\be\label{grpa}
\A^{\alpha}\A^{\del}h^{\beta\zeta}h^{\gam\varepsilon} \, 
\str(S_{\alpha}S_{\beta}S_{\gamma})\,\str(S_{\zeta}S_{\del} S_{\varepsilon}) 
\ee
Now we utilise the completeness relations in the forms \eq{comptree} and
\eq{comploop}.  We then find that (\ref{grpb})
and (\ref{grpa}) both equate to
\be
-{1 \over 4N}\, \A^{\alpha}\A^{\del}\left[ \, \str(S_{\alpha}\sig3S_{\del}) +
\str(\sig3S_{\alpha}S_{\del}) \, \right] 
= -{1 \over 2N} \, \str \left( \sig3 \A ^2 \right)
\ee
However, (\ref{Athree}) shows us that the  $\A^3$ Feynman
rule is antisymmetric in the exchange of any two momenta.  Consequently
there is a relative minus sign between the two contributions of
figure \ref{fig:propcor3}, but since they have the same momentum and group
theory parts at large loop momentum, these two topologies cancel in this
regime. The same argument applies to the diagrams with two internal $\C$s or
ghosts.  Similar calculations have been performed (using the FORM algebra
manipulation package) for all One-loop
Remainder Diagrams with just two or three external $\A$ lines and these
diagram  were again shown 
shown to be finite.

\section{Ward identities}\label{sec:ward}

The only diagrams that now remain to be tested  whether or not they are
finite are the one-loop diagrams with $3 < E_{\A} \le D$ and $E_{\C} =
E_{\eta} = E_{\bar{\eta}} = 0$, originating from the unbroken theory.  In
this section, we shall use gauge invariant arguments to demonstrate that
the regularisation works up to $D=8$.  The key to doing this is the BRST
construction \cite{BRST}.  We use only  the unbroken action
\eq{unbroken} as we have seen that all contributions from the broken sector
are finite by power counting. 

\subsection{BRST}\label{sec:brst}

We introduce the BRST parameter $\epsilon$ which is even under the group grading but
odd under ghost grading. The BRST algebra is given as 
\bea
\del \A &=& \epsilon \Lam^{D/2 -2}   \, [\nabla_{\mu}, \eta],  \label{abrst}\\
\del \C &=& -i \epsilon [\C,\eta], \label{cbrst}\\
\del \eta &=& ig^2 \epsilon \eta^2, \label{ghbrst}\\
\del \bar{\eta} &=& \epsilon \, \Lam^{D/2-2} \, \xi \, \ctil \partial_{\mu}
\A_{\mu}. \label{aghbrst} 
\eea
The unbroken action \eq{unbroken} is invariant under these transformations,
as is the na\"{\i}ve functional measure.

The next stage is to construct the Lee-Zinn-Justin identities \cite{zinn}.
We need to add source terms for the fields and non-linear BRST
transformations:
\be\label{brstsrc}
S_{sources} = - \str \!\!\int \!\! d^Dx \left( \J_{\mu} \A_{\mu} + \J \C + \bar{\zeta}
\eta + \bar{\eta} \zeta + \Lam^{{D \over 2}-2}\K_{\mu}[\nabla_{\mu}, \eta] - ig \H[\C,\eta] +
ig \L \eta^2 \right).
\ee 
Here $\J$ is an unconstrained superfield
\be
\J = \left( \begin{array}{cc} J^1 & K \\ \bar{K} & J^2 \end{array} \right), 
\ee
but $\J_{\mu}$ (distinguished from $\J$ by the Lorentz index) expands only
over\footnote{If the *bracket formalism is used, the expansion is just over
$S_{\alpha}$.} $S_{\alpha}$ and $\sig3$:
\be
\J_{\mu} = 2\J_{\mu}^{\alpha} \, S_{\alpha} + {1 \over 2N} \J_{\mu}^{\sigma}
\sig3, 
\ee
so that 
\be
\str \J_{\mu} \A_{\mu} = \J_{\mu}^{\alpha} \A_{\mu\alpha} +
\J_{\mu}^{\sigma} \A^{0}_{\mu},
\ee
and these same constraints apply to $\zeta$, $\bar{\zeta}$, $\H$, $\K$ and
$\L$.  We define the functional differential so as to extract the conjugate
from under the supertrace \cite{mor:erg2}, \ie we require
\be
{\del \over \del J} \, \str \dint \, \J \C = \C,
\ee
so we have
\be\label{defderiv}
{\del \over \del \J} := \left( \begin{array}{cc} {\del \over \del J^1} & -
{\del \over \del \bar{K}} \\
{\del \over \del K} & -{\del \over \del J^2} 
\end{array} \right),
\ee
with a similar definition for $\del / \del \C$. Analogously, we choose
\bea
{\del \over \del \J_{\mu}}  &:=& 2 S_{\alpha} {\del \over \del
\J_{\mu\alpha}} + \one {\del \over \del \J^{\sigma}_{\mu}}, \\[2mm]
{\del \over \del \Amu } &:=& 2 S_{\alpha} {\del \over \del
\A_{\alpha \mu}} + {\sig3 \over 2N} {\del \over \del \A_{\mu}^{0}},
\eea
and likewise for the other field and source differentials.

Viewing the BRST transformations \eq{abrst}--\eq{aghbrst} as changes in
integration variables, we find
the generator of connected Green's functions $W = \ln Z$ satisfies the
following relation 
\be
\xi \, \zeta \Lam^{D/2-2}\cdot \ctil \cdot \partial_{\mu} {\del W \over \del \J_{\mu}}
+ \str \dint \left( \J_{\mu} {\del W \over \del \K_{\mu}} + \J {\del W
\over \del \H} - \bar{\zeta} {\del W \over \del \L} \right) =0.
\ee
We then perform the Legendre transformation to obtain the equivalent
equation for the generator of 1PI diagrams
\be
\Gam + \xi \,\partial_{\mu} \A_{\mu} \cdot \chat \cdot \partial_{\nu}
\A_{\nu} = -W + \str \dint \left( \J_{\mu} \A_{\mu} + \J \C + \bar{\zeta}\eta +
\bar{\eta} \zeta \right),
\ee
where $\A_{\mu}$, $\C$ and $\eta$ must now be viewed as classical fields.
The gauge fixing term  has been extracted so that upon using the antighost
Dyson-Schwinger equation
\be
\str \, T_A \left({\del \Gam \over \del \bar{\eta}} -2\Lam^{2-D/2}
\chat\tilde{c} \, \partial_{\mu} {\del \Gam \over \del \K_{\mu}} \right) = 0,
\ee
the simplified Lee-Zinn-Justin identities are obtained
\be\label{LZJ}
\str \dint \left( {\del \Gam \over \del \Amu} {\del \Gam \over \del
\K_{\mu}} + {\del \Gam \over \del \C} {\del \Gam \over \del \H} + {\del
\Gam \over \del \eta} {\del \Gam \over \del \L} \right) = 0. 
\ee

\subsection{Finiteness of diagrams with BRST source insertions}\label{sec:brstdiag}

An issue which must now be addressed is the finiteness (or otherwise) of the
new diagrams introduced by the BRST sources $\K_{\mu}$, $\H$ and $\L$ in
\eq{brstsrc}. Fortunately, since these interactions do not involve higher
derivatives it is straightforward to adapt the arguments of section \ref{sec:power} to show such
diagrams are superficially finite by power counting.

We first note that \eq{secdod} remains the same, but \eq{euler}--\eq{geo:c}
now become
\bea
&&\hspace{-1.8cm} L= I_{\A} + I_{\C} + I_{\eta} + 1 - \sum_i V_{\A^i} - \sum_j V_{\A^j\C} -
\sum_k V_{\A^k \C^2} - V_{\eta^2 \A} - V_{\eta^2\C} - V_{\C^3} - V_{\C^4}
\nonumber \\
&& \hspace{6cm} 
- V_{\K\eta} - V_{\K \A \eta} - V_{\H \C \eta} - V_{\L \eta^2},
\label{euler2} \\[2mm]
&& \hspace{-1.8cm}E_{\A} = -2I_{\A} + \sum_i i V_{\A^j\C} +  \sum_j j V_{\A^j\C} + \sum_k k
V_{\A^k \C^2} + V_{\eta^2 \A} + V_{\K \A \eta},\label{geo:a2} \\[2mm]
&&\hspace{-1.8cm}E_{\C} = -2I_{\C} + \sum_j V_{\A^j\C} + 2 \sum_k V_{\A^k \C^2} +
V_{\eta^2\C} + 3V_{\C^3} +4  V_{\C^4} + V_{\H\C\eta}. \label{geo:c2}
\eea
The ghost equation \eq{geo:gh} in the desired form 
\be
E^{\A}_{\bar{\eta}} + E^{\C}_{\bar{\eta}} = - I_{\eta} + V_{\eta^2 \A} +
V_{\eta^2 \C}
\ee
is unchanged, while we also have the new (trivial) relations
\bea
&& E_{\K} = V_{\K\eta} - V_{\K \A \eta} \\
&& E_{\H} = V_{\H \C \eta} \\
&& E_{\L} = V_{\L \eta^2} 
\eea
The net result of this is that $\dgam$ in the form \eq{findg} picks up the
new terms
\be\label{powbrst}
-(2r + 3) E_{\K} - (r+ \rtil +3) E_{\H} - (2r+4)E_{\L}.
\ee
Proposition 1 still holds as do the sufficient conditions
\eq{r:great}--\eq{r&rtil} since  these ensure that \eq{powbrst} provides a
negative contribution to $\dgam$. The only set of diagrams that remain
unregularised by the covariant derivatives are exactly those defined before
as One-loop Remainder Diagrams.  Thus all diagrams involving BRST source
terms are finite in any dimension $D$.

\subsection{Finiteness of one-loop diagrams using Ward Identities}

We write $\Gam$ in terms of its classical and one-loop parts, $\Gam = \Gam^0
+ \hbar \Gam^1$.  In the unbroken theory we expand the one-loop pure $\A$
vertices as (similar to \eq{winexp} and the double supertrace result of
the previous section)
\bea
&&\hspace{-1.5cm}
{1 \over 2 !} \sum_{n,m=2} {1 \over nm} \int \!d^Dx_1 \cdots d^Dx_n d^Dy_1
\cdots d^Dy_m
\,\,
 \Gam^1_{\mu_1 \cdots \mu_n, \nu_1 \cdots \nu_m} (x_1,
\cdots, x_n; y_1,\cdots, y_m)  
\nonumber \\
&& \hspace{1cm}
\str \, [\A_{\mu_1}(x_1)\cdots \A_{\mu_n}(x_n)] \,\,
\str \, [\A_{\nu_1}(y_1)\cdots \A _{\nu_m}(y_m)]  \label{gamA}
\eea
The only $O(\hbar)$ terms in \eq{LZJ} with one $\eta$ and otherwise only
$\A$s come from 
\be\label{mystery}
\str \dint \left({\del \Gam^1 \over \del \Amu} \, {\del \Gam^0 \over \del
\K_{\mu}} + {\del \Gam^0 \over \del \Amu} \,  {\del \Gam^1 \over \del
\K_{\mu}} \right),
\ee
and so we can deduce that
\bea
&& \hspace{-0.6cm}
p^{\mu_1}_1 \Gam^1_{\mu_1, \cdots \mu_n, \nu_1 \cdots \nu_m}(p_1, \cdots,
p_n; q_1,\cdots, q_m) = 
\Gam^1_{\mu_2, \cdots \mu_n, \nu_1 \cdots \nu_m}(p_1+p_2,p_3, \cdots,
p_n; q_1,\cdots, q_m) 
\nonumber \\
&& \hspace{1cm}
-\Gam^1_{\mu_2, \cdots \mu_n, \nu_1 \cdots \nu_m}(p_2, \cdots,
p_{n-1},p_n+p_1; q_1,\cdots, q_m) + \, \mathrm{finite}, \label{wid}
\eea
where `finite' denotes parts arising  from the second term of \eq{mystery}
(finiteness following from the results of subsection \ref{sec:brstdiag}).
Similar Ward identities can be obtained using the cyclicity and invariance
under the exchange of the two sets of arguments implied by \eq{gamA}
[similar to 
\eq{exid} and \eq{ccinv}].  If we set $n=m=2$, \eq{wid} and its
counterparts will relate the longitudinal parts of the four-point vertex to
the unbroken three-point vertices, which we know from the preceeding
section vanish. 
Hence, we know that the longitudinal part of the four-point vertex is finite in
any dimension. 

Thus a divergence, if it is to exist, has to arise in the totally
transverse part.  However, this part of the four-point 
vertex must have a tensor structure involving at least four
external momenta. This means that the superficial degree of divergence has
been over estimated by  four as these powers of momentum are not available
for use as loop momentum, \ie instead of $\dgam = D -4$ we have $\dgam
=D-8$.  Thus we can infer that the one-loop four-point pure $\A$ vertex is
finite in all dimensions $D<8$.

This argument can be extended to show the finiteness of all the remaining
diagrams.  The longitudinal part of the five-point pure $\A$ vertex is
related to the difference of finite\footnote{For $D<8$.} four-point
vertices plus finite corrections, while
the transverse part actually has $\dgam = D-5-5$ and so is finite for all
$D<10$. Proceeding in this manner we see that for $D<8$, the remaining
$3<E_{\A}<D$ One-loop Remainder diagrams are finite.  Thus all 1PI diagrams
are finite in $D<8$ as a consequence of a combination of 
power counting, the supertrace mechanism and gauge invariance.

\section{Unitarity} \label{sec:uni}

It was noted earlier that the supertrace gives rise to the wrong sign
action for certain fields such as $A^2_{\mu}$.  The functional integrals
over these field that appear in
the partition function need to be analytically continued in a manner
consistent with the \SUNN symmetry in order for them to make sense.
Equivalently, the system could be defined through exact RG methods
\cite{mor:erg1, mor:erg2} which do not suffer from such problems of
definition. Rather than being a sign of instability, \ie the choice of Fock
vacuum leading to an unbounded Hamiltonian, covariant quantisation with
these wrong signs results in the appearance of negative norm states.  These
states are unphysical and lead to a non-unitary S-matrix.  A simple quantum
mechanics example which demonstrates this point is given below.  The
situation here is similar in many ways to the Gupta-Bleuler quantisation
procedure \cite{itz}, which also has to handle the wrong sign
action for time-like photons in quantum electrodynamics. Again,  choices of
vacua exist, but Lorentz covariant quantisation picks out the ones with
negative norm states.  Unfortunately, we have no  Gupta-Bleuler condition
to exclude 
unphysical states.  Instead we will find that in the continuum limit $\Lam
\rightarrow \infty$,  $A^1_{\mu}$ and $A^2_{\mu}$ fields decouple enabling
a unitary $SU(N)$ Yang-Mills theory to be recovered in the $A_1$ sector.

\subsection{$\boldsymbol{U(1|1)}$ quantum mechanics} \label{sec:qm}

We define the Hermitian superposition $\X$ to be
\be
\X = \left( \begin{array}{cc} x^1 & \vartheta \\ \bar{\vartheta} & x^2
\end{array} \right),
\ee
and consider the Lagrangian (in Minkowski space) of a simple harmonic
oscillator:
\be
L \, = \, \half \, \str \, \Xdot^2 - \half \, \str \, \X^2.
\ee
Classically this Lagrangian is invariant under $SU(1|1)$ transformations
$\del \X = i \, [\omega, \X]$, however we also get for free invariance
under $U(1|1)$. By Noether's theorem, these transformations are generated
by the charges
\be\label{noedef}
\Q = i \, [\X, \Xdot],
\ee
through the Poisson bracket with $\str \,\omega \Q$.  Note that the charge
for $\omega \sim \one$ 
vanishes which is a reflection of its trivial action on $\X$.  With a
supercovariant derivative  defined as in \eq{defderiv}, the supermomentum
is given by
\be
\P := {\partial \over \partial \Xdot} L = \Xdot.
\ee
This differs by some convenient signs from the usual set of definitions.
We can then write the Hamiltonian as
\be
H = \str \, \P \, \Xdot - L,
\ee
while quantisation is via the graded commutator:
\be
\left[ \, (\X)^a_{\gap b}, (\P)^c_{\gap d} \, \right]_{\pm} = i \,
(\sig3)^a_{\gap d} \, \del ^c_{\gap b}.
\ee
By including arbitrary constant supermatrices $U$ and $V$, we can easily
see that this respects $U(1|1)$:
\be
\left[ \, \str \, U \X, \str \, V \P \, \right] = i \, \str\, UV,
\ee
and this actually corresponds to the usual relations using the usual
definitions for momenta:
\bea
p^i &=& {\partial L \over \partial x^i}, \label{cmptmom1} \\[2mm]
p_{\vartheta} &=& {\partial L \over \partial \vartheta}, \\[2mm]
p_{\bar{\vartheta}} &=& {\partial L \over \partial
\bar{\vartheta}}. \label{cmptmom3}  
\eea
Care needs to be taken since the na\"{\i}ve ordering suggested by
\eq{noedef} will not leave $\Q$ supertraceless after quantisation.  This
problem can be cured by subtracting the supertrace which corresponds, as
we will see, to normal ordering:
\be\label{qnorm}
\Q = i \, [\X, \P] - {i \over 2} \, \sig3 \, \str \,[\X, \P] = i\, [\X, \P]
+ 2 \sig3.
\ee
The annihilation and creation operators are chosen to be
\bea
A &=& {(\X + i \P) \over \sqrt{2}},  \label{Ann} \\[2mm]
A^{\dagger} &=& {(\X - i \P) \over \sqrt{2}}, \label{Cre}
\eea
with the normalised vacuum defined to be $A \vac = 0$.  These operators
have the expected graded commutation relations:
\be\label{Agradcom}
\left[ \, (A)^a_{\gap b}, (A^{\dagger})^c_{\gap d} \, \right]_{\pm} = 
(\sig3)^a_{\gap d} \, \del ^c_{\gap b}.
\ee
The vacuum respects $U(1|1)$ since $\Q \vac =0$ and we also note that the
supercharges \eq{qnorm} may be written as $\Q = :  [A^{\dagger},A]  :\,
$ \ie we introduce normal ordering. 

We can rewrite \eq{Ann} and \eq{Cre} in terms of components using the usual
definitions of momenta contained in \eq{cmptmom1}--\eq{cmptmom3}.  We then
find that $x^1$ has the usual form of annihilation operator, namely $a^1 =
(x^1 + i p^1) /\sqrt{2}$, but the one  for $x^2$ actually contains a wrong
sign: $a^2 
=(x^2 - ip^2)/\sqrt{2}$.  This gives rise to a wrong sign commutation
relation $[a^2, a^{2\dagger}] = -1$ as can be easily seen from
\eq{Agradcom}. This sign is precisely what is needed to compensate for the
wrong sign of $a^{2\dagger}a^2$ in the Hamiltonian, $H= \str \, A^{\dagger}
A +2$, and ensuring that is is bounded from below.  However, it
also results in negative norms appearing in the `2' sector. With the
normalised ket vectors in this sector given by
\be
|n> = {1 \over \sqrt{n !} } (a^{2\dagger})^{n} \vac,
\ee
we find that
\be
<n|n> = (-1)^n.
\ee
Any attempt to rectify this by altering the sign in $a^2$ results in an
unbounded Hamiltonian and a $U(1|1)$ and $SU(1|1)$ violating vacuum.

\subsection{Recovery of unitarity in $\boldsymbol {A^1}$ sector}

We established in sections \ref{sec:power}, \ref{sec:strmech} and \ref{sec:ward} that covariant
derivative spontaneously broken \SUNN theory is finite in all dimensions
$D<8$.  However, more is required if this is to be a suitable regulating
method for $SU(N)$ Yang-Mills theory.   We also need to establish that in
the limit $\Lam \rightarrow \infty$, $SU(N)$ Yang-Mills theory can be
recovered from the \SUNN scheme.

Except for $A^i_{\mu}$,  all fields become infinitely heavy\footnote{The
fermionic $\eta^i$ fields also remain massless. Strictly speaking, we
should take into account the effects of the ghosts and BRST in the following
analysis. However these do not alter the conclusions of the  
Appelquist-Carazzone decoupling theorem \cite{kazama}.} in the $\Lam
\rightarrow \infty$ limit and consequently drop out of 
the spectrum, so at low energies the gauge group is just $SU(N) \times
SU(N)$. We  need to ascertain that there is no interaction
between the two $SU(N)$ gauge fields, thus  enabling us  to ignore
the $A^2$ sector in this limit. 

Such a problem is addressed by the Appelquist-Carazzone decoupling theorem
\cite{appel}.  This theorem states that for a renormalisable theory, as the
mass scale of the heavy sector tends to infinity, the effective
Lagrangian is given by a renormalisable one for the light fields with
corrections which vanish by inverse powers of the heavy scale which is
identified with the overall cutoff of the effective theory.\footnote{For
example, this theorem is used to justify the assumption that a spontaneously
broken Grand Unified Theory is equivalent to the Standard Model $SU(3)
\times SU(2) \times U(1)$ at low scales.}  Our case is actually even
simpler than this as the heavy mass  and cutoff scales have always been
identified so we need not concern ourselves with subtleties arising from
exchanges of limits of these scales. It must be stressed that the
Appelquist-Carazzone decoupling theorem is only applicable to initially
renormalisable theories.  The standard analysis for Yang-Mills theory
carries over to the supergroup case, so we know that spontaneously broken
\SUNN is renormalisable in $D\le 4$.  

We therefore conclude that in $D\le4$ dimensions, the effective  $SU(N)
\times SU(N)$ theory can be described by an effective Lagrangian containing 
just these fields with couplings $g_i \neq g$, and  with other interactions
weighted by appropriate powers of $\Lam$ as determined by dimensions.  If
such an interaction between the $A^1$ and $A^2$ fields  was to exist, it
would have to 
contain at least two traces, one for each sector.  The lowest dimension
interaction comes from a group theory structure $\tr A^1_{\mu} A^1_{\nu} \, 
\tr A^2_{\rho} A^2_{\sigma}$ with the Lorentz indices somehow contracted.
To be gauge invariant under $SU(N) \times SU(N)$ it must take the form (up
to $\ln \Lam$ corrections)
\be
\Lam^{-D} \, \tr F^1_{\mu\nu}F^1_{\rho\sigma} \, \tr F^2_{\alpha \beta}
F^2_{\gam \del},
\ee
where $F^i_{\epsilon \eta}$ is the field strength of $A^i$ and the Lorentz
indices are again contracted in some fashion.  This is irrelevant in any
dimension and since it is the minimal dimension interaction, we know that all
other  interactions are irrelevant and disappear as $\Lam
\rightarrow \infty$.\footnote{This is true in $D\le4$ dimensions. In $D>4$
the couplings $g_i$ are non-renormalisable and higher order interactions
are unsuppressed.  Thus not only is $D\le4$ sufficient, it is also
necessary.}

The fact that we have decoupled sectors as $\Lam \rightarrow \infty$ is
actually a statement that unitarity has been restored in the $A^1$ sector.
A non-unitary amplitude can only arise from contributions with internal
$A^2$ propagators. Cutkosky cutting such an amplitude will then result in a
non-vanishing amplitude connecting the $A^1$ and $A^2$ sector \cite{itz}
which we have shown cannot exist in the continuum limit.

\section{Summary and conclusions}

A method of regularising $SU(N)$ Yang-Mills theory in a gauge invariant
manner in fixed dimensions $D\le4$ has been established. By the use of a
higher covariant derivatives all but a small number of one-loop 1PI
diagrams of spontaneously broken \SUNN gauge theory were shown to be 
finite.  In turn, these troublesome diagrams were themselves shown to be
finite in $D<8$ either by cancellations caused by the supersymmetry through the
supertrace mechanism, or by appealing to gauge invariance arguments via Ward
Identities.  For the scheme to provide
regularisation for $SU(N)$ Yang-Mills theory it is necessary to be
able to recover it when the regularisation scheme is tuned to a certain
limit.  The first stage in this was to introduce spontaneous symmetry
breaking so that all fields except the field we wish to regulate and a
wrong sign copy, gain mass.  When these masses are taken to infinity they
decouple from the massless fields.  The last issue to address is whether
the remaining massless fields interact in the continuum limit; if so, the
embedded $SU(N)$ theory would violate unitarity.  Fortunately, the
Appelquist-Carazzone theorem, applicable in $D\le4$, guarantees no such
interactions can exist.  

There are a number of obvious applications and extensions to this work.
Since these ideas first arose within the context of the exact RG, it would
be very appealing to construct a fully \SUNN invariant flow equation.
Another attractive aspect of this work would be to 
investigate  large $N$ Yang-Mills theory.  The interest here lies with the
fact that at large $N$, the supertrace mechanism ensures that there are no
quantum corrections in the symmetric phase.  Of course these ideas would
also benefit from the introduction of quarks (and their bosonic
superpartners) so direct comparison with physically important theories such
as quantum chromodynamics would be  possible.

\appendix
\chapter{Proof of eq.\ \eq{superid}} \label{app:superid}  

This proof of \eq{superid} is based upon that of \cite{cornwell}.
Suppose we have a supermatrix of the special form
\be
{\bf N} = \left( \begin{array}{cc} {\bf A} & {\bf B} \\
                                  {\bf 0} & {\bf D} \end{array} \right).
\ee
Then it is evident that
\be\label{Nform1}
\exp  {\bf N} = \left( \begin{array}{cc} \exp {\bf A} & {\bf B^{\prime}} \\
                                  {\bf 0} & \exp {\bf D} \end{array}
                                  \right),
\ee
for some complicated ${\bf B^{\prime}}$ whose exact form is not required.
Then by using the definition of the superdeterminant \eq{sdetdef} we find
\be
\sdet {\bf N} = {\det (\exp {\bf A}) \over \det (\exp {\bf D})}.
\ee
Since ${\bf A}$ and ${\bf D}$ are ordinary matrices, we can further
conclude that 
\be
\sdet{\bf N} = {\exp \tr {\bf A} \over \exp \tr {\bf D}}.
\ee
\ie the result \eq{superid} holds if ${\bf N}$ is of the form \eq{Nform1}.
Similarly, it holds if ${\bf N}$ is of the form 
\be\label{Nform2}
{\bf N} = \left(\begin{array}{cc} {\bf A} & {\bf 0} \\
                                  {\bf C} & {\bf D} \end{array} \right).
\ee
Next we use the relation (which we shall not prove here) $\sdet ({\bf R \,
S}) = \sdet {\bf R} \, \sdet {\bf S}$ (for any supermatrices ${\bf R}$ and ${\bf S}$), to
deduce that 
\bea \hspace{2cm} &\ph{=}& \hspace{-4cm}
\sdet \left[ \exp\left(\begin{array}{cc} {\bf A} & {\bf B} \\
                                  {\bf C} & {\bf D} \end{array} \right)
\exp\left(\begin{array}{cc} {\bf -A} & {\bf -B} \\
                                  {\bf 0} & {\bf -D} \end{array} \right) 
\right] \label{sdetrel1}
\\[2mm]
&=& \sdet \left[ \exp\left(\begin{array}{cc} {\bf A} & {\bf B} \\
                                  {\bf C} & {\bf D} \end{array} \right)
\right]
\sdet \left[\exp\left(\begin{array}{cc} {\bf -A} & {\bf -B} \\
                                  {\bf 0} & {\bf -D} \end{array} \right) 
\right]
\\[2mm]
&=& \sdet \left[ \exp\left(\begin{array}{cc} {\bf A} & {\bf B} \\
                                  {\bf C} & {\bf D} \end{array} \right)
\right]
\exp \left[ - \str \left(\begin{array}{cc} {\bf A} & {\bf B} \\
                                  {\bf 0} & {\bf D} \end{array} \right)
\right]
\\[2mm]
&=& \sdet \left[ \exp\left(\begin{array}{cc} {\bf A} & {\bf B} \\
                                  {\bf C} & {\bf D} \end{array} \right)
\right]
\exp \left[ - \str \left(\begin{array}{cc} {\bf A} & {\bf B} \\
                                  {\bf C} & {\bf D} \end{array} \right)
\right]
\eea
However we could use the Campbell-Baker-Hasudroff formula to rewrite
\eq{sdetrel1} as
\be\label{fromCBH}
\sdet \left[ \,\,\exp \left\{ \, \left(\begin{array}{cc} {\bf 0} & {\bf 0} \\
                                  {\bf C} & {\bf 0} \end{array} \right)
+ {\bf M^{\prime}} \, \right\} \,\,\right],
\ee
where ${\bf M^{\prime}}$ is a set of commutators partitioned as
\be
{\bf M^{\prime}} =\left(\begin{array}{cc} {\bf A^{\prime}} & {\bf 0} \\
                 {\bf C^{\prime}} & {\bf D^{\prime}} \end{array} \right). 
\ee
Since the matrix to be exponentiated in \eq{fromCBH} is of the form
\eq{Nform2} (\ie of a form for which  we know \eq{superid} holds),
\eq{fromCBH} becomes  
\be
\exp \left[ \,\, \str \left\{ \left(\begin{array}{cc} {\bf 0} & {\bf 0} \\
                                  {\bf C} & {\bf 0} \end{array} \right)
            + {\bf M^{\prime}}      \right\} \,\,\right]
 = 1,
\ee
since the supertrace of commutators vanishes.
Thus we know 
\be
\sdet \left[ \, \exp\left(\begin{array}{cc} {\bf A} & {\bf B} \\
                                  {\bf C} & {\bf D} \end{array} \right)
\right] \,\,\,
\exp \left[ - \str \left(\begin{array}{cc} {\bf A} & {\bf B} \\
                                  {\bf C} & {\bf D} \end{array} \right)
\right] \,= \, 1 \, ,
\ee
and that \eq{superid} holds for general supermatrices.

\chapter{Completeness relations}

\section{\bSUNM} \label{app:nm}

The generators of \SUNM provide a complete set of supertraceless matrices
(as can be demonstrated by a simple counting argument).  Hence a general
(non-super) matrix  denoted ${\bf X}$ can be extended in these generators
supplemented by any matrix with an non-vanishing supertrace.  For the
purposes of this derivation we employ the identity $\one_{N+M} =
{\left(\!\begin{array} {cc} \one_N & 0 \\ 0 & \one_M \end{array}
\!\!\right)}$ to perform the latter r\^{o}le.  Thus we have 
\be\label{xexp}
{\bf X} = X^AT_A + \tilde{X}\one_{N+M},
\ee
where $X^A$ and $\tilde{X}$ are coefficients and there is an implied sum
over $A$. Furthermore,
\be
\begin{array}{c}
\ds X^A = 2 \, \str (T^A {\bf X}), \\
\ds \tilde{X} = {1 \over N-M}\,  \str({\bf X}).
\end{array}
\ee
Thus we can re-express \eq{xexp} as
\be
(X)^i_{\gap j} = 2 \,(\sig3)^k_{\gap l}\, (T^A)^l_{\gap m} \,(X)^m_{\gap k}\,
(T_A)^i_{\gap j} + {1 \over N-M} \,(\sig3)^k_{\gap l} \,(X)^l_{\gap k}\,
\del^i_{\gap j} \,.
\ee
Since ${\bf X}$ is an arbitrary matrix, this implies that
\be
\del^i_{\gap m} \, \del^k_{\gap j} = 2 \,(\sig3)^k_{\gap l} \,(T^A)^l_{\gap m}
\, (T_A)^i_{\gap j} + {1 \over N-M}\, (\sig3)^k_{\gap m} \,\del^i_{\gap j} ,
\ee
which, after re-arrangement and relabelling, returns the form of
\eq{sunmcomp}:
\be
(T^A)^i_{\gap j}\,(T_A)^k_{\gap l} = {1 \over 2} \, (\sig3)^i_{\gap l}
\,\del^k_{\gap j} - {1 \over 2(N-M)}\,\del^i_{\gap j}\,\del^k_{\gap l} \, .
\ee

\section{\bSUNN} \label{app:nn}

The $\ S_{\alpha}\ $ generators of \SUNN as defined above \eq{sunnmet} form a
complete  set of traceless and supertraceless matrices.  As a consequence, a
general (non-super) matrix ${\bf Y}$ can be expanded in terms of $S_{\alpha}$,
$\one_{2N}$ and $\sig3$, \ie
\be
{\bf Y} = Y^{\alpha}S_{\alpha} + \tilde{Y}\one_{2N} + \hat{Y}\sig3,
\ee
where there is a sum over $\alpha$ and the coefficients $Y^{\alpha}$,
$\tilde{Y}$ and $\hat{Y}$ are determined by
\be
\begin{array}{c} 
Y^{\alpha} = 2\,\str(S^{\alpha}{\bf Y}), \\
\tilde{Y} = {1\over 2N} \, \tr ({\bf Y}), \\
\hat{Y} = {1\over 2N} \, \str ({\bf Y}),
\end{array}
\ee
which means that
\be
(Y)^i_{\gap j} = 2 \, (\sig3)^k_{\gap l} \,(S^{\alpha})^l_{\gap m} \,
(Y)^m_{\gap k} \, (S^{\alpha})^i_{\gap j} + {1 \over 2N} \,(Y)^k_{\gap k} \,
\del^i_{\gap j} + {1 \over 2N} \,(\sig3)^k_{\gap l}\,(Y)^l_{\gap k}
\,(\sig3)^i_{\gap j} \, .
\ee
Using the fact that ${\bf Y}$ is an arbitrary matrix, we deduce that 
\be
\del^i_{\gap m} \, \del^k_{\gap j} = 2 \,(\sig3)^k_{\gap l}
\,(S^{\alpha})^l_{\gap m} \, (S_{\alpha})^i_{\gap j} + {1 \over 2N}
\,\del^k_{\gap m}\del^i_{\gap j} + {1 \over 2N} \,(\sig3)^k_{\gap m}
\,(\sig3)^i_{\gap j} \, ,
\ee
leading to the completeness relation for the $S_{\alpha}$ generators of
\SUNN \eq{sunncomp}
\be
(S^{\alpha})^i_{\gap j}\,(S_{\alpha})^k_{\gap l} = {1 \over 2} \,
(\sig3)^i_{\gap l} \,\del^k_{\gap j} - {1 \over 4N}\left[ (\sig3)^i_{\gap
j} \, \del^k_{\gap l} + \del^i_{\gap j}\,(\sig3)^k_{\gap l} \right]\, .
\ee

\chapter{Wine notation}\label{sec:wine}

\begin{figure}[tbh]
 \begin{picture}(100,100)(-20,0)
   \psfrag{u}{$u$}
   \psfrag{v}{$v$}
   \psfrag{=}{$=$}
   \psfrag{+}{$+$}
   \psfrag{dots}{$\cdots$}
   \psfrag{Amu}{\small $\A_{\mu}(p_1)$}
   \psfrag{Anu}{\small $\A_{\nu}(q_1)$}
   \includegraphics[scale=0.6]{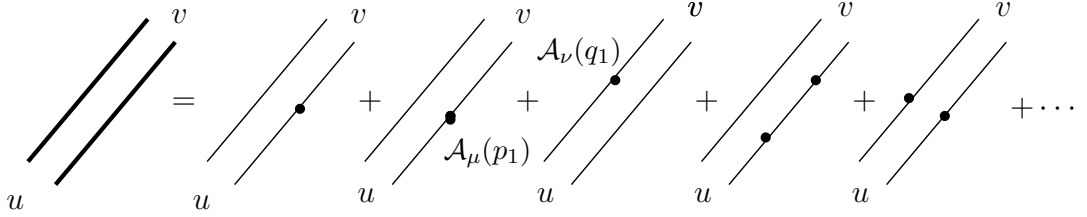}
  \end{picture}
\caption{Wine expansion, where the thick lines represent the full series.}
\label{fig:winexp}
\end{figure}

\begin{figure}[hbt]
 \begin{picture}(100,150)(-100,0)
   \psfrag{p1}{$p_1$}
   \psfrag{p2}{$p_2$}
   \psfrag{p3}{$p_3$}
   \psfrag{pn}{$p_n$}
   \psfrag{q1}{$q_1$}
   \psfrag{q2}{$q_2$}
   \psfrag{q3}{$q_3$}
   \psfrag{qn}{$q_n$}
   \psfrag{u1}{$\mu_1$}
   \psfrag{u2}{$\mu_2$}
   \psfrag{u3}{$\mu_3$}
   \psfrag{un}{$\mu_n$}
   \psfrag{v1}{$\nu_1$}
   \psfrag{v2}{$\nu_2$}
   \psfrag{v3}{$\nu_3$}
   \psfrag{vn}{$\nu_n$}
   \psfrag{r}{$r$}
   \psfrag{s}{$s$}
   \includegraphics[scale=0.35]{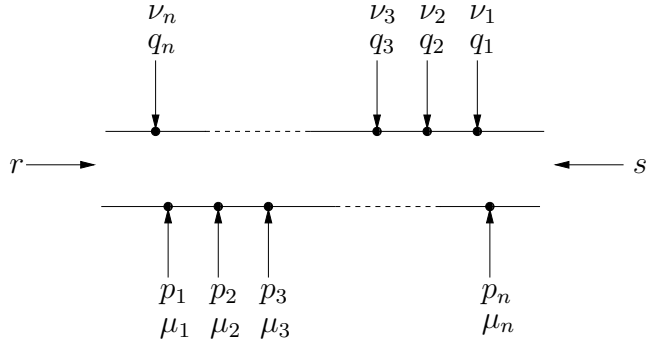}
  \end{picture}
\caption{Convention for wine labelling}
\label{fig:winelab}
\end{figure}

We introduce  the `wine' notation of refs \cite{mor:manerg,
mor:erg1, mor:erg2}.  Given a generic kernel $W(p^2/\Lam^2)$, we can
construct the wine ${}^{\phantom{x}i}_{{\bf x} l}\{W\}^k_{j{\bf y}}$.  This
functional is a gauge covariantization of the original kernel, and
incorporates parallel transport of the tensor representation.  If
$u^l_i(x)$ and $v^j_k(y)$  are $N \otimes \bar{N}$ representations of the
gauge group, we have
\be
{\bf u}\{W\}{\bf v}  := \int d^Dx d^Dy \, u^l_i(x) {}^{\phantom{x}i}_{{\bf x}
l}\{W\}^k_{j{\bf y}} v^j_k(y)
\ee
A wide choice exists for the exact form of the wine;  we shall only use
the following representation
\be
{\bf u}\{W\}{\bf v} = \str \dint \, u(x)W(-\nabla^2/\Lam^2)\cdot v(x),
\ee
where $\nabla_{\mu}$ is defined as in \eq{covder}.
The wine can be expanded in momentum space in terms of the gauge field [see
\eq{defa}] as 
\bea\label{winexp}
\lefteqn{{\bf u}\{W\}{\bf v} \equiv {\bf v}\{W\}{\bf u} = } \nonumber \\
& & \sum_{m,n=0}^{\infty}
\int d^D\!r \, d^D\!s \, d^D p_1\ldots d^D\!p_n \, d^D\!q_1\ldots d^D\!q_n
\, W_{\mu_1\ldots \mu_n,\nu_1\ldots \nu_m}(p_1,\ldots, p_n;q_1,\ldots,
q_n;r,s) \hspace{0.01cm} \nonumber \\
& & \hspace{2.5cm} \str \, [{\bf u}(r) \A _{\mu_1}(p_1)\ldots \A
_{\mu_n}(p_n){\bf v}(s) \A _{\nu_1}(q_1)\ldots \A _{\nu_m}(q_m)].
\eea
Such an expansion is represented graphically in figure \ref{fig:winexp},
with the labelling scheme shown in figure \ref{fig:winelab}.

We use the following shorthand to reduce the plethora of arguments
indices, commas and semi-colons. For the $n=0$ case of \eq{winexp}, we
replace the second string of $\A$ fields by the identity and define   
\be
W_{\mu_1\ldots \mu_n}(p_1,\ldots, p_n;r,s) \equiv
W_{\mu_1\ldots \mu_n,}(p_1,\ldots, p_n;\,;r,s),
\ee
while for the $m=n=0$ case we regain the original kernel
\be
W_p \equiv W_,(\,;\,;p,-p) = W(p^2/\Lam^2).
\ee

Evidently from the definition of a wine we have the exchange identity
\be\label{exid}
W_{\mu_1\ldots \mu_n,\nu_1\ldots \nu_m}(p_1,\ldots, p_n;q_1,\ldots, q_n;r,s)
=
W_{\nu_1\ldots \nu_m, \mu_1\ldots \mu_n}(q_1,\ldots, q_n;p_1,\ldots,p_n;s,r). 
\ee
Furthermore, charge conjugation invariance (arising from the symmetry $\Amu
\leftrightarrow - \Amu^{T}$) implies
\bea
\lefteqn{
W_{\mu_1\ldots \mu_n,\nu_1\ldots \nu_m}(p_1,\ldots, p_n;q_1,\ldots, q_n;r,s)
} 
\nonumber \\
& & \hspace{3cm}=
(-1)^{n+m}
W_{\mu_n\ldots \mu_1,\nu_m\ldots \nu_1}(p_n,\ldots, p_1;q_m,\ldots ,q_1;s,r).
\label{ccinv}
\eea

\chapter{Some Feynman rules for \bSUNN gauge theory}\label{sec:feyn}

Some of the Feynman rules for the broken action contained in \eq{bigact}
are given 
here. The Feynman rules were derived as follows: each rule
is the sum of all possible ways of assigning the relevant 
fields to points but maintaining the order within supertraces.  This
means that when it comes to calculating diagrams, care has to be taken
to ensure that all possible combinatorics and topologies are taken into
account.  
The following short hand is employed:
\bea
b^{\alpha\beta} &\equiv& {1\over 2}\,(h^{\alpha\beta} +
h^{\beta\alpha}),  \\ 
f^{\alpha\beta} &\equiv& {1\over 2}\,(h^{\alpha\beta} -
h^{\beta\alpha}), 
\eea
as well as the wine notation described in Appendix \ref{sec:wine}.
\newpage

The $\A$ propagator is 

\SetScale{1.0}
\begin{picture}(120,35)(-10,-15)
\Gluon(20,12)(100,12){4}{5}
\LongArrow(30,3)(90,3)
\put(0,12){${\cal A}^{\alpha}_{\mu}$}
\put(102,12){${\cal A}^{\beta}_{\nu}$}
\put(60,-3){$p$}
\end{picture}
\vspace{-.7cm}
\bea\label{Aprop}
\lefteqn{=
b^{\alpha\beta} \left \{ {g_{\mu\nu} \over {\c_pp^2}}
+ {p_{\mu}p_{\nu} \over {p^4}} {(\c_p - \xi\chat_p) \over
{\xi\c_p\chat_p}} \right \}   \hspace{4cm}}
\nonumber \\
& & 
+ f^{\alpha\beta} \left \{ {g_{\mu\nu} \over {\c_pp^2+4\Lam^2\ctil_p}}
+  {p_{\mu}p_{\nu} \over 
{(\c_pp^2+4\Lam^2\ctil_p)}}{(\c_p - \xi\chat_p) \over {(\xi\chat_pp^2+4\Lam^2\ctil_p)}}
\right\}. \hspace{1cm} 
\eea

In double index  notation we find
\bea
\lefteqn{\Big< (\A_{\mu})^i_{\gap j}(p)\, (\A_{\nu})^k_{\gap l}(-p) \Big>}
\nonumber \\[2mm]
& &=
{1 \over 4}\Bigg[ \left\{\del^i_{\gap l}\,(\sig3)^k_{\gap j}  + 
(\sig3)^i_{\gap l}\,\del^k_{\gap j} - {1 \over N}\left[\del^i_{\gap
j}\,(\sig3)^k_{\gap l} + (\sig3)^i_{\gap j}\del\,^k_{\gap l}\right]\right\}
\nonumber \\ & & \hspace{7cm}
\times \left \{ {g_{\mu\nu} \over {\c_pp^2}}
+ {p_{\mu}p_{\nu} \over {p^4}} {(\c_p - \xi\chat_p) \over
{\xi\c_p\chat_p}} \right \} \Bigg] 
\nonumber \\[2mm]
& & +{1 \over 4}\Bigg[
\left\{ \del^i_{\gap l}\,(\sig3)^k_{\gap j} - (\sig3)^i_{\gap
l}\,\del^k_{\gap j}\right\}  
\nonumber \\ & & \hspace{2cm} \times 
\left \{ {g_{\mu\nu} \over
{\c_pp^2+4\Lam^2\ctil_p}} + 
{p_{\mu}p_{\nu} \over  
{(\c_pp^2+4\Lam^2\ctil_p)}}{(\c_p - \xi\chat_p) \over
{(\xi\chat_pp^2+4\Lam^2\ctil_p)}} 
\right\} \Bigg].\hspace{1cm} 
\eea

\vspace{1cm}

\noindent
The $\C$ propagator is given by 
\be\label{Cprop}
\Big< \C^i_{\gap j}(p) \, \C^k_{\gap l}(-p) \Big>= {\tilde{c}_p \over
p^2}\del^i_{\gap l}\,(\sig3)^k_{\gap j} + {2 \xi \over \Lam^2}
{\tilde{c}^2_p  \over \hat{c}_p} \left( \del^i_{\gap l}\,(\sig3)^k_{\gap j}
- (\sig3)^i_{\gap l}\,\del^k_{\gap j} \right).
\ee

\vspace{0.5cm}

\noindent
The superghost propagator is found to be

\be\label{Etaprop}
\begin{array} {ll}
\begin{picture}(120,20)(10,5)
\DashLine(20,12)(100,12){5}
\LongArrow(30,3)(90,3)
\put(0,12){$\eta^{\alpha}$}
\put(104,12){$\bar{\eta}^{\beta}$}
\put(60,-3){$p$}
\end{picture}
\label{fig:Etaprop}
&
{\ds
=2 \left[ {b^{\alpha\beta} \over \chat_p \tilde{c}_p \, p^2} + {f^{\alpha\beta} \over
{\chat_p \tilde{c}_p \, p^2 + 4\Lam^2 
\xi^{-1} }} \right].  
}
\end{array}
\ee
Or equivalently
\bea
\Big<\eta^i_{\gap j}(p) \, \bar{\eta}^k_{\gap l}(-p) \Big> 
&=&{1 \over 2} \left[ \left\{\del^i_{\gap l}\,(\sig3)^k_{\gap j} +
(\sig3)^i_{\gap l}\,\del^k_{\gap j} - {1 \over N}\left[\del^i_{\gap
j}\,(\sig3)^k_{\gap l} + (\sig3)^i_{\gap j}\,\del^k_{\gap l}\right]\right\}  
{1 \over \chat_p \tilde{c}_p p^2} 
\right.  \nonumber \\ 
&\ph{=} & \hspace{2cm}
+ \left.
\Bigl\{\del^i_{\gap l}\,(\sig3)^k_{\gap j} - (\sig3)^i_{\gap l}\,\del^k_{\gap
j}\Bigr\}{1 \over {\chat_p \tilde{c}_pp^2  + 4\Lam^2 \xi^{-1}}}
\right]. 
\eea

\vspace{1cm}

\noindent The pure $\A^3$ interaction was found to have the Feynman rule

\vspace{0.5cm}

\be\label{Athree}
\begin{picture}(300,105)(45,-5)
\Gluon(20,50)(70,50){4}{4}
\Gluon(70,50)(95,93){4}{4}
\Gluon(70,50)(95,7){4}{4}
\LongArrow(30,43)(60,43)
\LongArrow(84,89)(69,63)
\LongArrow(97,18)(82,45)
\put(0,50){${\cal A}_{\mu}^{\alpha}$}
\put(93,95){${\cal A}_{\lam}^{\gam}$}
\put(93,1){${\cal A}_{\nu}^{\beta}$}
\put(45,35){$p$}
\put(69,75){$r$}
\put(93,33){$q$}
\put(100,60){$\ds
= 2g \Big[  
\c_{p}(p_{\nu}\del_{\lam\mu}-p_{\lam}\del_{\mu\nu}) 
+  
\c_{\nu}(q;r,p)(p_\lam r_\mu-p.r\del_{\lam\mu})  
$}
\put(120,40){$\ds
+\,\Lam^2\ctil_\nu(q;r,p)\,\del_{\mu\lam}   
+ \,2 \mathrm{\ cycles\ of\ } (p_\mu,q_\nu,r_\lam) 
\Big]  
$}
\put(280,20){$\ds
\times\,\str(S^{\alpha}S^{\beta}S^{\gam}). 
$} 
\end{picture}
\label{fig:Athree}
\ee

\vspace{0.5cm}

\noindent 
Three point interactions with inserted  $\sig3$s (arising from the symmetry
breaking) also occur.  The positioning of a $\sig3$ is indicated by a
wedge.

\be\label{Athreesig}
\begin{picture}(300,105)(50,5)
\Gluon(20,50)(70,50){4}{4}
\Gluon(95,93)(70,50){4}{4}
\Gluon(70,50)(95,7){4}{4}
\LongArrow(30,40)(60,40)
\LongArrow(84,89)(69,63)
\LongArrow(97,18)(82,45)
\Line(70,50)(66,60)
\Line(70,50)(62,56)
\Line(62,56)(66,60)
\Line(70,50)(66,40)
\Line(70,50)(62,44)
\Line(66,40)(62,44)
\put(0,50){${\cal A}_{\mu}^{\alpha}$}
\put(93,95){${\cal A}_{\lam}^{\gamma}$}
\put(93,1){${\cal A}_{\nu}^{\beta}$} 
\put(45,32){$p$}
\put(69,75){$r$}
\put(93,33){$q$}
\put(100,60){$\ds
= 2 g\Lam^2  \Big[ \ctil_\nu(q;p,r)\,\del_{\mu\lam} + 
\ctil_\mu(p;r,q)\,\del_{\nu\lam} +  \ctil_\lam(r;q,p)\,\del_{\mu\nu} \Big]
$}
\put(180,40){$\ds
\times \, \str(\sig3S^{\alpha}\sig3S^{\beta}S^{\gam}).
$}
\end{picture}
\label{fig:Athreesig}
\ee

(We refrain from including the double index  versions of these vertices.)

\newpage
Some more three point Feynman rules are given below. 

\vspace{-1cm}

\bea
\lefteqn{\Big< (\Amu)^i_{\gap j}(p) \, \C^k_{\gap l}(q) \, \C^m_{\gap n}(r)
\Big>} \nonumber \\ 
& & \hspace{1cm}
= {g \over 4}\left[ \ctil_rr_\mu - \ctil_qq_\mu + \ctil_\mu(p;r,q)
\,q\cdot 
r \right]  \nonumber \\
& & \hspace{3cm}
\times
(\sig3)^k_{\gap n} \left[ \del^i_{\gap l} (\sig3)^m_{\gap j} - {1 \over 2N}
\left\{ (\sig3)^i_{\gap j} \del^m_{\gap l} + \del^i_{\gap j} (\sig3)^m_{\gap
l}\right\} \right].
\eea


\be\label{Aetaetabar}
\begin{picture}(120,105)(150,-5)
\Gluon(20,50)(70,50){4}{4}
\DashArrowLine(70,50)(95,93){5}
\DashArrowLine(95,7)(70,50){5}
\LongArrow(30,43)(60,43)
\LongArrow(84,89)(69,63)
\LongArrow(97,18)(82,45)
\put(0,50){${\cal A}_{\mu}^A$}
\put(96,96){$\bar{\eta}^C$}
\put(96,0){$\eta^B$}
\put(45,35){$p$}
\put(69,75){$r$}
\put(93,33){$q$}
\put(110,50){$\ds 
= - g \chat_r \tilde{c}_r \,r_\mu \,\str(S^{\alpha}S^{\beta}S^{\gam})
$}
\end{picture}
\ee


\be\label{Aetabareta}
\begin{picture}(120,105)(150,-5)
\Gluon(20,50)(70,50){4}{4}
\DashArrowLine(95,93)(70,50){5}
\DashArrowLine(70,50)(95,7){5}
\LongArrow(30,43)(60,43)
\LongArrow(84,89)(69,63)
\LongArrow(97,18)(82,45)
\put(0,50){${\cal A}_{\mu}^A$}
\put(96,96){$\eta^C$}
\put(96,0){$\bar{\eta}^B$}
\put(45,35){$p$}
\put(69,75){$r$}
\put(93,33){$q$}
\put(110,50){$\ds 
=  g \chat_q \tilde{c}_q \,q_\mu \,\str(S^{\alpha}S^{\beta}S^{\gam})
$}
\end{picture}
\ee


\end{document}